\documentclass[a4paper,11pt]{article}
\usepackage{jheppub}
\usepackage{lineno}
\usepackage{bm}
\usepackage{placeins}
\usepackage[]{graphicx}
\usepackage{amsfonts}
\usepackage{float}
\usepackage{multirow}
\usepackage{color,amsmath,amssymb}
\usepackage{array}
\usepackage[utf8]{inputenc}
\usepackage{tikz-feynman}
\usepackage{makecell}
\usepackage{booktabs}

\usepackage{xcolor}

\usetikzlibrary{arrows.meta}
\usetikzlibrary{decorations.markings}

\usepackage[colorlinks=true, linkcolor=blue, citecolor=blue, urlcolor=blue]{hyperref}

\newcommand{\eg}{\textit{e.g.}}

\newcommand{\ie}{\textit{i.e.}}
\newcommand{\dd}{\mathrm{d}}

\newcommand{\gsim}{\gtrsim}
\newcommand{\barb}{\bar{b}}
\newcommand{\raa}{R_{\rm AA}}
\newcommand{\npart}{N_{\rm part}}

\newcommand{\Treg}{T_{\rm reg}}
\newcommand{\Tmelt}{T_{\rm melt}}
\newcommand{\Tavg}{T_{\rm avg}}
\newcommand{\Tf}{T_{\rm f}}
\newcommand{\Thad}{T_{\rm h}}
\newcommand{\tauform}{\tau_{\rm form}}

\newcommand{\pT}{p_T}

\newcommand{\bbb}{b\bar{b}}

\title{\boldmath Bottomonium transport in the sQGP at RHIC and the LHC}

\author[a,b]{Biaogang Wu,}
\author[b]{Jacob Boyd,}
\author[a]{Ralf Rapp}
\affiliation[a]{Cyclotron Institute and Department of Physics and Astronomy, Texas A$\&$M University, College Station, TX 77843-3366, United States}
\affiliation[b]{Department of Physics, Kent State University, Kent, OH 44242, United States}

\emailAdd{bwu8@kent.edu}
\emailAdd{jboyd29@kent.edu}
\emailAdd{rapp@comp.tamu.edu}

\abstract{
    Bottomonium transport is studied in heavy-ion collisions at RHIC and the LHC by implementing a kinetic rate equation into (3+1)D viscous hydrodynamic simulations of an expanding quark-gluon plasma (QGP). The two main transport parameters are the inelastic reaction rates and equilibrium limits for each individual bottomonium state, $Y$. The former are taken from the thermodynamic $T$-matrix formalism with recent constraints from lattice-QCD and including interference effects and in-medium binding energies, resulting in large rates characteristic of a strongly coupled QGP.
    The equilibrium limits are evaluated from pertinent in-medium bottomonium and bottom-quark masses. The calculation of observables includes a total of nine $Y$ states (up to 3$S$ and 2$P$) with a feed-down matrix estimated from vacuum branching fractions.
At the LHC, the large reaction rates rapidly suppress the initial population of excited states, rendering regeneration their main source even in rather peripheral collisions, while for the more strongly bound ground state $\Upsilon(1S)$, a significant primordial component survives in central collisions. On the other hand, at RHIC energies, regeneration is overall a smaller effect. Together with effects from nuclear absorption, this offers an explanation for the experimental observation that $\Upsilon(1S)$ production at RHIC and the LHC is of comparable magnitude despite the significantly higher temperatures reached in the QGP at the LHC.}

\begin{document}
\maketitle

\section{Introduction}
\label{sec:intro}
Ultra-relativistic heavy-ion collisions (URHICs) create extended volumes of deconfined quark-gluon plasma (QGP),
and the production of heavy quarkonia in such collisions is a well-established, albeit rather complex, probe of the partonic medium~\cite{Matsui:1986dk,Rapp:2008tf,Braun-Munzinger:2009dzl,Mocsy:2013syh,Zhou:2016vwq}.
While the systematics of charmonium production show a clear trend from a suppression-dominated regime at the SPS and RHIC to a relative enhancement at the LHC, bottomonia (which we generically refer to as $Y$ states) remain strongly suppressed even at the LHC, especially for the excited states. This has been interpreted as the originally proposed sequential suppression signature referring to a hierarchy in the various $Y$ dissociation temperatures. However, within kinetic transport models, the role of regeneration in the observed $Y$ yields is still controversial: While quantum transport approaches essentially attribute the observed production pattern to suppression reactions, semiclassical simulations vary in their predictions for the magnitude of the regeneration component~\cite{Andronic:2024oxz}. Even if there is only a single $b\bar b$ pair in the QGP fireball, the fact that the bottom quark and antiquark are produced in association implies that their density is constrained to a rather small sub-volume of the fireball, which notably increases the probability to (re-)form a bound state. This possibility is expected to be exacerbated in a strongly coupled QGP (sQGP), since (a) the $b$ and $\bar b$ have a small spatial diffusion coefficient and thus are slowed down compared to a vacuum environment~\cite{Young:2008he}, and (b) large collision rates of the $b$ quarks enhance the chemical reaction rates to (re-)form $Y$ states. 
Furthermore, given the low production rate of $Y$ states of approximately 0.1\% relative to the open $\bbb$ yield (compared with $\sim 1\%$ for charmonia), even a small regeneration probability can contribute very significantly to the observed $\raa$ of $Y$ states; \eg, if only one in $\sim$1000 $\bbb$ pairs produced in nucleus-nucleus (AA) collisions regenerates a $Y$ state, it contributes of order one to its $\raa$.

To date, a broad variety of transport models have been able to reproduce $Y$ data from the ALICE, ATLAS, and CMS collaborations in Pb-Pb collisions at the LHC~\cite{CMS:2018zza,ALICE:2018wzm,ALICE:2020wwx,ATLAS:2022exb,CMS:2023lfu}.
A somewhat surprising feature is found in Au-Au collisions at RHIC~\cite{STAR:2022rpk}, where the level of observed $\Upsilon(1S)$ suppression is very similar to that at the LHC, despite significantly lower initial temperatures at collision energies of 0.2\,TeV compared to $\sim$5\,TeV at the LHC. This has posed a challenge to theoretical models that rely solely on suppression mechanisms, and even modest regeneration components cannot resolve this tension~\cite{Du:2017qkv}.

In the present work, we expand on our previous study~\cite{Wu:2025lcj},
where we implemented recent calculations of nonperturbative bottomonium reaction rates into a viscous hydrodynamic evolution (aHydro). The reaction rates are adopted from thermodynamic $T$-matrix amplitudes whose input potential has been constrained by state-of-the-art Wilson line correlators (WLCs) computed in lattice QCD~\cite{Tang:2023tkm}. The constraints also encompass the lattice-QCD (lQCD) equation of state (EoS) which was computed from thermal-parton interactions using the same underlying potential in a self-consistent numerical iteration procedure using the Luttinger-Ward-Baym formalism~\cite{Liu:2017qah}. In this way, the microscopic quarkonium reaction rates are calculated in a QGP medium whose EoS, reproducing lQCD data,  governs the macroscopic dynamics of the hydrodynamic evolution. Key outcomes of the quantum many-body approach are large collisional widths of the partonic degrees of freedom (in excess of 0.5\,GeV for both thermal light partons and heavy quarks) which are at the core of the strong-coupling properties of the sQGP (such as its transport parameters).
The applications to basic $\raa$ observables for $\Upsilon(1S,2S,3S)$ states~\cite{Wu:2025lcj}, \ie, their centrality and transverse-momentum ($\pT$) dependence,  were carried out and led to fair agreement with available LHC data in Pb-Pb (5.02\,TeV) collisions, although some discrepancies at high $\pT$ were identified. Furthermore, while the rates and the hydrodynamic evolution did not involve new parameters, model uncertainties remained, \eg, in the calculation of the equilibrium limit (which is non-trivial for a strongly coupled system), the correlation volume (which simulates the aforementioned restriction of the effective volume explored by a $\bbb$ pair after its point-like production), and formation-time effects in the primordial bound-state evolution after initial hard production (where, in principle, a quantum-mechanical evolution of a wave packet should be considered).
The main finding of Ref.~\cite{Wu:2025lcj} was a large regeneration component for the $\Upsilon(1S)$ and, especially, the excited states in semi/-central collisions -- substantially higher than in any previous transport calculation,  but a direct consequence of the large reaction rates in the sQGP.

In the meantime, the same framework has been applied to proton--nucleus (pA) collisions~\cite{Thapa:2025jua}. Cold-nuclear matter (CNM) effects and the reduced space-time volume of a hydrodynamic medium lead to somewhat different bottomonium dynamics, although hot-matter effects with both suppression and regeneration remain considerable, especially for excited states, and yield a fair agreement with data.

Notable advances of the present paper over our earlier work~\cite{Wu:2025lcj}
include a more detailed discussion of the suppression and regeneration mechanisms, more comprehensive comparisons to experimental data, and an extension to RHIC energies, where we address the aforementioned tension in a combined interpretation with LHC data.

Our paper is organized as follows.
In Sec.~\ref{sec:supp} we introduce the kinetic rate equation and focus on bottomonium suppression in a hydrodynamic medium evolution.
In Sec.~\ref{sec:reg} we detail the solution of the kinetic rate equation for bottomonium regeneration within hydrodynamics incorporating its spatial temperature profiles.
In Sec.~\ref{sec:exp} we first examine the time evolution of the bottomonium yields following from the trajectories sampled in the hydrodynamic fireball and then turn to comparisons to experimental data at the LHC and RHIC, including a benchmark comparison using perturbative reaction rates.
We summarize our main findings and discuss future prospects in Sec.~\ref{sec:concl}.

\section{Bottomonium suppression in hydrodynamics}
\label{sec:supp}
In  Sec.~\ref{ssec:rate_eq} we first recall the kinetic rate equation and its transport parameters and then focus on the hot-medium suppression of $Y$ states,
thereby defining the appropriate transformation from the lab into the rest frame of the hydrodynamic medium.
In  Sec.~\ref{ssec:hydro} we briefly summarize the main features of the (3+1)D anisotropic hydrodynamics that we employ to sample $Y$ trajectories and record local temperatures along them (which are subsequently used to compute $Y$ suppression).
In Sec.~\ref{ssec:form} we recall our schematic implementation of formation time effects in the evolution from the hard (point-like) $\bbb$ production to the fully formed bound state. In Sec.~\ref{ssec:cnm} we summarize our treatment of cold-nuclear-matter (CNM) effects, most notably nuclear shadowing, while Sec.~\ref{ssec:feeddown} describes late-time
feed-down from excited states needed to assess the ``prompt'' production yields as measured in experiment.

\subsection{Rate equation}
\label{ssec:rate_eq}
The kinetic rate equation for the number, $N_{Y}(\tau)$, of a bottomonium state, $Y$, at time $\tau$ reads~\cite{Zhao:2010nk}
\begin{equation}
   \begin{aligned}
       \frac{\dd N_{Y}(\tau)}{\dd\tau} = -\Gamma_{Y}[T(\tau)] \ \Bigl[N_{Y}(\tau) - N_{Y}^{\rm eq}[T(\tau)]\Bigr] \,,
   \end{aligned}
\label{eq:rate_eq}
\end{equation}
where $N_{Y}^{\rm eq}\left(T(\tau)\right)$ is
the equilibrium limit,
and $\Gamma_{Y}[T(\tau)]$ is the reaction rate at the corresponding temperature.
Focusing on the suppression part for now, the solution of the rate equation \eqref{eq:rate_eq} for each state reads
\begin{equation}
N_{Y}(\tau)
= N_{Y}(\tau_0)\,
\exp\!\left[-\!\!\int_{\tau_0}^{\tau} \Gamma_{Y} [T(\tau')] \dd\tau'\right]\,,
\label{eq:rate_eq_solution}
\end{equation}
which characterizes the survival probability against medium‐induced breakup. Here, $\tau_0$ denotes the aHydro initialization time at which the hydrodynamic evolution starts (typically $\tau_0$=0.25\,fm$/c$).
In the present work we focus on the transport in the QGP, which we carry out for all considered $Y$ states down to a decoupling (or hadronization) temperature of $\Thad$=170\,MeV. Little is currently known about $Y$ reaction rates in the hadronic phase, and neglecting them should be a good approximation up to the $3S$ and $2P$ states whose vacuum binding energies are larger than $\Thad$. 

The reaction rates depend on the in-medium $Y$ binding energies and radii, as well as on the masses of the bottom quarks and their coupling to the QGP medium. These quantities are obtained from a self-consistent, non-perturbative $T$-matrix calculation where the input potential is constrained by lQCD results for Wilson line correlators~\cite{Tang:2023tkm}.
Specifically, an analysis of the pertinent $Y$ spectral functions in the complex energy plane has been carried out to quantitatively extract the masses and widths from the pole positions, as well as the melting temperatures from the disappearance of the poles~\cite{Tang:2025ypa,Wu:2025hlf}.
Since in Eq.~\eqref{eq:rate_eq_solution} the temperature-dependent quarkonium reaction rates are defined in the rest frame of the medium,
one has to boost the bottomonium four-momentum from the lab frame into the thermal frame.  
\begin{figure}[t]
    \centering
    \includegraphics[width=0.7\textwidth]{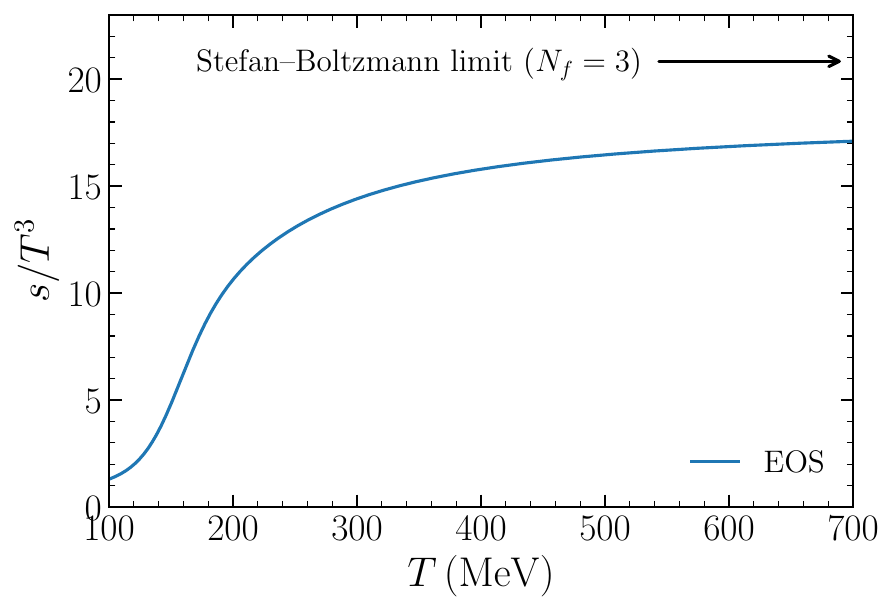}
    \caption{Normalized entropy density, $s/T^3$, as a function of temperature, fit to lattice QCD~\cite{HotQCD:2014kol}.
    }
    \label{fig:s}
\end{figure}
We denote the four-momentum of the bottomonium pair in the laboratory frame by
\begin{equation}
p^\mu_{\rm lab}
= \bigl(p^0,\mathbf{p}\bigr)
= \bigl(m_T\cosh y,\;\pT\cos\phi,\;\pT\sin\phi,\;m_T\sinh y\bigr) \ ,
\end{equation}
with transverse mass $m_T=\sqrt{m_Y^2+\pT^2}$, rapidity $y$, and azimuthal angle $\phi$. Furthermore, denoting the four-velocity of the hydro cell at the position 
of the pair by $ u^\mu = \gamma\,(1,\mathbf{v})$, we can obtain the momentum in the rest frame as 
\begin{equation}
p^\mu_{\rm rest}
= \Lambda^\mu{}_{\!\nu}\,p^\nu_{\rm lab}\,,
\end{equation}
with a Lorentz-boost matrix,
\begin{equation}
\Lambda =
\begin{pmatrix}
\gamma & -\gamma\,\beta_x & -\gamma\,\beta_y & -\gamma\,\beta_z\\
-\gamma\,\beta_x & 1 + (\gamma-1)\,\frac{\beta_x^2}{\beta^2}
                 & (\gamma-1)\,\frac{\beta_x\beta_y}{\beta^2}
                 & (\gamma-1)\,\frac{\beta_x\beta_z}{\beta^2}\\
-\gamma\,\beta_y & (\gamma-1)\,\frac{\beta_y\beta_x}{\beta^2}
                 & 1 + (\gamma-1)\,\frac{\beta_y^2}{\beta^2}
                 & (\gamma-1)\,\frac{\beta_y\beta_z}{\beta^2}\\
-\gamma\,\beta_z & (\gamma-1)\,\frac{\beta_z\beta_x}{\beta^2}
                 & (\gamma-1)\,\frac{\beta_z\beta_y}{\beta^2}
                 & 1 + (\gamma-1)\,\frac{\beta_z^2}{\beta^2}
\end{pmatrix},
\end{equation}
For simplicity, we set $\beta_z=0$ in the hydrodynamic model. 

\begin{figure}[t]
   \centering
   \includegraphics[width=1\textwidth]{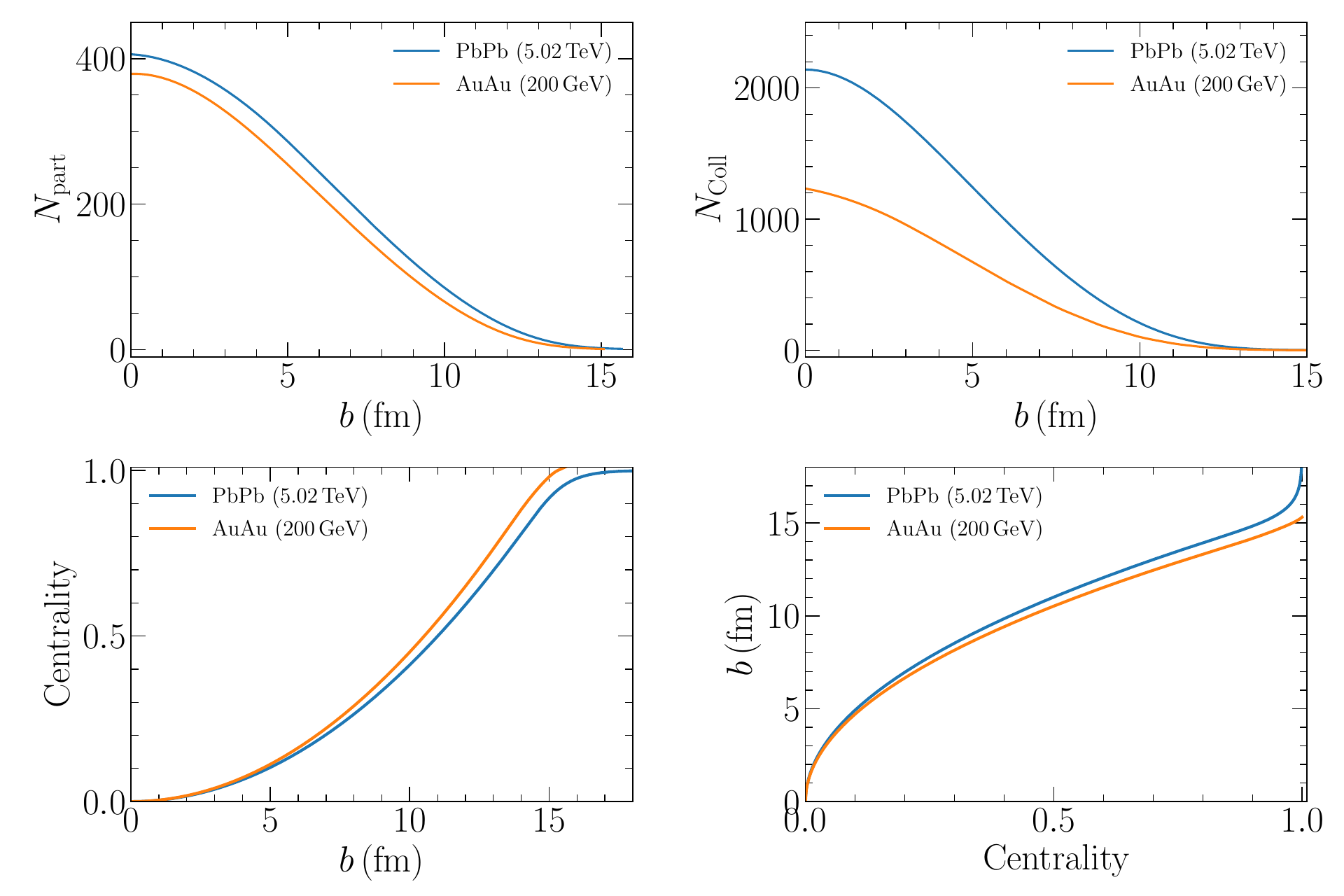 }
   \caption{The number of participants (upper left) and binary collisions (upper right), and the centrality percentile (lower left) as a function of impact parameter, and the impact parameter (lower right) as a function of centrality for Pb-Pb ($5.02$\,TeV) (blue curves) and Au-Au ($200$\,GeV) collisions (orange curves).
}
   \label{fig:centrality}
\end{figure}

\subsection{Hydrodynamic evolution and bottomonium trajectories}
\label{ssec:hydro}
\begin{figure}[t]
  \centering
  \includegraphics[width=0.48\textwidth]{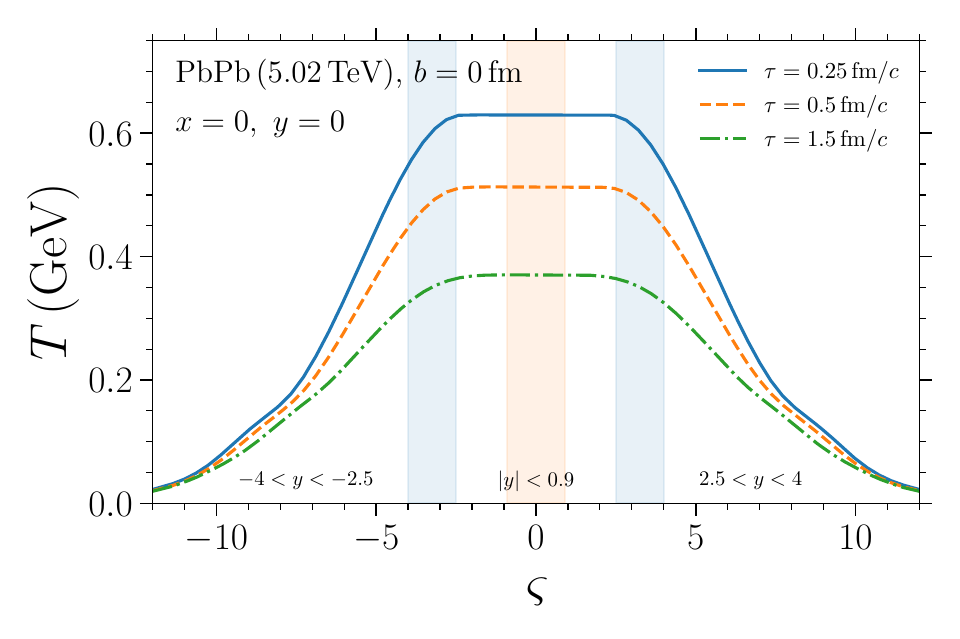}
  \includegraphics[width=0.48\textwidth]{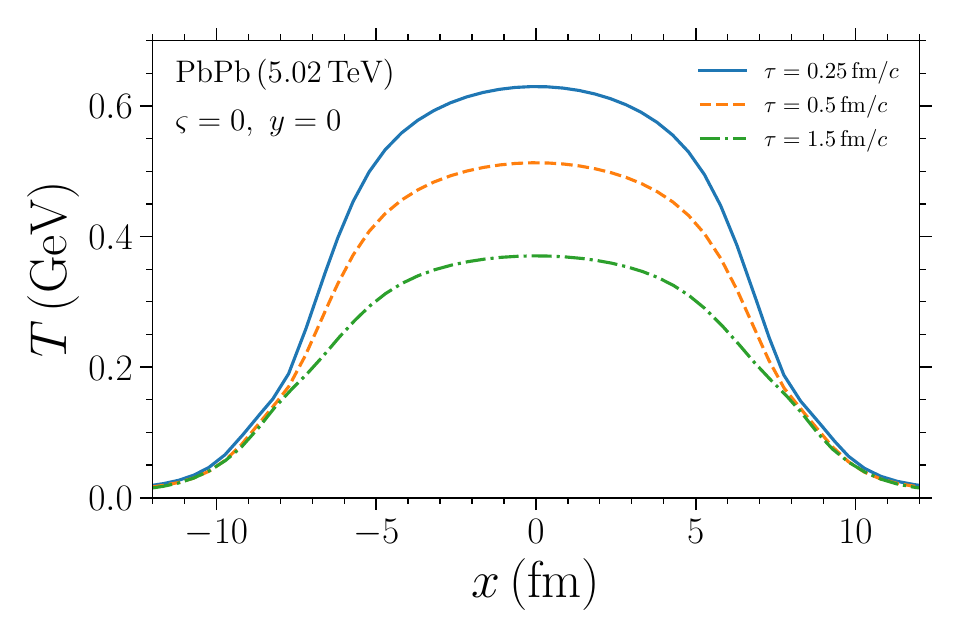}
  \includegraphics[width=0.48\textwidth]{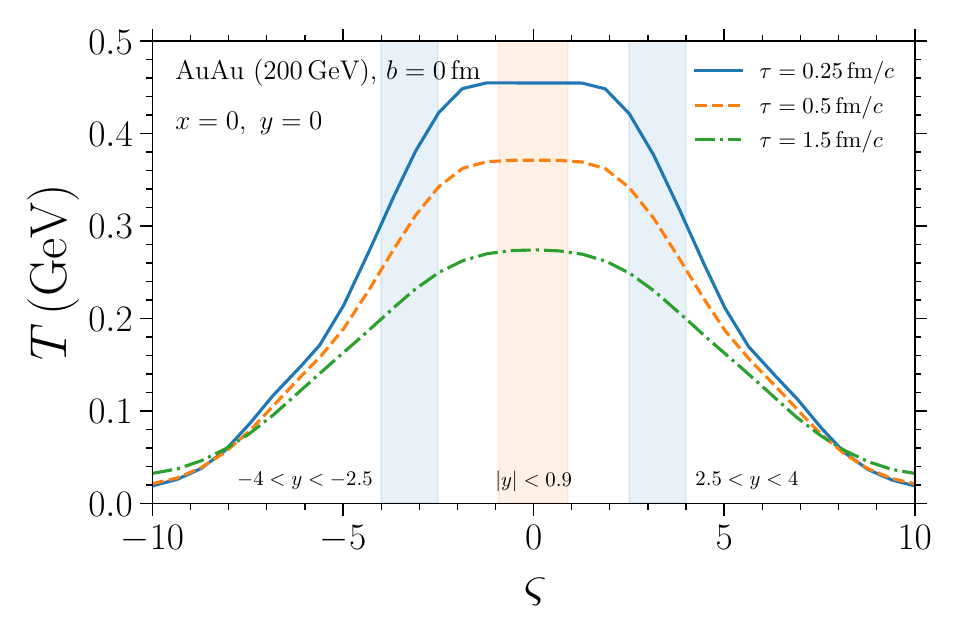}
  \includegraphics[width=0.48\textwidth]{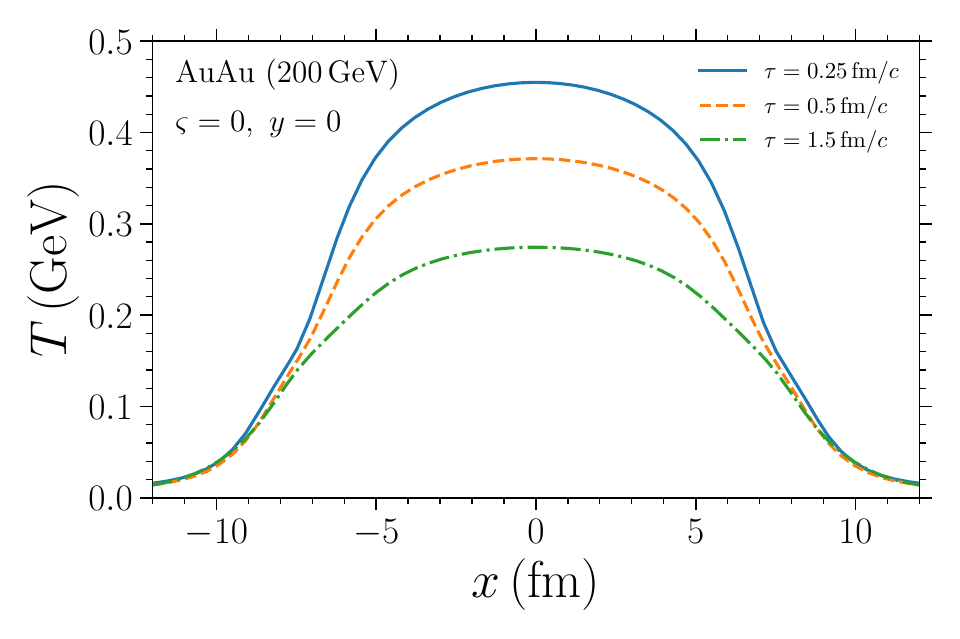}
  \caption{
      Evolution of temperature distributions obtained using aHydro as a function of spatial rapidity $\varsigma$ at fixed transverse position $x = 0$ and $y = 0$ (left panels) and as a function of transverse position $x$ (right panels) for central Pb-Pb ($5.02$\,TeV) (upper panels) and Au-Au ($200$\,GeV) (lower panels) collisions. The solid, dashed, and dot-dashed curves represent time snapshots at proper times $\tau = 0.25$\,fm$/c$, 0.5\,fm$/c$, and 1.5\,fm$/c$, respectively.
  }
  \label{fig:temp-evo-rap}
\end{figure}

We utilize (3+1)D anisotropic hydrodynamics (aHydro) to describe the expansion of the QGP liquid in AA collisions, with parameters tuned to reproduce experimentally observed soft-hadron spectra, charged-hadron multiplicities, elliptic flow,
and Hanbury-Brown-Twiss (HBT) radii at both RHIC and LHC
energies~\cite{Almaalol:2018gjh,Alqahtani:2020paa}.
The simulations employ an EoS consistent with lQCD computations~\cite{Bazavov:2013txa}, with a temperature-dependent entropy density depicted in  Fig.~\ref{fig:s}, as well as second- and higher-order transport coefficients (including non-conformal effects such as bulk viscosity), which are self-consistently computed within a quasiparticle model.
The latter's parameters are fixed to obtain a temperature-independent shear viscosity to entropy density ratio of 
$\eta/s = 0.179$, as part of the fit to soft observables.
A smooth optical Glauber initial condition is used for the initial spatial energy density profile of the QGP as a function of the collision impact parameter; see Fig.~\ref{fig:centrality} for the resulting number of participants, numbers of binary collisions, and centrality percentile as functions of the impact parameter, as well as the impact parameter versus centrality for Pb-Pb ($5.02$\,TeV) and Au-Au ($200$\,GeV) collisions.
Figure~\ref{fig:temp-evo-rap} shows snapshots, at various proper times of the QGP, of the temperature distribution for central collisions at both RHIC and the LHC as a function of spatial rapidity, $\varsigma$, and transverse coordinate.
At 5.02\,TeV, the fits of aHydro to experimental data yield an initial central temperature of $T_0 = 630\,\text{MeV}$ at an initial longitudinal proper time of $0.25\,\text{fm}/c$.

\begin{figure}[t]
  \centering
  \includegraphics[width=0.48\textwidth]{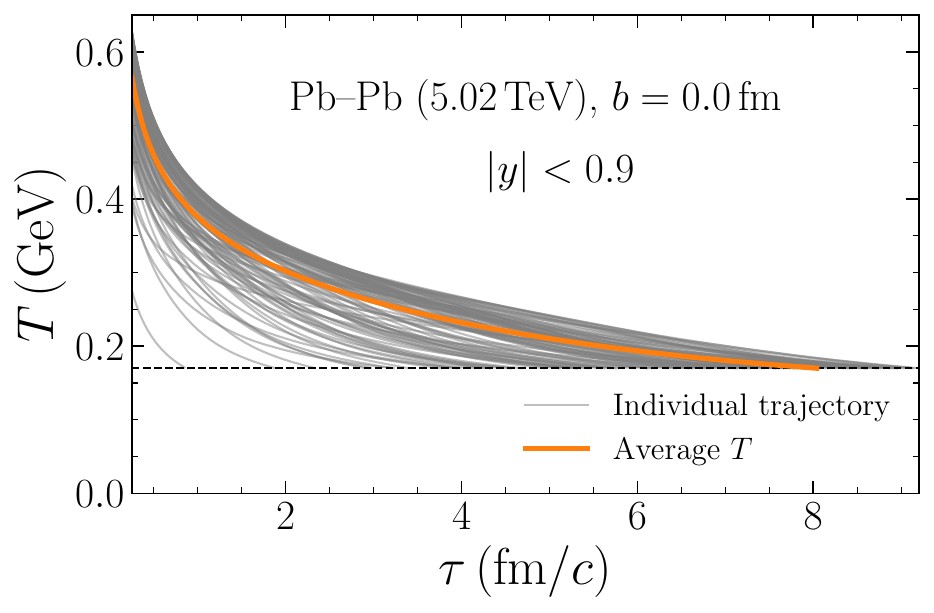}
  \includegraphics[width=0.48\textwidth]{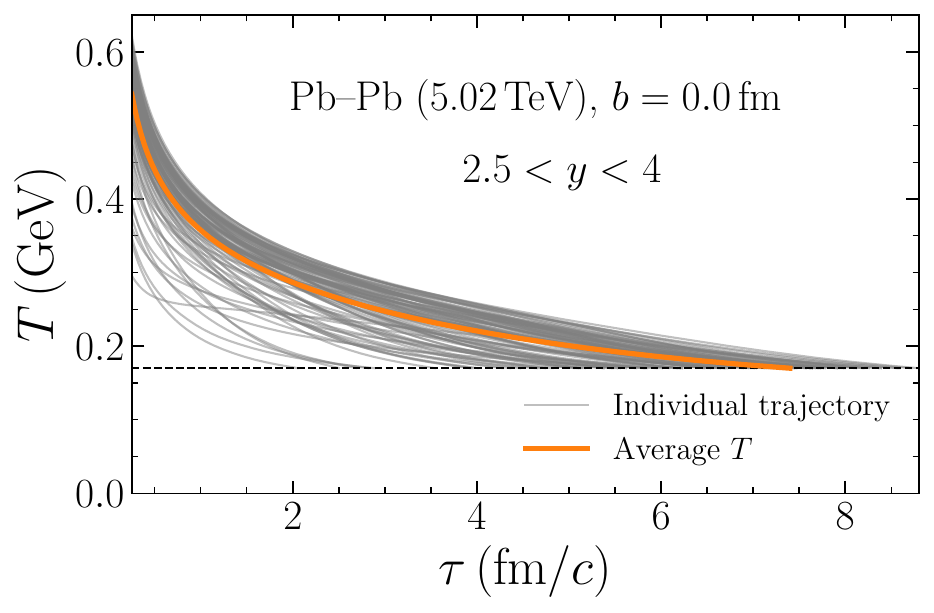}
  \includegraphics[width=0.48\textwidth]{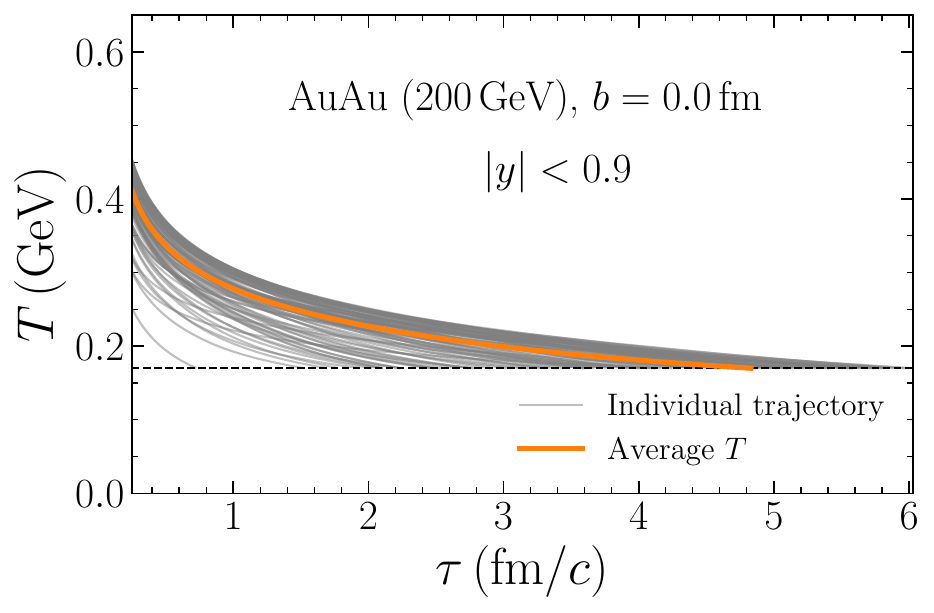}
  \includegraphics[width=0.48\textwidth]{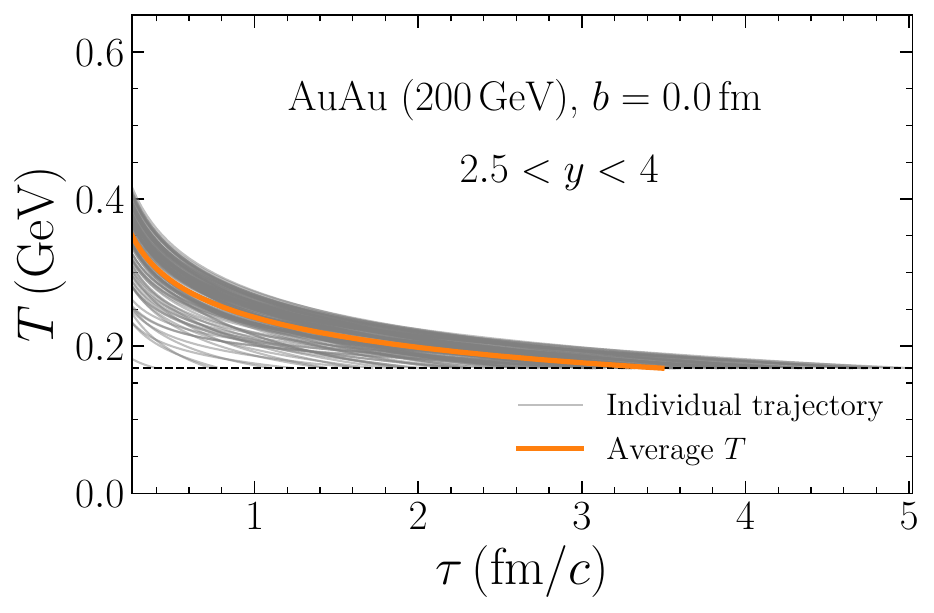}
  \caption{Time evolution of temperature along individual $Y$ trajectories (gray lines) and their average (orange line) for maximally central Pb-Pb (5.02\,TeV) (upper panels) and Au-Au (200\,GeV) collisions (lower panels)
  at mid‐(left) and forward rapidity (right).}
  \label{fig:traj-temp}
\end{figure}
Turning now to the trajectories of bottomonia, following Ref.~\cite{Strickland:2023nfm},
we sample their 3D motion for a given centrality according to their momentum distribution (in both $\pT$ and rapidity) and record the encountered temperature along each path within the aHydro medium; we then compute the survival probabilities by averaging over a large ensemble.
Due to the large masses of the $Y$ mesons, we neglect the effects of diffusion (elastic scattering) and assume that they travel along straight-line trajectories.
We describe bottomonium trajectories in Milne coordinates $(\tau,\vec{x}_\perp,\varsigma)$, where $\tau$ is the proper time, $\vec{x}_\perp$ the transverse position, and $\varsigma$ the spatial rapidity.
We initialize each trajectory in Cartesian space, with the position at lab time $t$ given by
\begin{equation}
\vec{x}(t) = \vec{x}_0 + \vec{v}\,(t - t_0)\,,\quad \vec{v}=\frac{\vec{p}_0}{E}\,,
\end{equation}
where $\vec{x}_{0}$ is the initial position, $\vec{p}_{0}$  the initial momentum, and $E = \sqrt{\vec{p}_{0}^2 + m_Y^2}$ the $Y$'s energy,
and then transform $(t,\vec{x})\to(\tau,\vec{x}_\perp,\varsigma)$ via
\begin{equation}
   t=\tau\cosh\varsigma\,,\quad z=\tau\sinh\varsigma \ .
\end{equation}
Noting that $v_z = \tanh y$ (with $y$ being the momentum rapidity), one obtains
\begin{eqnarray}
\vec{x}_\perp &=& \vec{x}_{\perp,0} + \vec{v}_\perp \left(\tau \cosh\varsigma - \tau_{0} \cosh\varsigma_{0}\right) \,, \nonumber \\
\tau \sinh\varsigma &=& \tau_{0} \sinh\varsigma_{0} + \tanh y \, \left(\tau \cosh\varsigma - \tau_{0} \cosh\varsigma_{0}\right) \,,
\end{eqnarray}
where $\varsigma_0$ is the spatial rapidity at $\tau_0$ and $\vec{v}_\perp$ is the transverse velocity.
The last equation can be solved for $\varsigma$, yielding
\begin{equation}
\varsigma = y + \log({\cal G}) \,,
\end{equation}
where
\begin{equation}
{\cal G} \equiv \sqrt{1 + \frac{\sinh^2(\varsigma_0 - y)}{\bar\tau^2}} + \frac{\sinh(\varsigma_0 - y)}{\bar\tau} \,,
\end{equation}
and
\begin{equation}
\bar\tau \equiv \frac{\tau}{\tau_0} \,.
\end{equation}
As $\bar\tau\to\infty$, one finds ${\cal G}\to1$, while for $\bar\tau\to1$, we have ${\cal G}\to e^{\,(\varsigma_{0}-y)}$.
Moreover, if all production occurs at $\tau_{0}\to0$, then $\varsigma=\varsigma_{0}=y$ at all times.

\begin{figure}[t]
   \centering
\includegraphics[width=0.75\linewidth]{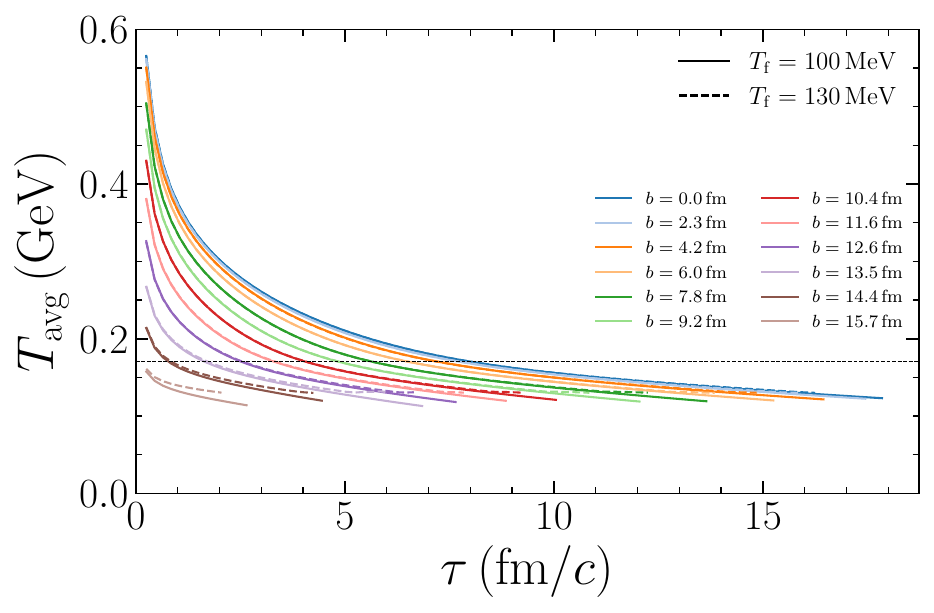}
\caption{Time evolution of the average temperature, $\Tavg$, of $Y$ trajectories in the hydrodynamic medium for different impact parameters, $b$, at LHC energies. The solid and dashed curves use different terminal temperatures of $\Tf=100$ and 130\,MeV, respectively. 
The horizontal dotted line indicates $\Thad=170$\,MeV.}
\label{fig:Tcut}
\end{figure}

The initial transverse production points are taken to scale with the binary overlap density, $N_{\rm AA}^{\rm bin}(x,y)$, from the Glauber model.
The $p_{T}$ spectrum for all states is parameterized as~\cite{Du:2017qkv}
\begin{equation}
   \frac{\dd N^{pp}_{Y}}{\dd \pT^2}=\frac{N}{\left(1+\left({\pT}/{A}\right)^2\right)^n} \,,
   \label{p_T}
\end{equation}
where the three parameters are $N=0.017$, $A=4.9$ and $n=2.3$ at both mid- and forward rapidity for the $\Upsilon(1S)$, and $N=0.012$, $A=5.9$ and $n=2.3$ for all excited states at the LHC.
At RHIC, the three parameters are $N=0.023$, $A=5.3$ and $n=3.0$ for all states.
The azimuthal angle $\phi$ is drawn uniformly from $\left[0,2\pi\right)$.
The initial rapidity, $\varsigma_{0}=y_{0}$,
is sampled from a Gaussian of width 2.3, centered at $y_{0}=0$.
As our default, we terminate the trajectories when the temperature reaches $\Tf=100$\,MeV and average over 65536 samples per centrality bin (weighted over the $Y$ momentum distributions). 

A typical sample and the pertinent average temperatures for central AA collisions at RHIC and the LHC are shown in Fig.~\ref{fig:traj-temp}.  
\begin{figure}[t]
  \centering
  \includegraphics[width=0.48\textwidth]{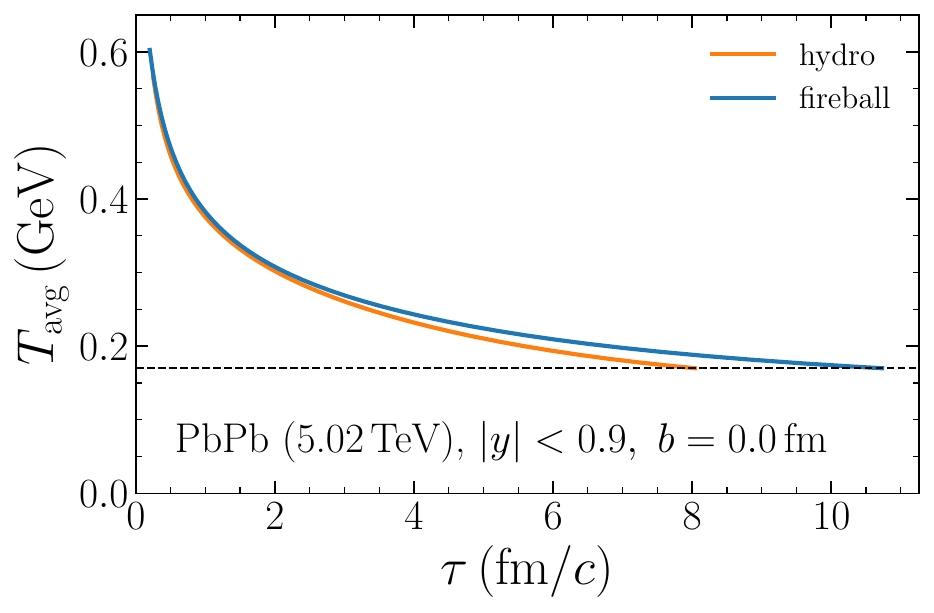}
  \includegraphics[width=0.48\textwidth]{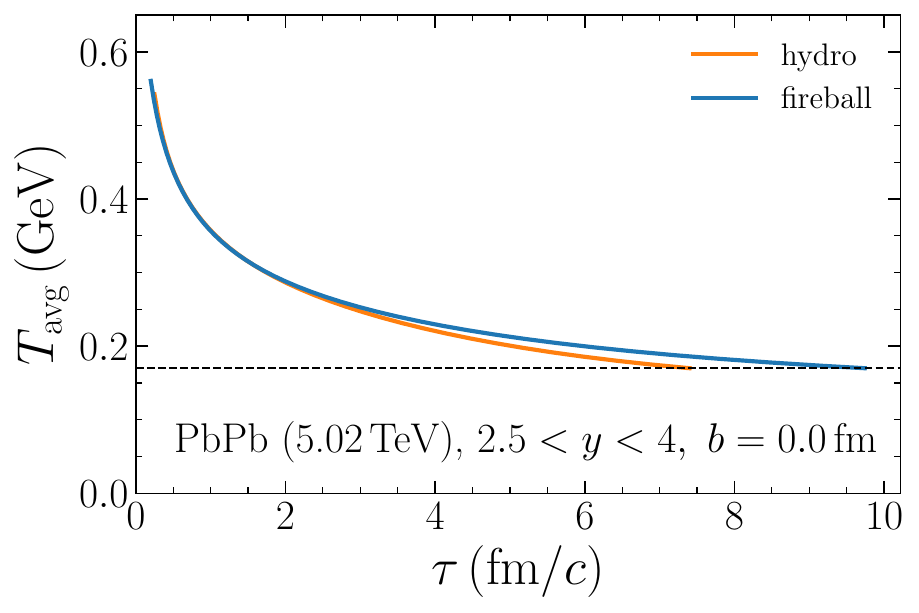}
  \includegraphics[width=0.48\textwidth]{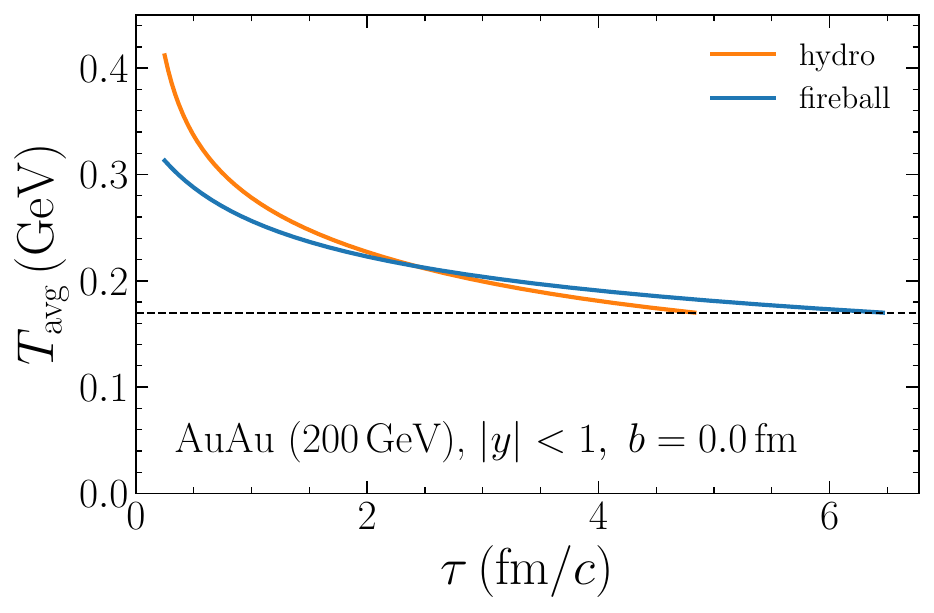}
  \includegraphics[width=0.48\textwidth]{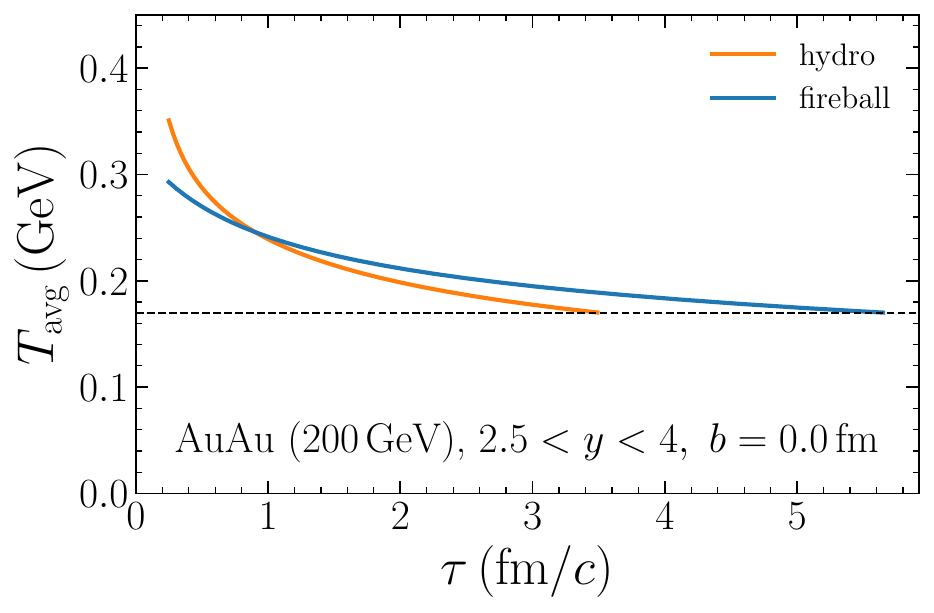}
  \caption{Time evolution of the trajectory-averaged temperature, $\Tavg(\tau)$ (orange), from hydrodynamics for maximally central Pb-Pb (5.02\,TeV) (upper panels) and Au-Au (200\,GeV) collisions (lower panels)
  at mid‐ (left) and forward rapidity (right), compared to the expanding-blast-wave model (blue) of Ref.~\cite{Du:2017qkv}.}
  \label{fig:temp-fb}
\end{figure}
In Fig.~\ref{fig:Tcut} we display the time dependence of the trajectory-averaged temperature, $\Tavg(\tau)$, for different centralities at the LHC. As expected, the time spent above the hadronization temperature varies substantially with centrality, from about 8\,fm$/c$ in central collisions down to about 1--2\,fm$/c$ when the impact parameter approaches the nuclear diameter. We also find that an increase in the terminal temperature from 100\,MeV to $\Tf=130$\,MeV leads to a slight increase in the average temperature below $\Thad$, along with some reduction in the lifetime. However, essentially no difference occurs above $\Thad$ where the $Y$ transport is active.

Finally, we compare in Fig.~\ref{fig:temp-fb} the trajectory-averaged temperature profiles to those from the fireball model used in Ref.~\cite{Du:2017qkv} (which is basically an expanding blast-wave model whose inputs are also adjusted to soft-hadron observables), for central Pb-Pb (5.02\,TeV) and Au-Au (200\,GeV) collisions at mid- and forward rapidity. The average temperatures from the hydrodynamic evolution tend to lie above the fireball temperatures at small times but fall below later, leading to shorter lifetimes; this is due to surface effects in the hydro which feature a continuous freeze-out, which in turn also amplify escape effects on the $Y$ trajectories.

\subsection{Bottomonium formation times}
\label{ssec:form}
Compared to the near-instantaneous formation of a $b\bar{b}$ pair, after a time of $\approx 1/(2m_b)$, the development into a $Y$ state (say, in a pp collision) takes significantly longer~\cite{Blaizot:1987ha,Karsch:1987uk}. In principle, this should be described by a quantum-mechanical wave packet evolution. However, in the past, this has been widely described in a formation-time approximation where the main effect for the purpose at hand is a reduced dissociation rate (or ``cross section''). We incorporate this effect following earlier works~\cite{Farrar:1988me} by reducing the dissociation rate in proportion to the transverse size of the nascent $Y$ state (\ie, cross-sectional area), assumed to grow linearly in time (contrary to the classical expectation of a quadratic dependence). We account for this by a multiplicative factor $\tau/\tau_{\rm form}$ for $\tau \leq \tau_{\rm form}$. 
We estimate the quantum formation times associated with different $Y$ states as in our previous work~\cite{Wu:2025lcj} by a schematic criterion based on the inverse vacuum binding energy, \ie, $\tauform=C/E_B$, where we take a range of $C=1$--$2$ as a rough uncertainty estimate that will be reflected in our comparisons to observables. This ansatz is likely to break down for weakly bound states, but in our calculations, where the $\Upsilon(3S)$ has the smallest binding energy at $\sim$0.2\,GeV, this should not be a problem.

\begin{table}[t]
\centering
\begin{tabular}{|c|c|c|c|}
\hline
state &$E_B\,$(GeV) & Radius\,(fm) & \makecell{$\tauform$\,(fm/$c$)} \\
\hline
$1S$ & 1.10  & 0.146 & 0.18--0.36 \\
\hline
$1P$ & 0.66  & 0.232 & 0.30--0.60 \\
\hline
$2S$ & 0.54  & 0.315 & 0.37--0.73 \\
\hline
$2P$ & 0.30  & 0.467 & 0.66--1.32 \\
\hline
$3S$ & 0.20 & 0.587 & 0.99--1.97 \\
\hline
\end{tabular}
\caption{Vacuum binding energies, radii, and estimated formation-time ranges of bottomonia explicitly treated in this work.
}
\label{tab:Y}
\end{table}

\subsection{Cold-nuclear-matter effects}
\label{ssec:cnm}
Cold-nuclear-matter (CNM) effects in particle production commonly comprise nuclear shadowing of the parton distribution functions (PDFs) in the nucleon, nuclear absorption, \ie, a dissociation of the nascent quarkonium bound state by the incoming nucleons, as well as energy loss and nuclear $k_T$ broadening (referred to as the Cronin effect) of the incoming partons prior to the hard production process of the $\bbb$ pair. We briefly outline our implementation of these in the following.

\subsubsection{Nuclear shadowing}
\label{ssec:shadow}
\begin{figure}[t]
   \centering
\includegraphics[width=0.45\linewidth]{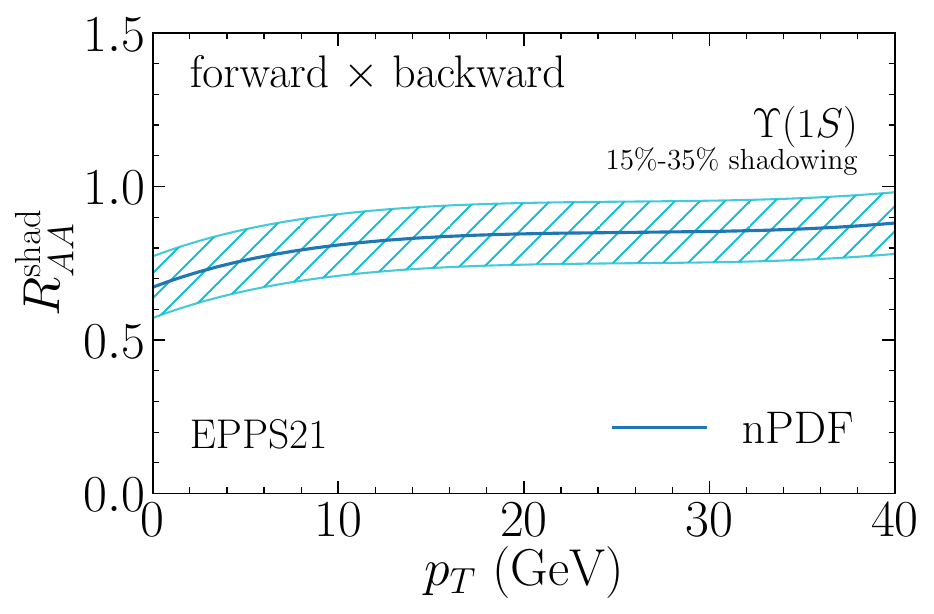}
\includegraphics[width=0.45\linewidth]{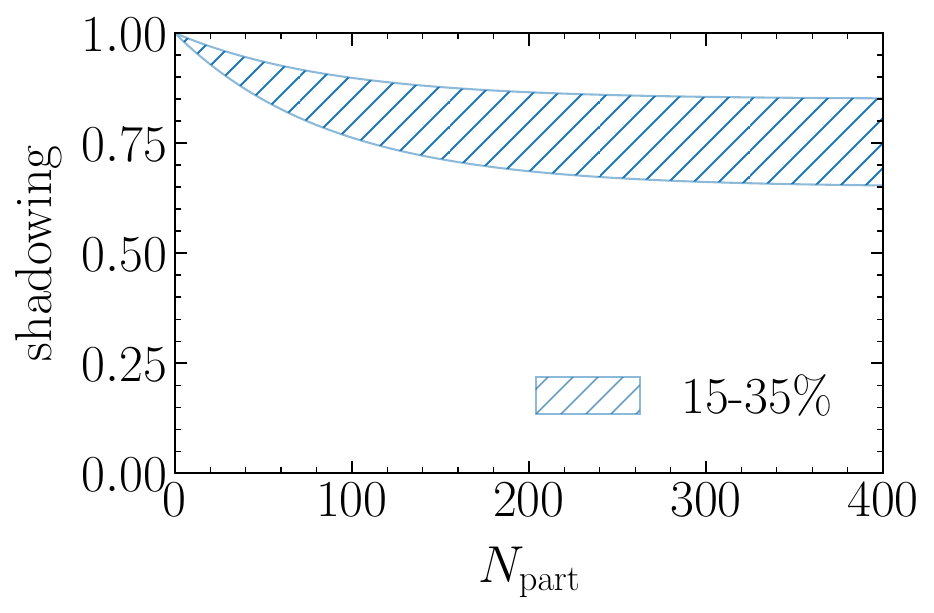}
\caption{Left panel: the product of the $Y$ nuclear modification factors, $R_{\rm pPb} (5.02\,{\rm TeV})$, at forward- and backward-rapidity, resulting from a fit to EPPS21 nPDFs~\cite{Eskola:2021nhw}. 
Right panel: the centrality dependence of the nuclear-shadowing factor in Pb-Pb (5.02\,TeV) collisions at mid-rapidity, with a suppression of up to 15-35\% in the most central collisions.}
\label{fig:cnm}
\end{figure}

At LHC energies, we employ a momentum-dependent shadowing correction for bottomonium production in Pb-Pb collisions, constructed as the product of the forward- and backward-rapidity nuclear modification factors obtained from our previous p+Pb study in Ref.~\cite{Thapa:2025jua} based on EPPS21 nPDFs~\cite{Eskola:2021nhw}.
We adopt a suppression of up to 15-35\% in central collisions on the integrated bottomonium production at mid-rapidity and 20-40\% at forward rapidity, with a centrality dependence shown in Fig.~\ref{fig:cnm}. This encompasses earlier studies~\cite{Du:2017qkv} where up to a 25\% suppression at mid-rapidity and 30\% at forward rapidity was used, and is largely consistent with more recent calculations that also include energy loss and $k_t$ broadening of the incoming partons~\cite{Thapa:2025jua}.
For Au-Au collisions at RHIC the situation is more uncertain~\cite{Adamczyk:2013poh,Ye:2017fwv} and further complicated by the interplay with nuclear absorption effects discussed in the following section; we assume up to 10-30\% shadowing in central Au-Au (200\,GeV) collisions, with the same centrality dependence as for the LHC. 

We also need the shadowing correction to the open-bottom cross section that enters the regeneration contribution. At both RHIC and the LHC we assume the same values as for bottomonium production.

\subsubsection{Nuclear absorption}
\label{ssec:nuc-abs}
At LHC energies, nuclear absorption of pre-resonant $Q\bar Q$ pairs is expected to be negligible due to the large Lorentz contraction of the colliding nuclei, which significantly dilates the quarkonium formation time beyond the nuclear size. As a consequence, the breakup probability
resulting from nuclear absorption is believed to be strongly suppressed.
At RHIC energies, however, the Lorentz boost is smaller and the formation time of the pre-resonant $Q\bar Q$ state becomes comparable to the nuclear traversal time. 
Nuclear absorption may therefore lead to a non-negligible suppression of the quarkonium yield. We account for this through an effective absorption cross section $\sigma_{\rm abs}$, 
folded over the nuclear overlap functions~\cite{Gerschel:1988wn} based on a Woods--Saxon density distribution parametrized as
\begin{equation}
  \rho(r,z)
  =
  \frac{\rho_{0}}
       {1+\exp\!\left[
       \left(\sqrt{r^2+z^2}-R\right)/d
       \right]}\,,
  \label{eq:glauber:woodsaxon}
\end{equation}
where $\rho_{0}$, $R$, and $d$ denote the central density, nuclear radius, and skin depth, respectively. In the present calculations, we use
$\rho_{0}=0.17\,{\rm fm}^{-3}$ and $d=0.54\,{\rm fm}$, while $R$ is chosen
according to the collision system. The corresponding normalized densities for a nuclear mass number $A$ are defined as
\begin{equation}
  \hat\rho(r,z)=\frac{\rho(r,z)}{A} \ ,
  \qquad
  \int\!\dd^2r\,\dd z\,\hat\rho(r,z)=1 \ .
  \label{eq:glauber:normalized}
\end{equation}
At a fixed impact parameter $\mathbf b$ for two colliding nuclei $A_1$ and $A_2$, the transverse position of the
production point is denoted by $\mathbf r$, with
\begin{equation}
  r_{A_1}=\left|\mathbf r+\frac{\mathbf b}{2}\right|\,,
  \qquad
  r_{A_2}=\left|\mathbf r-\frac{\mathbf b}{2}\right|\,.
\end{equation}
The transverse spatial density of
quarkonia produced within a rapidity interval $\Delta y$ is
\begin{equation}
  f(\mathbf r,b)
  =
  \Delta y\,\frac{\dd\sigma_{\mathcal Q}^{\rm pp}}{\dd y}\;A_1\,A_2
  \int\!\dd z_1\,\dd z_2\;
  \hat\rho(r_1,z_1)\,
  \hat\rho(r_2,z_2)\,.
  \label{eq:glauber:spatial}
\end{equation}
Upon including nuclear absorption, the transverse
spatial density becomes
\begin{equation}
    f_{\rm abs}(\mathbf r,b)
    =
  f(\mathbf r,b)
    \exp\!\Biggl\{
      -\sigma_{\rm abs}
      \Biggl[
      (A_1-1)\int\limits_{z_1}^{\infty}\!\dd z'\,
      \hat\rho(r_1,z')
      +
      (A_2-1)\int\limits_{z_2}^{\infty}\!\dd z'\,
      \hat\rho(r_2,z')
      \Biggr]
    \Biggr\}\,.   
    \label{eq:glauber:absorption}
\end{equation}
Here, the longitudinal coordinates $z_1$ and $z_2$ are defined along the
outgoing propagation direction in the respective nuclei, such that the
remaining path length through each nucleus is represented by the integral from the production point to $+\infty$.
The corresponding nuclear-absorption survival factor at a fixed impact
parameter $b$ is
\begin{equation}
  S_{\rm abs}(b)
  =
\frac{\displaystyle\int\!\dd^2r\;f_{\rm abs}(\mathbf r,b)}
    {\displaystyle\int\!\dd^2r\;f(\mathbf r,b)}\,.
  \label{eq:glauber:suppression}
\end{equation}
For the nuclear absorption cross section, we use a range of $\sigma_{\rm abs}$=0--3\,mb. Note that bottomonium suppression due to nuclear absorption (which is a final-state effect, and does not affect open-bottom production) is on top of nuclear shadowing (which is an initial-state effect). 
Experimental measurements of inclusive $\Upsilon(1S+2S+3S)$ production in d+Au and p+Au collisions at RHIC yield a suppression of approximately $0.7\pm0.2$ around midrapidity~\cite{Adamczyk:2013poh,Ye:2017fwv}.
If they were entirely associated with shadowing, one would have to square them to obtain the shadowing in Au+Au. On the other hand, there could also be final-state effects, both from nuclear absorption and/or comovers. Thus our assumption of a 10-30\% shadowing combined with a moderate nuclear absorption seems fair.

\subsection{Production cross sections and late-time feed-down from excited states}
\label{ssec:feeddown}
Once the abundance of each state has been evaluated at fireball decoupling, referred to as ``direct'' production, feed-down contributions from the decays of excited states need to be accounted for.
Following Refs.~\cite{Islam:2020gdv,Islam:2020bnp}, we employ a feed-down matrix $\hat F$ which is built from experimental branching fractions from the Particle Data Group~\cite{ParticleDataGroup:2024cfk} to obtain the experimental inclusive production cross sections from the direct production cross sections as 
\begin{equation}
\vec\sigma_{\rm exp} = \hat F\,\vec\sigma_{\rm direct}\,, 
\end{equation}
where the vector
\begin{equation}
   \begin{aligned}
\vec\sigma = \bigl\{\sigma[\Upsilon(1S)],\,\sigma[\Upsilon(2S)],\,\sigma[\chi_{b0}(1P)],\,\sigma[\chi_{b1}(1P)],\,\sigma[\chi_{b2}(1P)],\,\\
\sigma[\Upsilon(3S)],\,\sigma[\chi_{b0}(2P)],\,\sigma[\chi_{b1}(2P)],\,\sigma[\chi_{b2}(2P)]\bigr\}
\end{aligned}
\end{equation}
assembles the cross sections per unit rapidity over a given range (\eg, $|y|\le2.4$).
The matrix elements $\hat F_{ij}$ are the branching ratios for the $i \to j$ transition of $Y_i \to Y_j +X$, explicitly given by
\begin{equation}
\hat F = \left(
\begin{array}{ccccccccc}
	1 & 0.2645 & 0.0194 & 0.352 & 0.18 & 0.0657 & 0.0038 & 0.1153 & 0.077 \\
	0 & 1 & 0 & 0 & 0 & 0.106 & 0.0138 & 0.181 & 0.089 \\
	0 & 0 & 1 & 0 & 0 & 0 & 0 & 0 & 0 \\
	0 & 0 & 0 & 1 & 0 & 0 & 0 & 0.0091 & 0 \\
	0 & 0 & 0 & 0 & 1 & 0 & 0 & 0 & 0.0051 \\
	0 & 0 & 0 & 0 & 0 & 1 & 0 & 0 & 0 \\
	0 & 0 & 0 & 0 & 0 & 0 & 1 & 0 & 0 \\
	0 & 0 & 0 & 0 & 0 & 0 & 0 & 1 & 0 \\
	0 & 0 & 0 & 0 & 0 & 0 & 0 & 0 & 1 \\
\end{array}
\right)\,.
\label{eq:fdm}
\end{equation}
The direct cross sections, which are the input to our calculations, can be obtained as
\begin{equation}
\vec{\sigma}_\text{direct} = \hat F^{-1} \vec{\sigma}_\text{exp} \,,
\end{equation}
where experimental measurements at mid-rapidity 
\begin{equation}
\vec\sigma_{\rm exp}
= \{57.6,\,19.0,\,3.72,\,13.69,\,16.1,\,6.8,\,3.27,\,12.0,\,14.15\}\,\text{nb}\,,
\end{equation}
are taken from Refs.~\cite{CMS:2018zza,ParticleDataGroup:2024cfk,LHCb:2014ngh}.
For forward rapidity we follow Ref.~\cite{Du:2017qkv} and take half the mid-rapidity values for all states.

With the above ingredients, we have all information in place to evaluate bottomonium suppression in heavy-ion collisions, which is commonly quantified via the nuclear modification factor
\begin{equation}
\raa^Y(N_{\rm part}) = \frac{N_Y^{\rm AA}(N_{\rm part})}{N_Y^{\rm pp}\,N_{\rm coll}(N_{\rm part})}\,,
\label{eq:raa}
\end{equation}
where:
\begin{itemize}
    \item $N_Y^{\rm AA}(\npart)$ is the $Y$ yield measured in nucleus--nucleus (AA) collisions for a given collision centrality as characterized by the number of participating nucleons, $\npart$;
    \item $N_Y^{\rm pp}$ is the corresponding yield in proton--proton (pp) collisions at the same collision energy and rapidity;
    \item $N_{\rm coll}(\npart)$ is the number of binary nucleon--nucleon collisions determined for a given centrality.
\end{itemize}

\section{Bottomonium regeneration in a hydrodynamic background}
\label{sec:reg}
In this section we describe the evaluation of bottomonium regeneration in the hydrodynamic background. In terms of the rate equation, this corresponds to its inhomogeneous solution, which can be written as
\begin{equation}
N_{Y}^{\text{reg}} (\tau) = \int_{\tau_{\text{reg}}}^{\tau} \Gamma_{Y} (T(\tau')) N_Y^{\text{eq}}(\tau') \exp \left(- \int_{\tau'}^{\tau} \Gamma_{Y} (T(\tau'')) \dd\tau'' \right)  \dd\tau' \,.
\label{eq:reg}
\end{equation}

In Sec.~\ref{ssec:eq}, we first introduce the equilibrium limit in a homogeneous fireball.
In Sec.~\ref{ssec:escape} we discuss how the escape of $b\bar b$ pairs from the fireball reduces the effective number of pairs available for regeneration, and in
Sec.~\ref{ssec:local_eq} we extend the equilibrium-limit calculation to a space-time-dependent medium as present in hydrodynamic simulations.

\subsection{Bottomonium equilibrium in a homogeneous fireball}
\label{ssec:eq}
\begin{figure}[t]
   \centering
\includegraphics[width=0.7\linewidth]{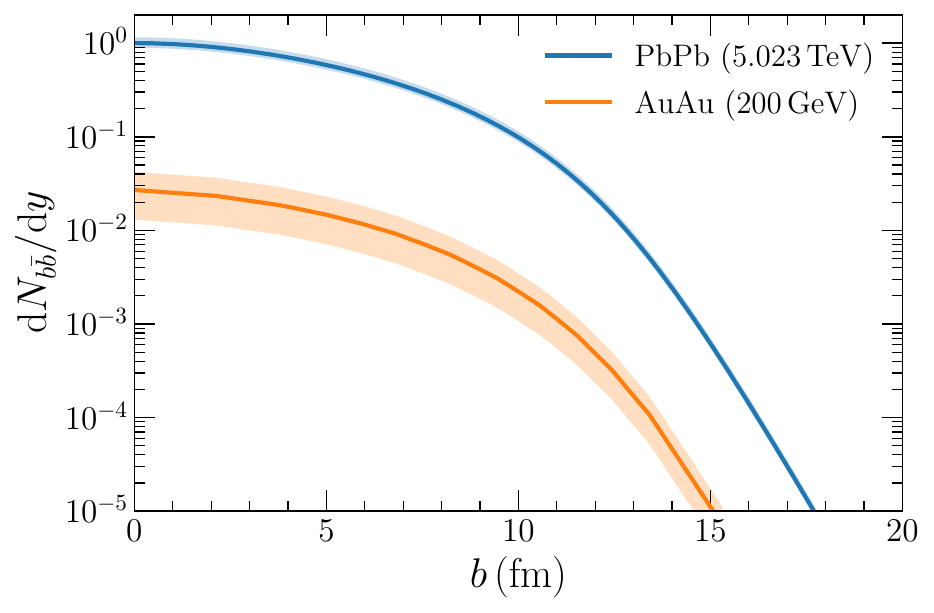}
\caption{
      Number of $b\bar b$ pairs per unit rapidity at mid-rapidity as a function of impact parameter for Pb--Pb collisions at $\sqrt{s_{\mathrm{NN}}}=5.023\,$TeV (blue) and Au--Au collisions at $\sqrt{s_{\mathrm{NN}}}=200\,$GeV (orange). Solid curves show the central values of the binary-collision-scaled yields, $\dd N_{b\bar b}/\dd y=N_{\rm coll}(b)\,\sigma_{b\bar b}/\sigma_{\rm pp}^{\rm inel}$, while shaded bands show the propagated asymmetric total uncertainties of the corresponding $b\bar b$ production cross sections, $\sigma_{b\bar b}=34.5^{+5.3}_{-3.8}\,\mu\mathrm{b}$ (Pb--Pb) and $0.92^{+0.52}_{-0.48}\,\mu\mathrm{b}$ (Au--Au). The Pb--Pb input is from ALICE~\cite{ALICE:2021mgk}, and the Au--Au input is from PHENIX~\cite{PHENIX:2009dpd}.
}
\label{fig:Nbb}
\end{figure}
For a homogeneous fireball of volume $V_{\rm FB}$ at temperature $T$, the equilibrium number of a state $Y$ is given by the Boltzmann distribution,
\begin{equation}
N_{Y}^{\rm eq}=d_{Y} V_{\rm FB}\gamma_b^2  \int \frac{\dd^3 \mathbf{p}}{\left(2 \pi\right)^3} e^{-\sqrt{\mathbf{p}^2+m_{Y}^2}/T}
   =\gamma_b^2 V_{\rm FB}\frac{d_{Y}}{2 \pi^2} T m_{Y}^2  K_2\left(\frac{m_{Y}}{T}\right)\,,
   \label{eq:density}
\end{equation}
where $K_2$ is the modified Bessel function of the second kind and
$d_Y=2J_Y+1$ is the spin degeneracy.
The fugacity $\gamma_b$ is fixed by bottom-antibottom conservation~\cite{Braun-Munzinger:2009dzl},
\begin{equation}
N_{b \bar{b}}=\frac{1}{2} \gamma_b n_{\mathrm{op}} V_{\mathrm{FB}} \frac{I_1\left(\gamma_b n_{\mathrm{op}} V_{\text {corr }}\right)}{I_0\left(\gamma_b n_{\mathrm{op}} V_{\text {corr }}\right)}+\gamma_b^2 n_{\mathrm{hid}} V_{\mathrm{FB}}\,,
\label{eq:fugacity}
\end{equation}
where $n_{\rm op}$ and $n_{\rm hid}$ denote the total open- and hidden-bottom densities, respectively, and
$I_0$ and $I_1$ are the modified Bessel functions of the first kind. The correlation volume~\cite{Hamieh:2000tk,Grandchamp:2003uw}
\begin{equation}
V_{\rm corr}
= \tfrac{4}{3}\pi\bigl(r_0 + \langle v_b\rangle\,\tau\bigr)^3 \, ,
\label{eq:corrV}
\end{equation}
enforces local $\bbb$ conservation after the initial point-like production, where $r_0\simeq1$\,fm is a typical strong-interaction range and $\langle v_b\rangle=0.65\,c$ is an average $b$ quark recoil velocity estimated from $B$-meson $\pT$ spectra~\cite{Du:2017qkv}.
The open-bottom density, $n_{\rm op}=n_b + n_{\bar b}$, is the key quantity to convert the number of $\bbb$ pairs into the bottom fugacity whose square enters into the computation of the $Y$ equilibrium limits. In addition to the freely moving $b$ and $\bar b$ quarks in the QGP, we include ground ($J=0$) and first excited ($J=1$) $B$ mesons, similar to what has been done in the charm sector in Ref.~\cite{Zhao:2010nk}. This accounts for the onset of hadronic bound-state formation when the temperature drops toward the hadronization temperature, which is borne out by both lQCD~\cite{Bazavov:2023xzm} and $T$-matrix calculations~\cite{Liu:2021rjf}.
The ground states represent the most robust mesonic open-bottom channels. We neglect contributions from bottom baryons and higher excited open-bottom states which are expected to be approximately offset by a reduced pole strength of broadened in-medium $B$-meson states.  
Their contribution is smoothly switched off toward high temperatures using a hyperbolic-tangent parametrization centered around $T=1.3\,\Thad$, where the $B$-meson contribution is reduced to half of its low-temperature value. Consequently, the open-bottom density reduces to the bottom-quark density in the high-temperature QGP.
The $Y$ states are negligible for the calculation of the fugacity.

The total bottom number is determined by the $b\bar b$ production cross section in pp collisions and the number of binary collisions $N_{\rm coll}$ in an AA collision over a rapidity interval $\Delta y=1.8$, corresponding to the effective rapidity coverage of a thermal fireball (over which all inputs are evaluated), 
\begin{equation}
N_{b\bar b}
= \frac{\dd\sigma_{b\bar b}}{\dd y}\;\frac{N_{\rm coll}}{\sigma^{\rm inel}_{\rm pp}}\;\Delta y \ .
\end{equation}
For the total inelastic cross section, we use $\sigma^{\rm inel}_{\rm pp}=67.6(42)$\,mb at 5.02(0.2)\,TeV.
In Fig.~\ref{fig:Nbb} we display $N_{b\bar b}$ per unit rapidity at mid-rapidity versus impact parameter $b$ for Pb-Pb (5.02\,TeV) and Au-Au (200\,GeV) collisions.

\begin{figure}[t]
   \centering
\includegraphics[width=0.7\linewidth]{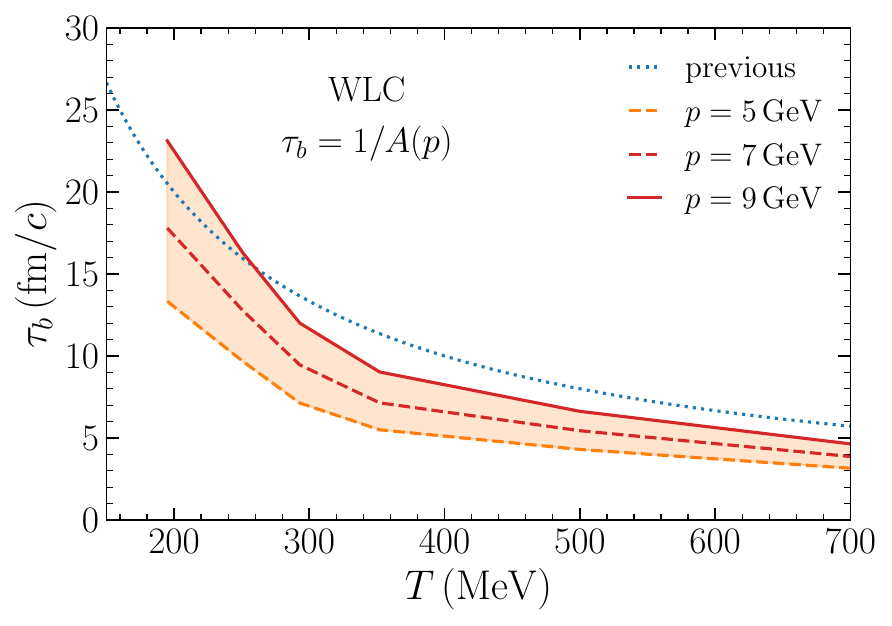}
\caption{
Thermal relaxation time $\tau_b=1/A(p)$ of bottom quarks as a function of temperature. Results are shown for different momenta $p$, compared with the relaxation time previously used in Ref.~\cite{Du:2017qkv}
(dotted line).
}
\label{fig:taub}
\end{figure}
One also has to account for incomplete thermalization of $b$ quarks in the QGP which suppresses the phase space for $Y$ production. We apply a thermal-relaxation time correction via a factor~\cite{Grandchamp:2002wp,Song:2012at}
\begin{equation}
\mathcal{R}(\tau)
= 1 - \exp\!\Bigl[-\!\int_{\tau_0}^{\tau}\frac{\dd\tau'}{\tau_b}\Bigr]\,,
\end{equation}
that effectively reduces the equilibrium bottomonium yield as
\begin{equation}
   N_{Y}^{\rm eq}(\tau)\to N_{Y}^{\rm eq}(\tau)\,\mathcal{R}(\tau)\,.
\end{equation}
The bottom-quark relaxation time is related to the momentum- and temperature-dependent bottom-quark 
drag coefficient, $\tau_b(p, T)=1/A(p,T)$. Within the same nonperturbative framework used for the $Y$ reaction rates, \ie, constrained by WLCs~\cite{Tang:2023tkm}, the resulting relaxation times are shown in Fig.~\ref{fig:taub} for several $b$ quark momenta and compared to the input previously used in Ref.~\cite{Du:2017qkv}.
In the present calculation, we approximate
\begin{equation}
    \tau_b(T)=\frac{1}{A(\langle p\rangle,T)},
    \qquad \langle p\rangle=7~{\rm GeV},
\end{equation}
as a representative relaxation time for $b$ quark kinetic  equilibration.

\begin{figure}[t]
   \centering
\includegraphics[width=0.48\linewidth]{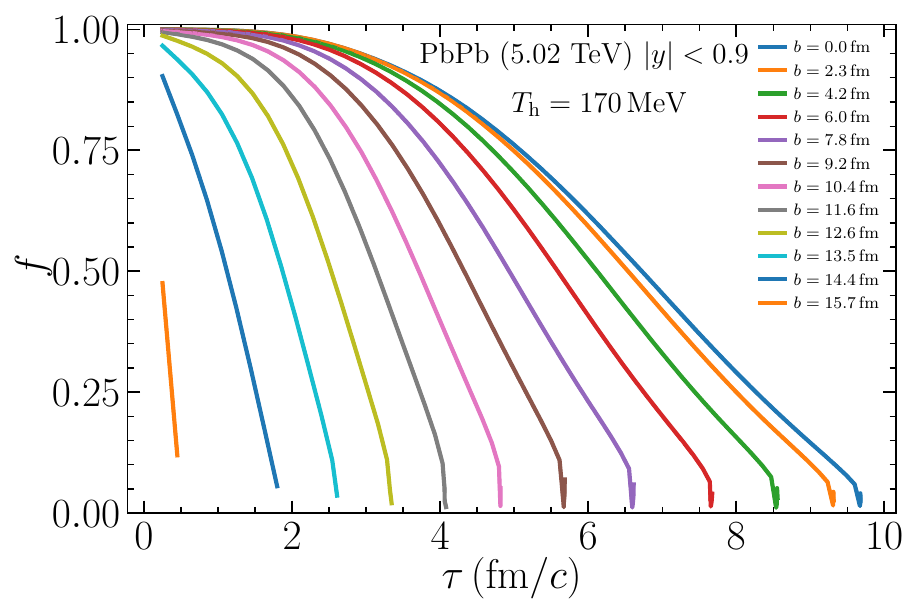}
\includegraphics[width=0.48\linewidth]{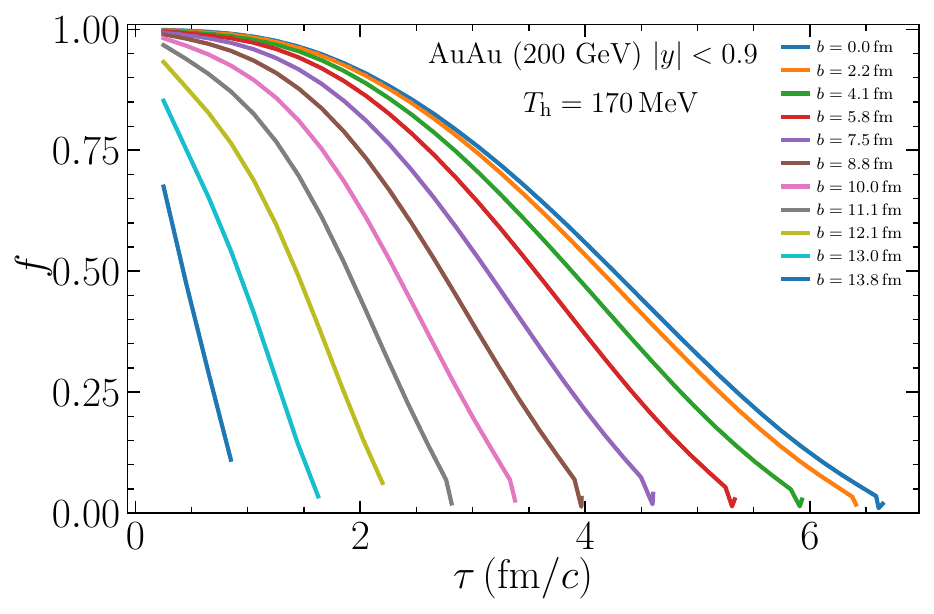}
\caption{
    Fraction of $b\bar b$ pairs remaining inside the fireball volume as a function of proper time for various impact parameters in Pb-Pb (5.02\,TeV) collisions (left panel) and for Au-Au (200\,GeV) collisions (right panel).
}
\label{fig:Nbb_escape}
\end{figure}

\subsection{Escape of $b\bar{b}$ pairs from the fireball}
\label{ssec:escape}
Bottom-antibottom quark pairs are produced upon initial impact in a  heavy-ion collision and subsequently propagate through the QGP.
Some of the $b$ quarks may escape the evolving fireball volume over time.
To estimate the fraction of $b\bar{b}$ pairs that remains within the fireball, we sample their initial positions according to a Glauber-model distribution, assuming point-like production at a common time and location. The $b$ quarks are initialized at given rapidity with a $p_T$ sampled from a fixed order plus next-to-leading logarithms (FONLL) distribution~\cite{Cacciari:2015fta}
assuming back-to-back kinematics.
In each time step of the fireball evolution, we count the number of $b$ and $\bar{b}$ quarks that remain within the fireball volume, denoted as $N_{b\bar{b}}^{\rm med}(\tau)=(N_b+N_{\bar b})/2$.
The ratio of this number to the total number of initially produced $b\bar{b}$ pairs defines the time-dependent in-medium fraction,
\begin{equation}
f_{b\bar{b}}^{\rm med}(\tau) = \frac{N_{b\bar{b}}^{\rm med}(\tau)}{N_{b\bar{b}}^{\rm tot}}\,.
\end{equation}
Figure~\ref{fig:Nbb_escape} shows
this fraction as a function of time at various impact parameters in Pb-Pb (5.02\,TeV) (left) and Au-Au (200\,GeV) (right) collisions at mid-rapidity. Here, the active in-medium 
region is defined by fluid cells satisfying 
$T(\tau,\mathbf{x})>\Thad =170$\,MeV.
In central and semi-central collisions, essentially all initially produced 
$b\bar b$ pairs are located within this region at the initial time. 
As the medium expands and cools, the fraction decreases because the 
bottom quarks propagate out of the active region where fluid cells 
have cooled below $\Thad$.
In very peripheral collisions, even the initial in-medium fraction can be smaller than unity, since only a part of the initially produced medium satisfies the condition $T(\tau,\mathbf{x})>\Thad$ at the initial time while a fraction of the initially produced $b\bar b$ pairs can 
be outside this region relevant for regeneration.
The effective number of bottom pairs available for regeneration is then 
obtained by applying this in-medium fraction to the total number of 
initially produced bottom pairs,
\begin{equation}
    N_{b\bar b}^{\rm med}(\tau)
    =
    f_{b\bar b}^{\rm med}(\tau)\,
    N_{b\bar b}^{\rm tot}.
    \label{eq:Nbb_med}
\end{equation}

\subsection{Bottomonium equilibrium in a space-time-dependent medium}
\label{ssec:local_eq}
\begin{figure}[t]
\includegraphics[width=1\textwidth]{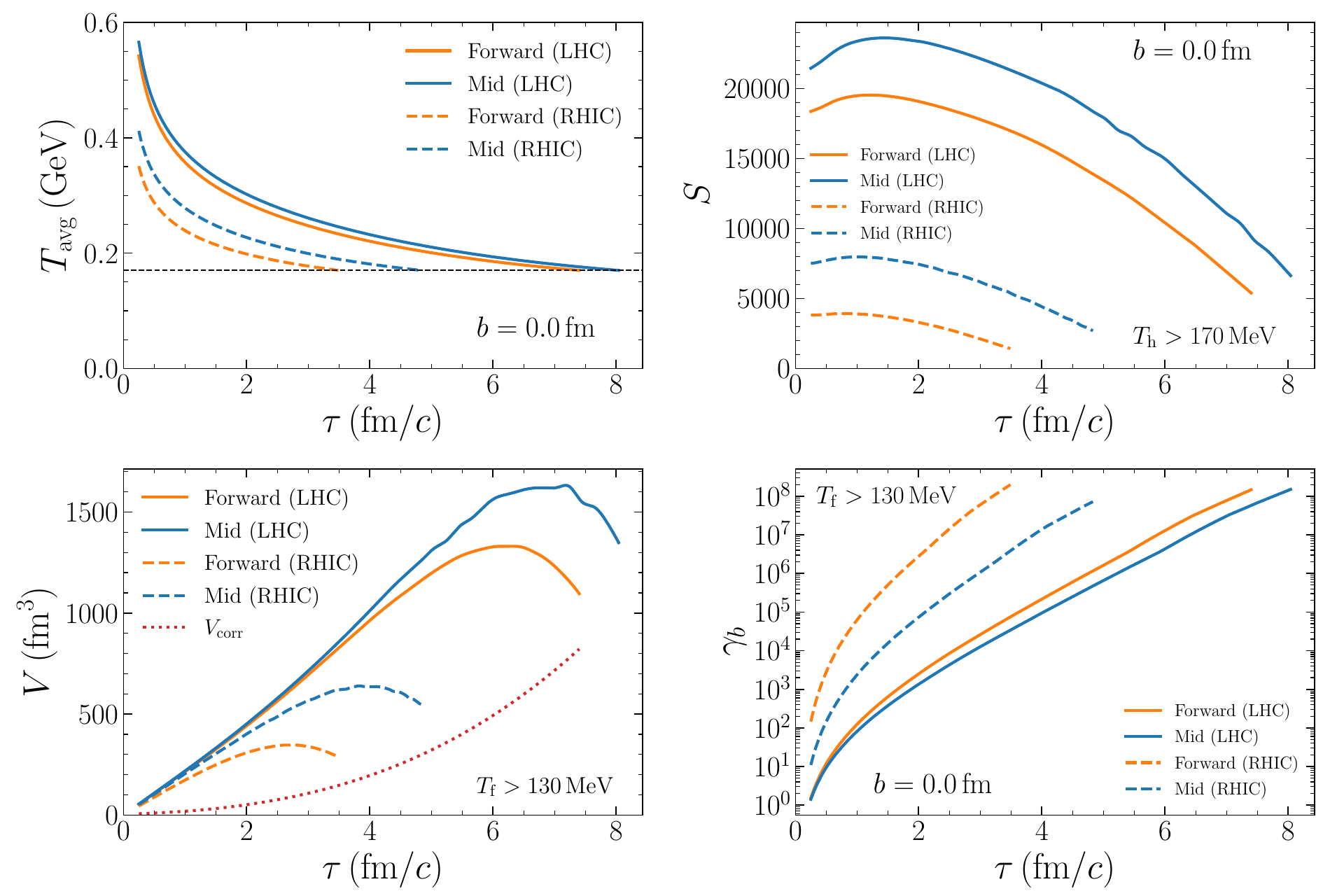}
\caption{Time evolution of the trajectory-averaged temperature, $\Tavg(\tau)$ (top left; corresponding to the orange curves in Fig.~\ref{fig:temp-fb})
        , active entropy (top right), effective volume (bottom left), and fugacity (bottom right) for maximally central Pb-Pb (5.02\,TeV) (solid lines) and Au-Au (200\,GeV) (dashed lines) collisions
    at mid-rapidity (blue) and forward rapidity (orange).
}
   \label{fig:thermo}
\end{figure}
To evaluate bottomonium regeneration in a hydrodynamic background with space-time-dependent temperature and flow,
we use the same trajectory-averaged temperature, $\Tavg(\tau)$, introduced in Sec.~\ref{ssec:hydro} and shown in Fig.~\ref{fig:temp-fb}, together with the active volume and total entropy within it.
For a given $T_{\rm avg}(\tau)$ we define an effective ``regeneration volume'' by matching its entropy to the total entropy 
contained in hydrodynamic cells that are above the hadronization temperature,
\begin{equation}
V_{\rm eff}(\tau)
= \frac{S(\tau)}{s\bigl[T_{\rm avg}(\tau)\bigr]}\,,
\end{equation}
with
\begin{equation}
S(\tau)
= \int_{T(\tau,\mathbf{x})>\Thad}\!\dd^3x\; s\bigl[T(\tau,\mathbf{x})\bigr] \, ,
\end{equation}
where $s[T(\tau,\mathbf{x})]$ is the local entropy density throughout the active volume, which comprises all cells with temperature above $\Thad$=170\,MeV.
In Fig.~\ref{fig:thermo} we display $\Tavg(\tau)$, $S(\tau)$, and $V_{\rm eff}(\tau)$ for both mid- and forward rapidity in central Pb-Pb (5.02\,TeV) and Au-Au (200\,GeV) collisions. As expected, the RHIC fireballs are cooler and shorter-lived than their LHC counterparts, the same is true at forward rapidity relative to midrapidity. As the system expands and cools, cells whose temperature falls below $\Thad$ are excluded from the entropy integral, which causes $S(\tau)$ to decrease at later times.
Despite the smaller entropy at forward rapidity, the corresponding 
effective volume remains comparable to that at mid-rapidity.
The effective volume initially increases as a 
result of the medium expansion but decreases near the end of the 
evolution when only a part of the region remains above 
$\Thad$. For comparison, the correlation volume $V_{\rm corr}$ defined in~\eqref{eq:corrV} is also shown in the bottom-left panel of Fig.~\ref{fig:thermo}.
The bottom-quark fugacity $\gamma_b$ (bottom right) increases rapidly as the 
system cools. This increase compensates for the strong reduction of the 
thermal open-bottom density at decreasing temperature while conserving 
the initially produced number of bottom pairs (modulo escape effects). The larger values of 
$\gamma_b$ at forward rapidity and at RHIC energies reflect the lower 
temperatures under these conditions.

If the number of initially produced bottom-quark pairs is much smaller than 1 (see Fig.~\ref{fig:Nbb}),
the fugacity in equation~\eqref{eq:fugacity} simplifies according to
\begin{equation}
\frac{I_1\!\bigl(\gamma_b\,n_{\rm op}\,V_{\rm corr}\bigr)}
     {I_0\!\bigl(\gamma_b\,n_{\rm op}\,V_{\rm corr}\bigr)}
\simeq
\tfrac12\,\gamma_b\,n_{\rm op}\,V_{\rm corr} \ ,
\end{equation}
corresponding to the canonical limit.
Upon neglecting the hidden-bottom contribution one has
\begin{equation}
N_{b\bar b}
\simeq
\tfrac14\,\gamma_b^2\,n_{\rm op}^2\,V_{\rm eff}\,V_{\rm corr} \ .
\end{equation}
Consequently, the coefficient of the equilibrium $Y$ yield in Eq.~\eqref{eq:density},
\begin{equation}
N_{Y}^{eq}
   = \gamma_b^2 V_{\rm eff} n_Y
   \propto
   N_{b\bar b}\,
   \frac{n_Y}{n_{\rm op}^2 V_{\rm corr}}\, .
   \label{eq:Ydensity}
\end{equation}
is controlled by the conserved bottom-pair abundance $N_{b\bar b}$ and the correlation volume $V_{\rm corr}$, both of which are independent of the hadronization temperature $\Thad$. Therefore, the resulting equilibrium yields are robust against reasonable variations of $\Thad$.

\subsection{Transverse-momentum spectra}
\label{ssec:pt}
Our current approach does not account for an explicit three-momentum dependence of the regeneration component. This can, in principle, be done in a coupled Langevin-Boltzmann framework, as recently elaborated in the charm sector for a uniform fireball in Ref.~\cite{Fu:2026ous}. For now we defer its implementation into hydrodynamics to a future study and instead take recourse to an instantaneous coalescence model (ICM) utilizing transported $b$ quark distributions at an average regeneration temperature, $\Treg$~\cite{Wu:2023djn}.
The ICM has been widely used in both light- and heavy-quark sectors; see, \eg, Ref.~\cite{Fries:2025jfi} for a recent review.
We deploy it in its basic form which assumes global quark distributions in coordinate space and is given by
\begin{equation}
  \frac{\dd^{3} N_{Y}^{\mathrm{coal}}\left(\mathbf{p}\right)}{\dd^{3} \mathbf{p}}={C_{\mathrm{reg}}}g_{Y}\int \dd^{3} \mathbf{p}_{b} \dd^{3} \mathbf{p}_{\barb} \frac{\dd^{3} N_{b}}{\dd^{3} \mathbf{p}_{b}} \frac{\dd^{3} N_{\barb}}{\dd^{3} \mathbf{p}_{\barb}}\delta^{(3)}\left(\mathbf{p}-\mathbf{p}_{b}-\mathbf{p}_{\barb}\right)w\left( \mathbf{k} \right)\, ,
\label{coal-eq}
\end{equation}
where the normalization, $C_{\mathrm{reg}}$, is obtained to match the yield resulting from the rate equation calculation in Eq.~\eqref{eq:reg}, and
$\frac{\dd^{3} N_{b}}{\dd^{3} \mathbf{p}_{b}}$ and $\frac{\dd^{3} N_{\barb}}{\dd^{3} \mathbf{p}_{\barb}}$ are the momentum distributions of the $b$ and $\bar b$ quarks, respectively.
The initial-state averaged and final-state summed degeneracy factors are $g_{Y}=1/12$ for $S$ states (with spin degeneracy 3) and $g_Y=1/3$
for $P$ states (with total spin degeneracy 12), accounting for the probability of forming a colorless meson of given spin from the underlying quark colors and spins.
The coalescence probability of the $b$ and $\bar b$ quarks is governed by the Wigner distribution~\cite{Sun:2017ooe}, 
\begin{equation}
\begin{aligned}
w(\mathbf{k}) &=\frac{\left(4 \pi \sigma^{2}\right)^{\frac{3}{2}}}{V_{\rm FB}} \frac{\left(2 \sigma^{2} \mathbf{k}^{2}\right)^{l}}{(2 l+1) ! !} e^{-\sigma^{2} \mathbf{k}^{2}} \, ,
\end{aligned}
\end{equation}
for a $Y$ state with angular momentum $l$; $k$ denotes the relative momentum of the two quarks and $\sigma$ is estimated from the mean-square radii, 
\begin{equation}
\begin{aligned}
\sigma^2\left( nS \right)&=\frac{2}{3}\frac{\left( m_b+m_{\barb} \right)^2}{m_b^2+m_{\barb}^2}\langle r^2_{nS}\rangle\,,\\
\sigma^2\left( nP \right)&=\frac{2}{5}\frac{\left( m_b+m_{\barb} \right)^2}{m_b^2+m_{\barb}^2}\langle r^2_{nP}\rangle \, ,
\end{aligned}
\end{equation}
where $\langle r^2_{nS}\rangle$ and $\langle r^2_{nP}\rangle$ correspond to $nS$ and $nP$ states and are taken from the self-consistent $T$-matrix calculations~\cite{Wu:2025hlf}; they are listed in Table~\ref{table:coal}.
The Wigner distribution is normalized as~\cite{Greco:2003mm}
\begin{equation}
\int \dd^3\mathbf{x}\dd^3\mathbf{k} w\left(\mathbf{k}\right)= (2\pi)^3\, ,
\label{eq:WignerNorm}
\end{equation}
and the relative momentum is~\cite{Sun:2017ooe}
\begin{equation}
\mathbf{k}=\sqrt{2}\frac{m_{\barb} \mathbf{p}_{b}-m_b \mathbf{p}_{\barb}}{m_b+m_{\barb}}\,,
\end{equation}

We evaluate Eq.~(\ref{coal-eq}) to obtain the  $\pT$ spectra of the $Y$ states at mid-rapidity
using $\dd p_{z}=E_{Y} \dd y$. 
For the heavy quark spectra we employ the results of relativistic Langevin simulations~\cite{He:2012xz} at average temperatures, $\Treg$, at which, based on the time evolution of the yields, most of the regeneration for the $Y$ state in question occurs, cf.~Table~\ref{table:coal}.

\begin{table}[t]
\centering
\begin{tabular}{c|ccccccc}
\hline
State & 1S & 1P & 2S & 2P & 3S & 3P & 4S \\
\hline
$\Treg$ (MeV) & 278 & 220 & 201 & 189 & 183 & 170 & 170 \\
\hline
 $r_{nL}$ (fm) & 0.17 & 0.24 & 0.32 & 0.46 & 0.58 & 0.79 & 0.98 \\
 \hline
\end{tabular}
\caption{
State-dependent temperatures at which regeneration predominantly occurs and the corresponding in-medium radii used for the coalescence calculation of the $Y$ $\pT$ spectra.
}
\label{table:coal}
\end{table}

\section{Time evolution and comparisons to experiment}
\label{sec:exp}

We are now in a position to examine our final results for $Y$ suppression and regeneration in the evolving medium and compare them with experimental observables.
In Sec.~\ref{ssec:time_evo}, we study the time evolution of bottomonium yields at the $\raa$ level throughout the hydrodynamic background.
In Sec.~\ref{ssec:lhc} we turn to comparisons with measurements at LHC energies; this includes centrality dependencies (including a benchmark against previous results with perturbative medium coupling) in Sec.~\ref{sssec:npart} and transverse-momentum spectra
in Sec.~\ref{sssec:pT}.
In Sec.~\ref{ssec:rhic}, we extend our analysis to RHIC energies, where we discuss both the centrality dependence and transverse-momentum dependence and compare with available STAR data.

\subsection{Time evolution of bottomonium yields}
\label{ssec:time_evo}
\begin{table}[b]
\centering
\begin{tabular}{cccccccc}
\toprule
State
& 1S & 1P & 2S & 2P & 3S & 3P & 4S \\
\midrule
$T_{\mathrm{melt}}$
& $>$700 & $>$700 & 700 & 500 & 350 & 250 & 250 \\
$T(E_B=0)$
& $>$700 & 450 & 355 & 233 & 214 & 195 & --- \\
\bottomrule
\end{tabular}
\caption{Melting temperatures (in MeV) from the vanishing of the bound-state pole in the $T$-matrix and temperatures at which the binding energy vanishes for bottomonia in the QGP.}
\label{tab:melt}
\end{table}
While the suppression of the primordially produced states commences right after the aHydro initialization time at $\tau=\tau_0$ (including formation time effects), the onset of regeneration depends on the ``melting'' temperature, $\Tmelt$, of each individual bound state; that is, once the temperature of the ambient medium drops below $\Tmelt$, regeneration commences (if the initial temperature in the hydro evolution is below $\Tmelt$, the regeneration rates are also subject to the same formation time effects, as dictated by detailed balance). 
In previous calculations~\cite{Zhao:2010nk,Wu:2023djn}, the melting temperature was assumed to coincide with the vanishing of the binding energy, $E_B=0$, of each state. However, in recent work a rigorous criterion has been introduced based on the vanishing of the bound-state pole in the complex-energy plane~\cite{Tang:2025ypa}. It was found that, in a strongly coupled QGP, bottomonium states can survive at temperatures beyond the vanishing of their binding energy, \ie, as resonance states above their nominal mass threshold. This obviously leads to larger melting temperatures, as summarized in Table~\ref{tab:melt}.
In our present calculations, for the onset of regeneration, corresponding to the lower limit, $\tau_{\rm reg}$, of the time integral in the solution to the inhomogeneous rate equation (\ref{eq:reg}), we adopt the rigorous pole criterion. 
However, due to the very large reaction rates in the sQGP, particularly those of the excited states, their yields are quickly driven toward equilibrium, and thus the $E_B=0$ criterion gives very similar results, \ie, 
the precise choice of the onset time is immaterial.

\begin{figure}[t]
   \includegraphics[width=0.32\textwidth]{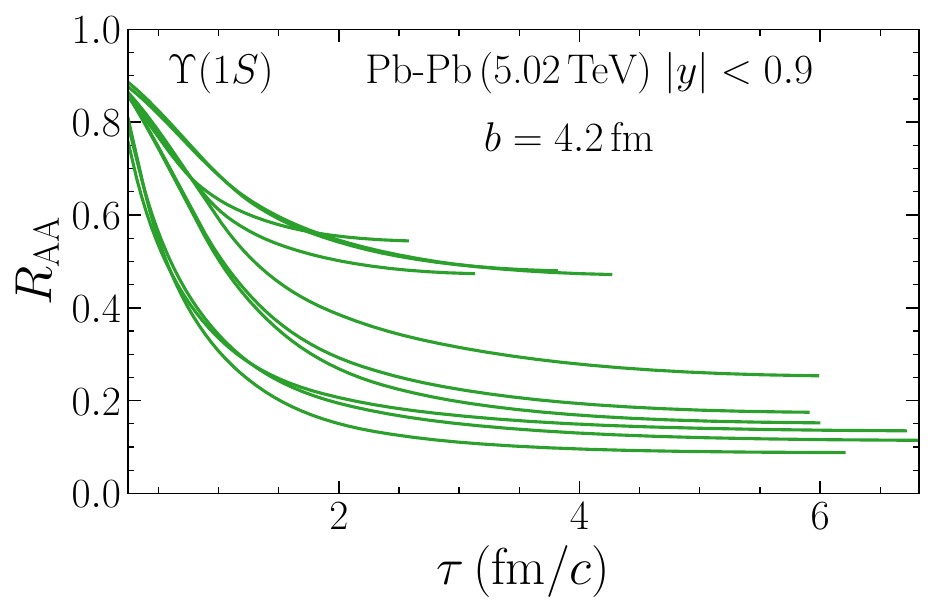}
   \hfill
   \includegraphics[width=0.32\textwidth]{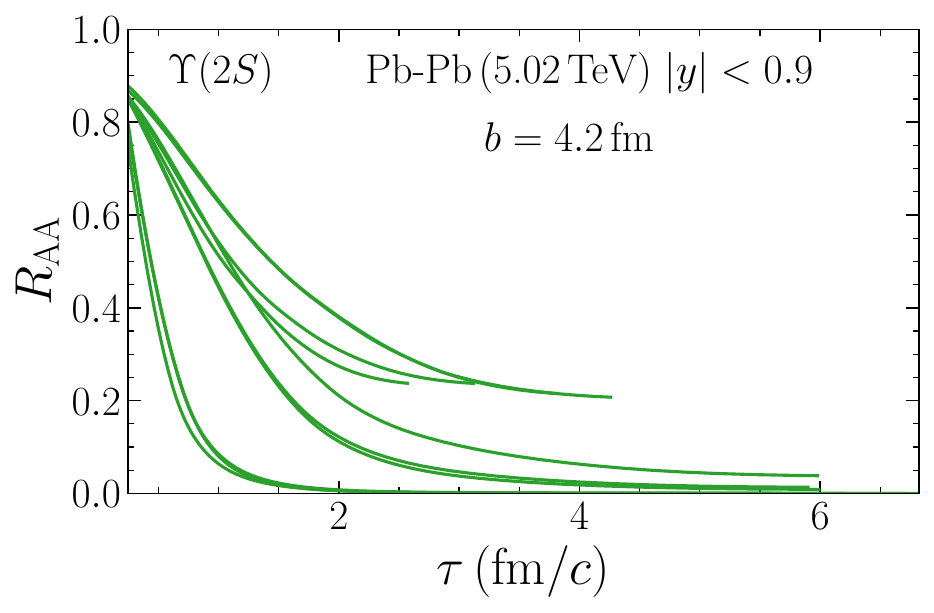}
   \hfill
   \includegraphics[width=0.32\textwidth]{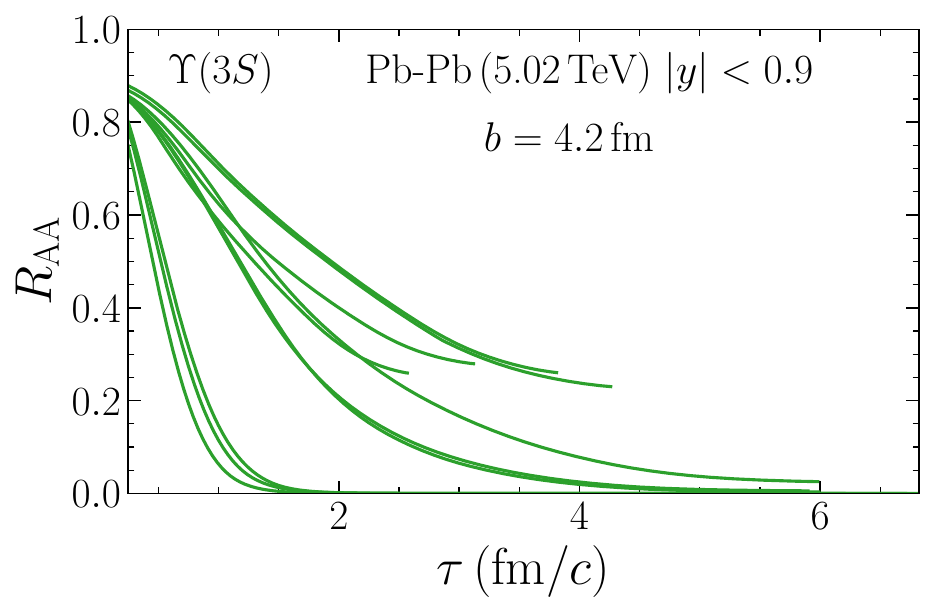}
   \\
   \includegraphics[width=0.32\textwidth]{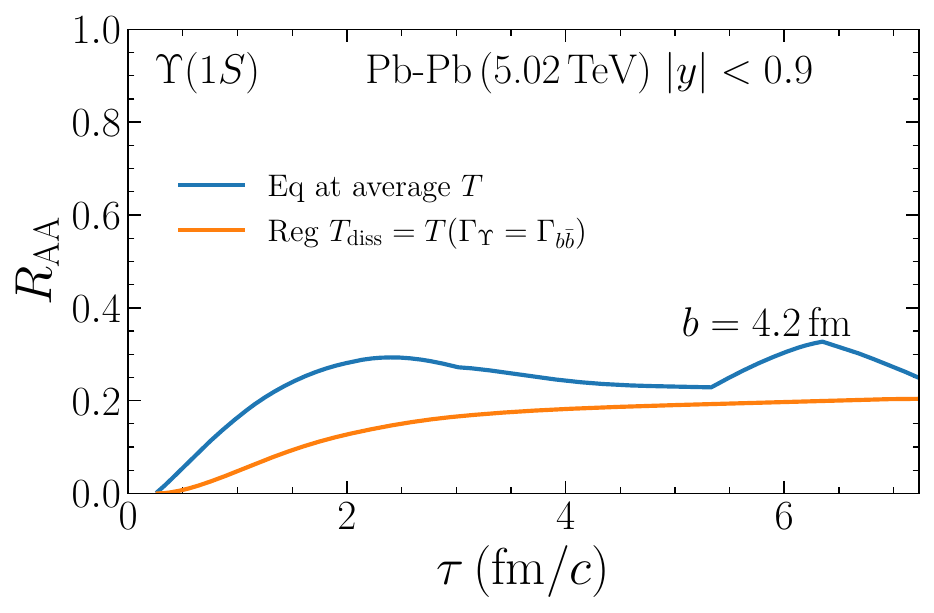}
   \hfill
   \includegraphics[width=0.32\textwidth]{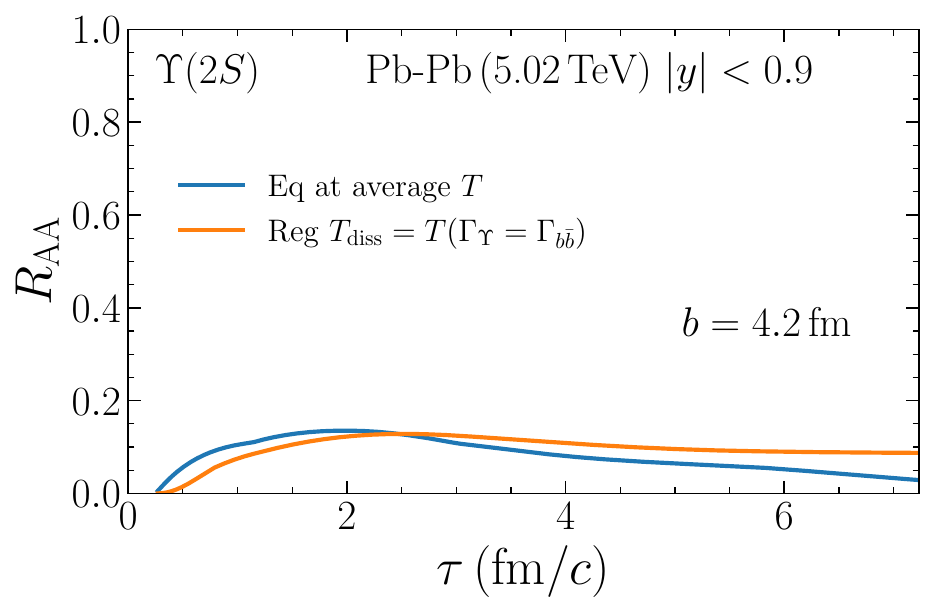}
   \hfill
   \includegraphics[width=0.32\textwidth]{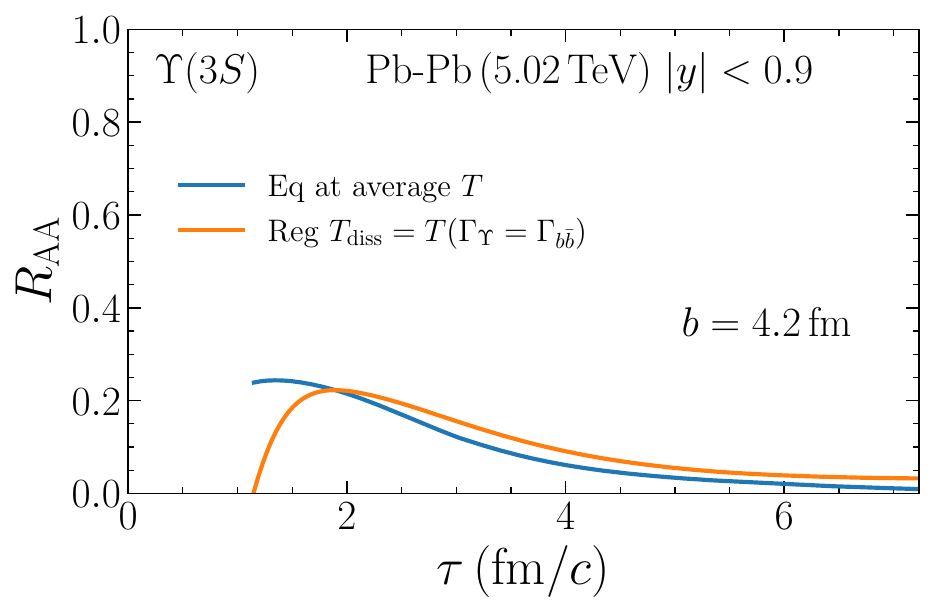}
   \caption{Time evolution of the $\raa$ for bottomonium suppression (top row) and regeneration (bottom row) for $\Upsilon(1S)$ (left panels), $\Upsilon(2S)$ (center panels), and $\Upsilon(3S)$ (right panels) in $b=4.2\,$fm Pb-Pb (5.023\,TeV) collisions at mid-rapidity with 20\% shadowing. In the lower panels, the blue curves show the equilibrium abundance (with the thermal relaxation factor) computed using the trajectory‐averaged temperatures; the orange curves correspond to regeneration starting at $\tau(T=T_{\rm melt})$.}
     \label{fig:LHC_time_evo_b4}
\end{figure}

\begin{figure}[t]
   \includegraphics[width=0.32\textwidth]{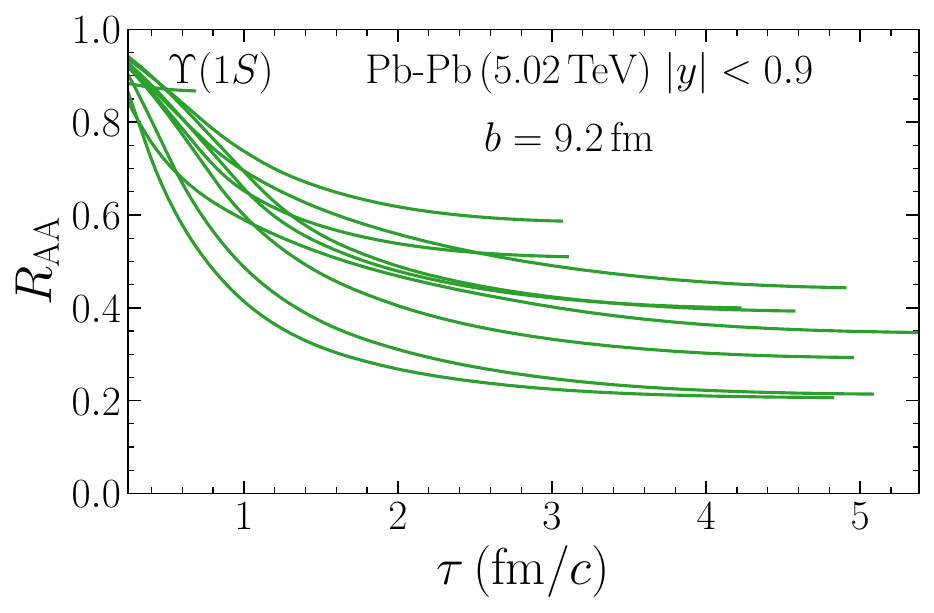}
   \hfill
   \includegraphics[width=0.32\textwidth]{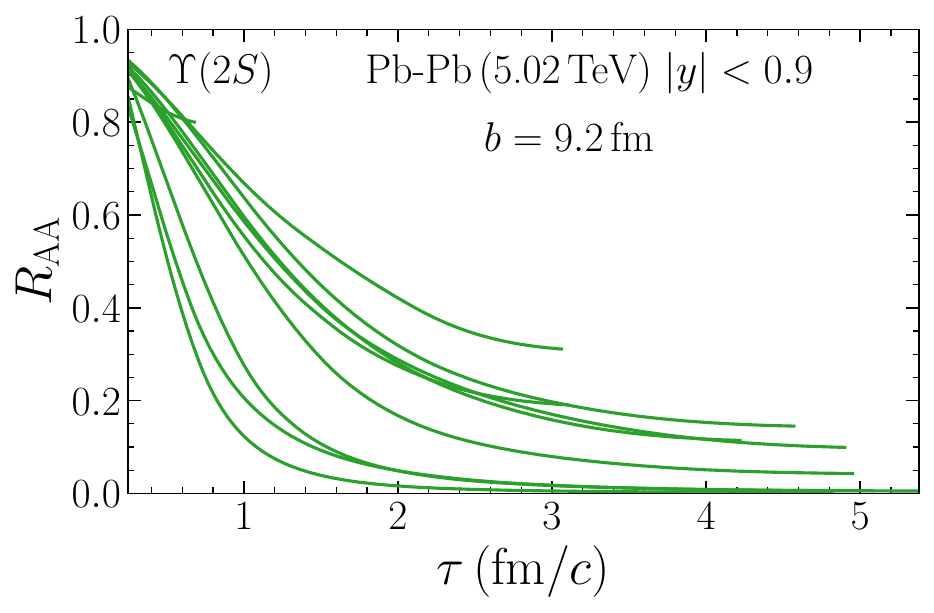}
   \hfill
   \includegraphics[width=0.32\textwidth]{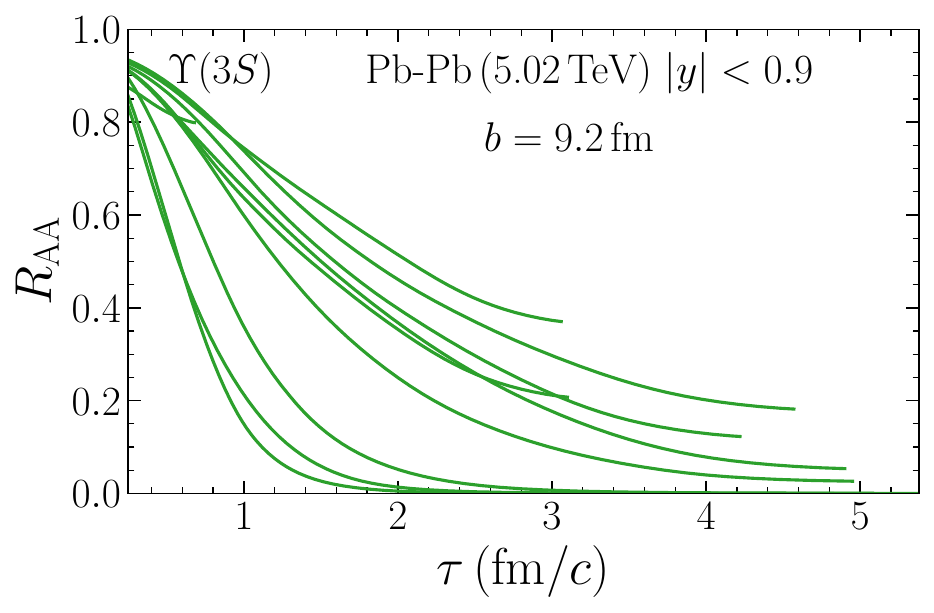}
   \\
   \includegraphics[width=0.32\textwidth]{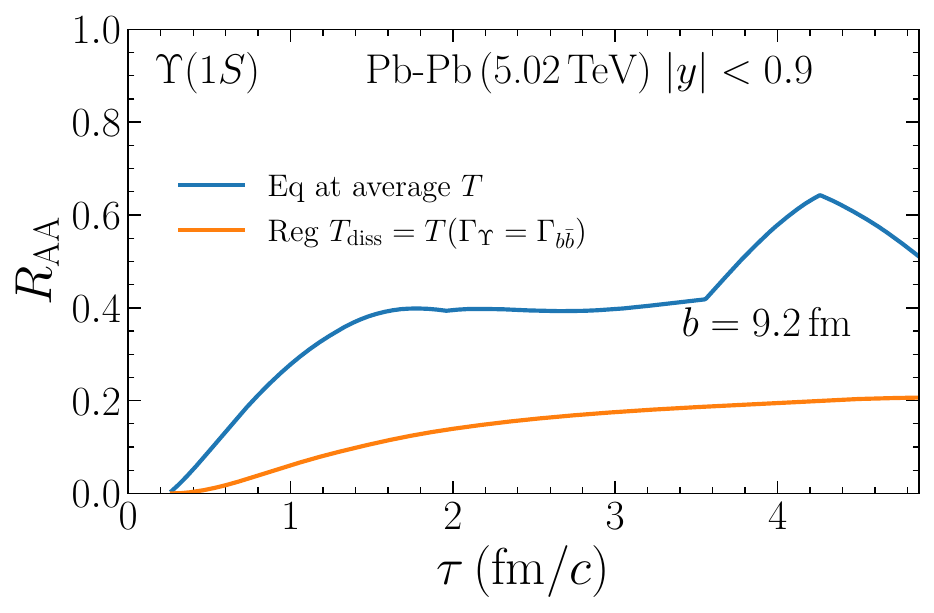}
   \hfill
   \includegraphics[width=0.32\textwidth]{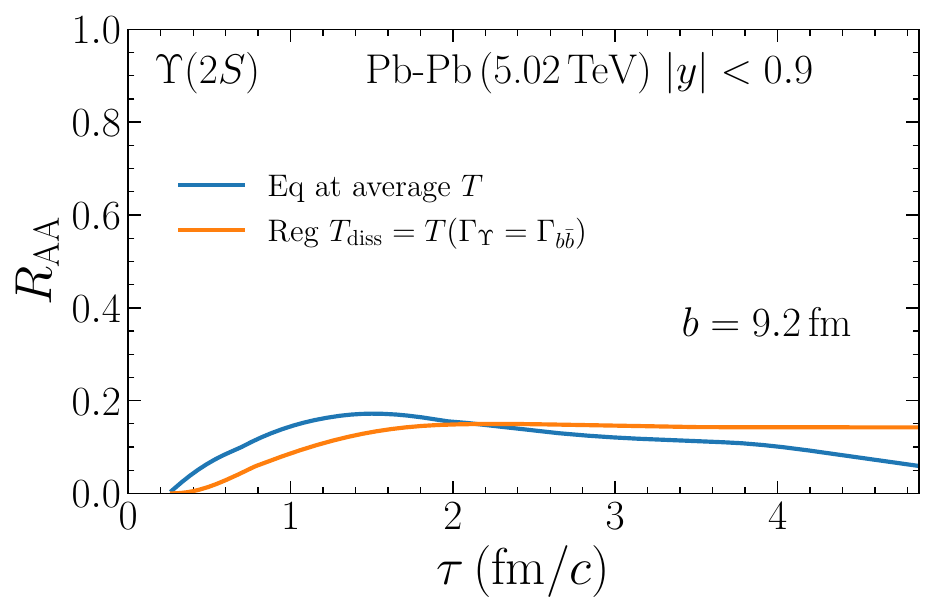}
   \hfill
   \includegraphics[width=0.32\textwidth]{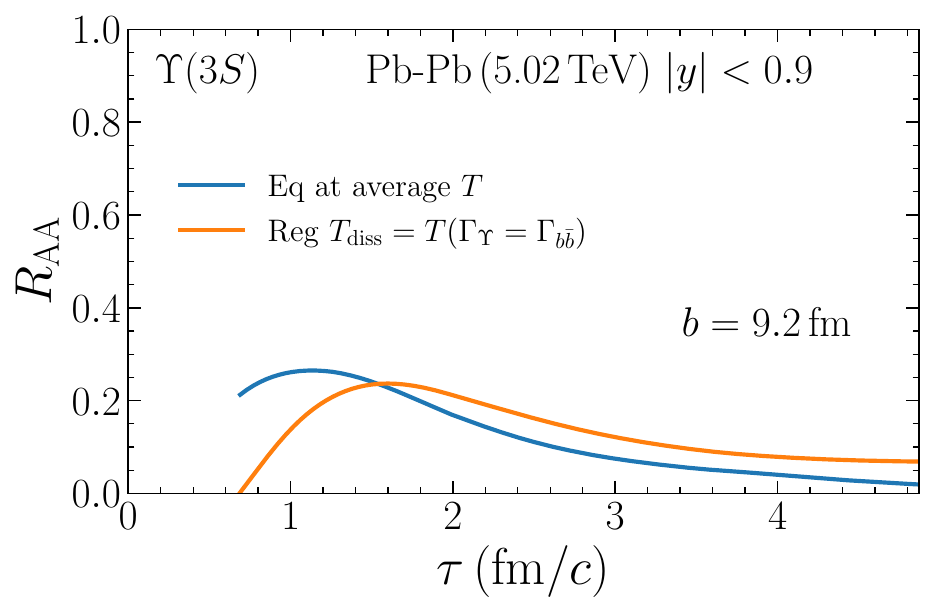}
   \caption{Same as Fig.~\ref{fig:LHC_time_evo_b4} but for $b=9.2\,$fm with 15\% shadowing.
   }
     \label{fig:LHC_time_evo_b9}
\end{figure}

In Figs.~\ref{fig:LHC_time_evo_b4} and~\ref{fig:LHC_time_evo_b9}, we summarize the time evolution of primordial and regenerated components for Pb-Pb collisions at the LHC at two different impact parameters, $b=4.2\,$fm and 9.2\,fm.
The upper panels show the survival probability along trajectories for the $1S$, $2S$, and $3S$ states (same trajectories for each state at each collision centrality; the final suppression factor is given by the average of the trajectory endpoints).
The lower panels illustrate the regeneration based on the trajectory-averaged temperature, $\Tavg(\tau)$, using the equilibrium limit constructed in Sec.~\ref{ssec:local_eq}.
As expected, the primordial $1S$ state has significantly higher survival probabilities than the $2S$ and $3S$ states, but the spread from different trajectories is rather large for all states.
The regeneration of the $1S$ state is coupled to an appreciable equilibrium population at the $\raa$ level, but it saturates well before the end of the evolution due to small reaction rates in the late stages.
By contrast, the $2S$ and $3S$ states have smaller equilibrium numbers and higher reaction rates, quickly reaching and remaining near their equilibrium limits until relatively late in the evolution.


\subsection{Bottomonium observables in Pb-Pb collisions at the LHC}
\label{ssec:lhc}
Section~\ref{sssec:npart} contains the centrality dependence of $1S$, $2S$, and $3S$ $\raa$'s and  their ratios at mid- and forward rapidity, while Sec.~\ref{sssec:pT} focuses on $\pT$ dependencies.

\subsubsection{Centrality dependence}
\label{sssec:npart}
In the comparisons to experimental data presented below, we include two 
sources of model uncertainty: CNM effects and different values for the bottomonium formation times. 
We denote the uncertainty associated with CNM effects by
$\Delta \raa^{\rm CNM}$, while the uncertainty associated with the nuclear
modification factor obtained from our transport calculation in the evolving
medium, including both suppression and regeneration, $\Delta \raa^{\rm QGP}$,
is estimated solely from variations of the bottomonium formation time. 
Since the two sources of uncertainty are uncorrelated, we evaluate the combined uncertainty in quadrature as
\begin{equation}
    \Delta \raa
    =
    \sqrt{
        \left(\Delta \raa^{\rm CNM}\right)^2
        +
        \left(\Delta \raa^{\rm QGP}\right)^2
     }\,,
\end{equation}
and display pertinent bands in our comparisons to data. 
Other possible sources of theoretical uncertainty, such as those related to the construction of the equilibrium limit or the initial-production spectra,  
are not included in the present uncertainty bands.
The two main components, \ie, the hydrodynamic evolution and the nonperturbative reaction rates, have been constrained independently and as such do not contain tunable parameters.

\begin{figure}[t]
        \includegraphics[width=0.325\textwidth]{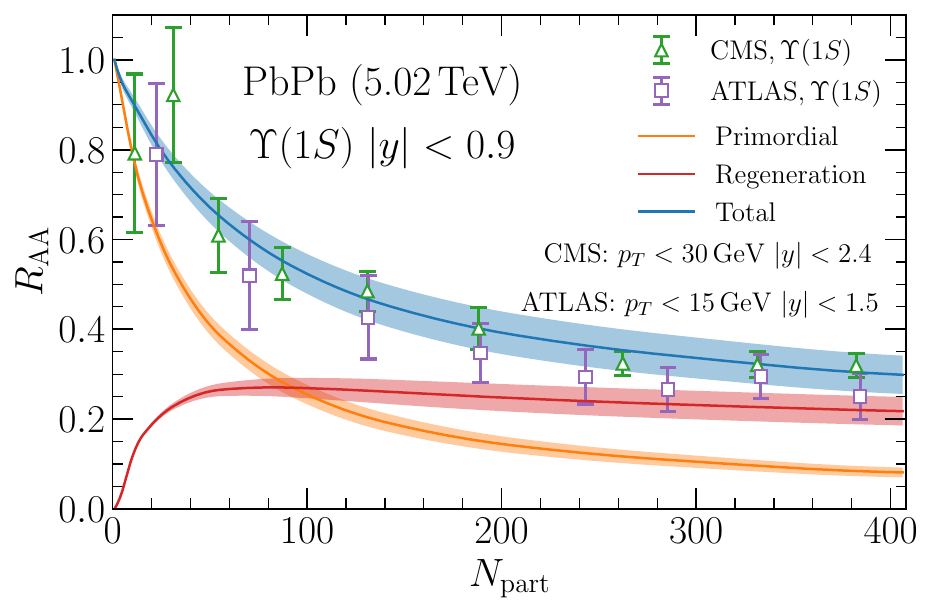 }
        \hfill
        \includegraphics[width=0.325\textwidth]{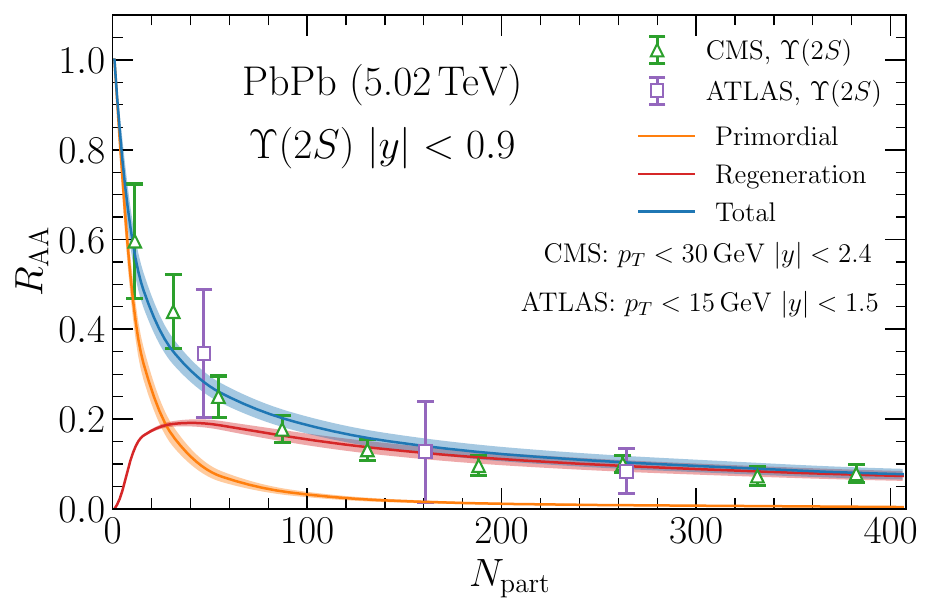 }
        \hfill
        \includegraphics[width=0.325\textwidth]{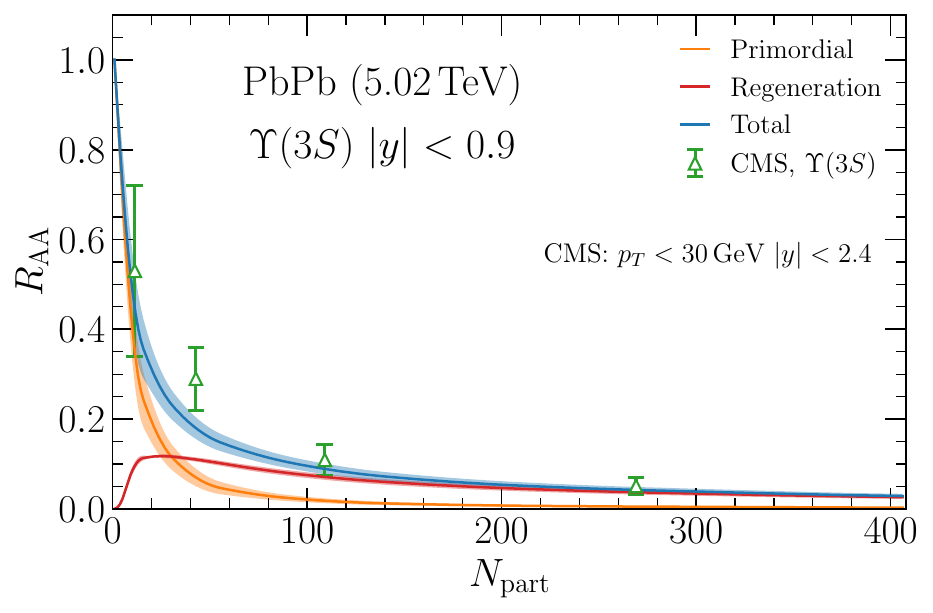 }
        \hfill
        \includegraphics[width=0.325\textwidth]{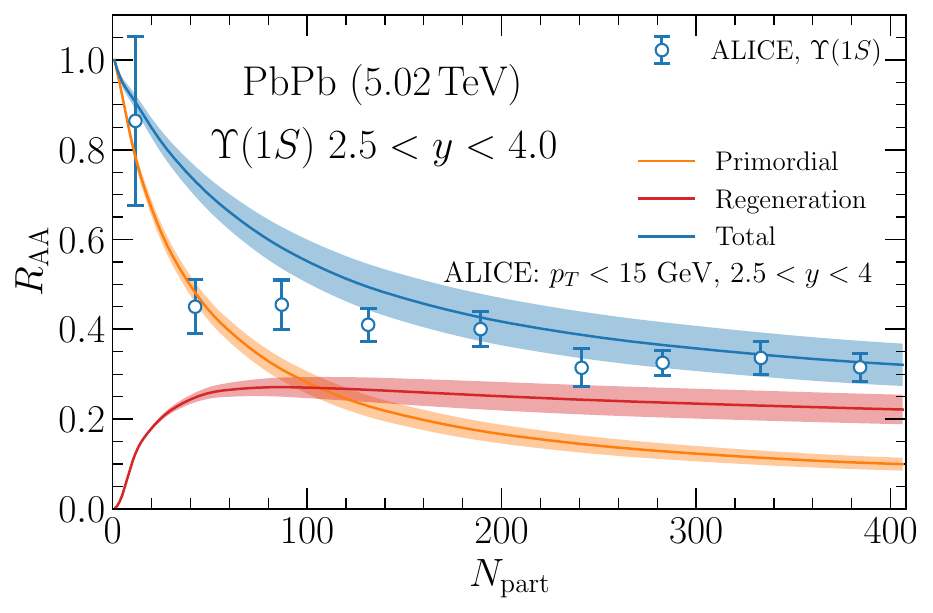 }
        \hfill
        \includegraphics[width=0.325\textwidth]{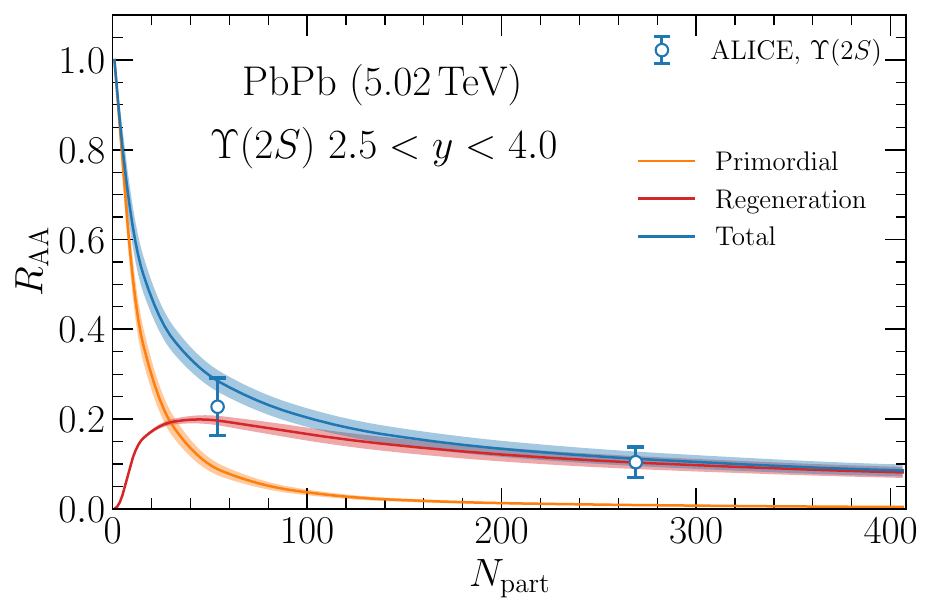 }
        \hfill
        \includegraphics[width=0.325\textwidth]{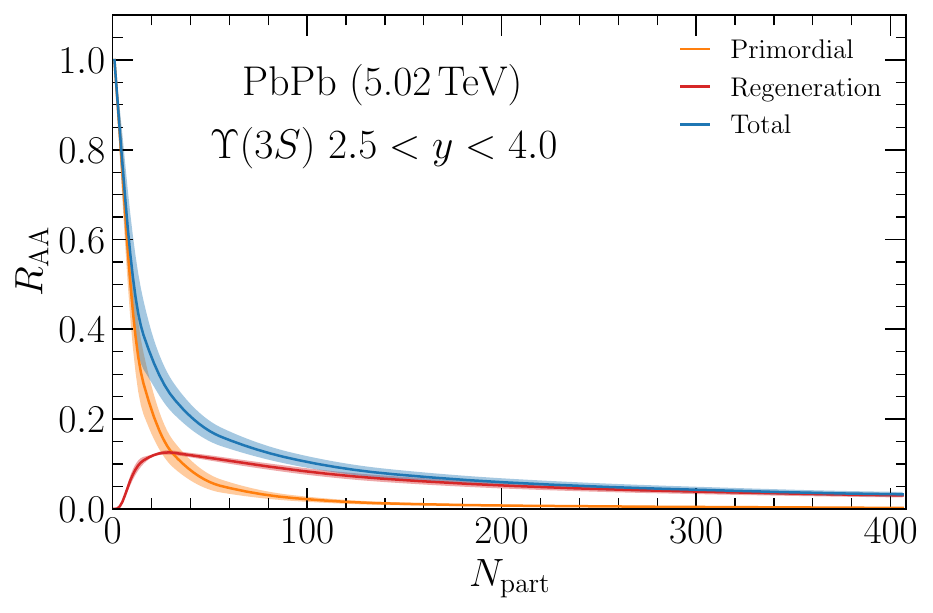 }
        \caption{
           Centrality dependence of the  nuclear modification factor of $\Upsilon(1S)$ (left panels), $\Upsilon(2S)$ (middle panels), and $\Upsilon(3S)$ (right panels) at mid-rapidity (upper row) and forward rapidity (lower row), compared to data from CMS~\cite{CMS:2018zza,CMS:2023lfu}, ATLAS~\cite{ATLAS:2022exb}, and ALICE~\cite{ALICE:2018wzm,ALICE:2020wwx}. A cut of $\pT<30(15)$\,GeV is applied in the upper (lower) panels to match the CMS (ALICE) data, while the ATLAS data have a 15\,GeV cut.
   The bands encompass an up to 15-35\% shadowing at mid-rapidity and 20-40\% at forward rapidity.}
        \label{fig:raa-npart-lhc}
     \end{figure}

In Fig.~\ref{fig:raa-npart-lhc}, we collect our results for the $\raa$ of $\Upsilon(1S)$, $\Upsilon(2S)$, and $\Upsilon(3S)$
as a function of the number of participants, in comparison to mid- and forward rapidity data from
CMS~\cite{CMS:2018zza,CMS:2023lfu}, ATLAS~\cite{ATLAS:2022exb}, and ALICE~\cite{ALICE:2018wzm,ALICE:2020wwx}.
At mid-rapidity, our predictions agree fairly well with the experimental measurements from ATLAS and CMS across the entire centrality range.
Notably, the much larger reaction rates compared to previous studies cause markedly stronger suppression and regeneration. In particular, the regeneration contribution to the $\raa$ for $\Upsilon(1S)$ exceeds the primordial one already for mid-central collisions, while for the  $\Upsilon(2S)$ and $\Upsilon(3S)$, regeneration starts to exceed the primordial component already in rather peripheral collisions and dominates the yield in central collisions.
We also note that in the upper panels the calculations and the CMS data include a $\pT<30$\,GeV cut, while the ATLAS data are for $\pT<15$\,GeV. When applying the latter (instead of the former) to our calculations at mid-rapidity, both primordial and regeneration  contributions are essentially unaffected.
At forward rapidity (lower panels in Fig.~\ref{fig:raa-npart-lhc}), the results are very similar to those at mid-rapidity. A slight hint of less suppression can be discerned, caused by the somewhat smaller temperatures and lifetimes of the fireball, somewhat mitigated by 5\% stronger shadowing. 
At the same time, the regeneration contribution is also very similar to the mid-rapidity results. The agreement with ALICE data for the $\Upsilon(1S)$ and $\Upsilon(2S)$ is fair, except for the $1S$ in peripheral collisions.

\begin{figure}[t]
        \includegraphics[width=0.32\textwidth]{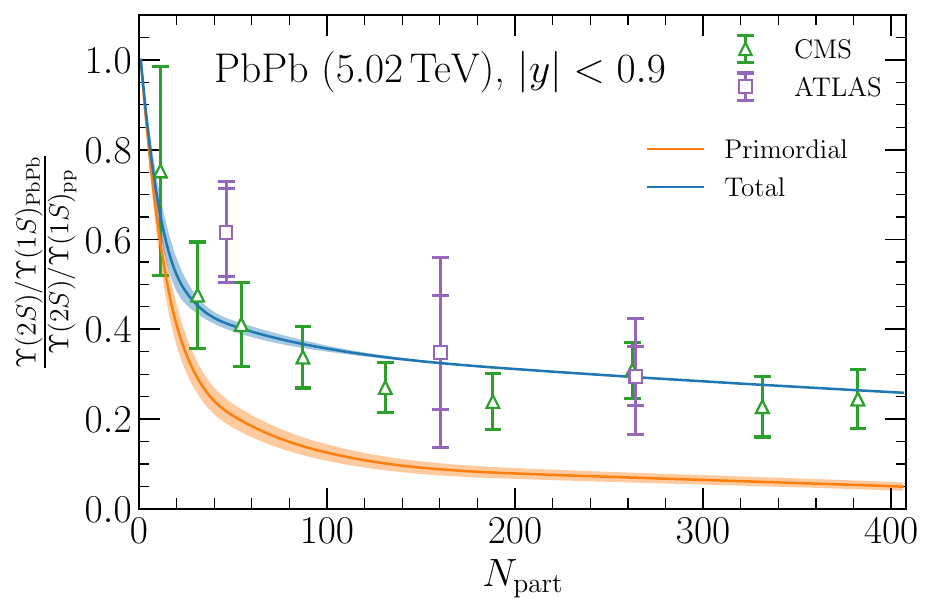 }
        \hfill
        \includegraphics[width=0.32\textwidth]{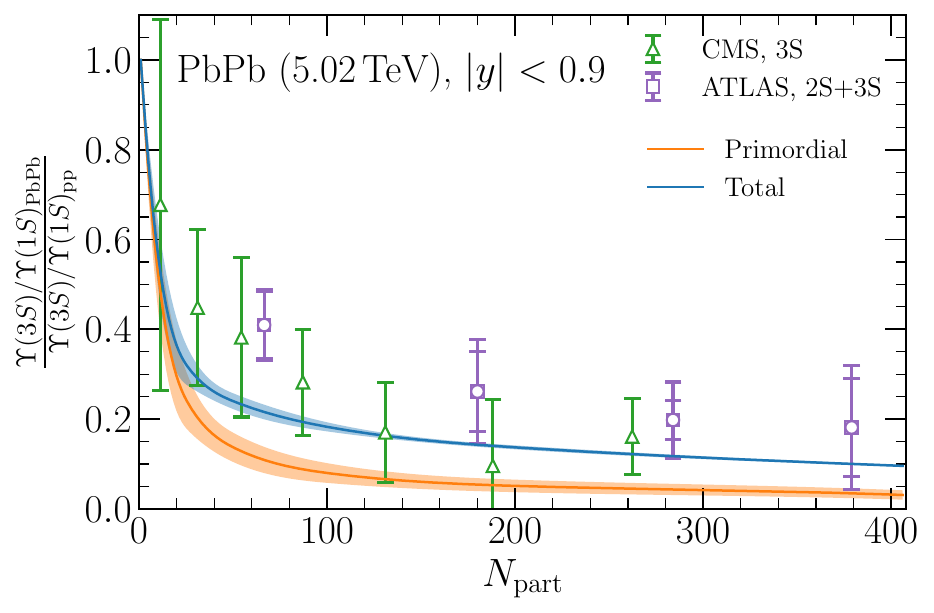 }
        \hfill
        \includegraphics[width=0.32\textwidth]{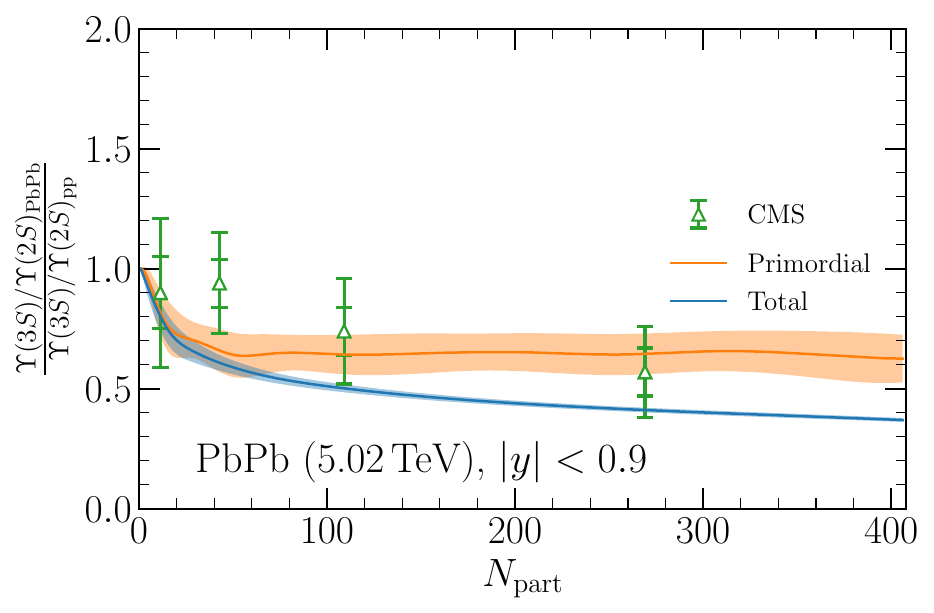 }
        \hfill
        \includegraphics[width=0.32\textwidth]{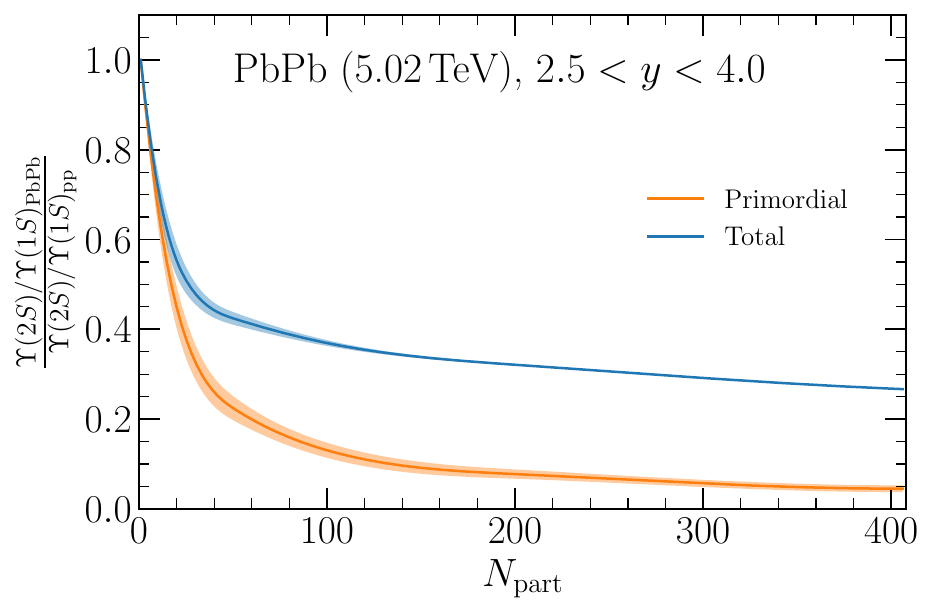 }
        \hfill
        \includegraphics[width=0.32\textwidth]{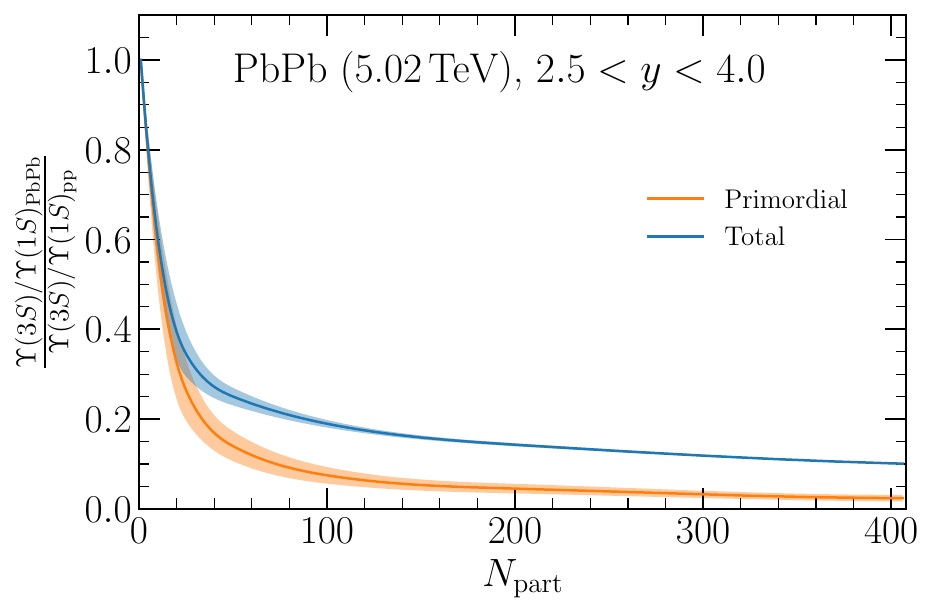 }
        \hfill
        \includegraphics[width=0.32\textwidth]{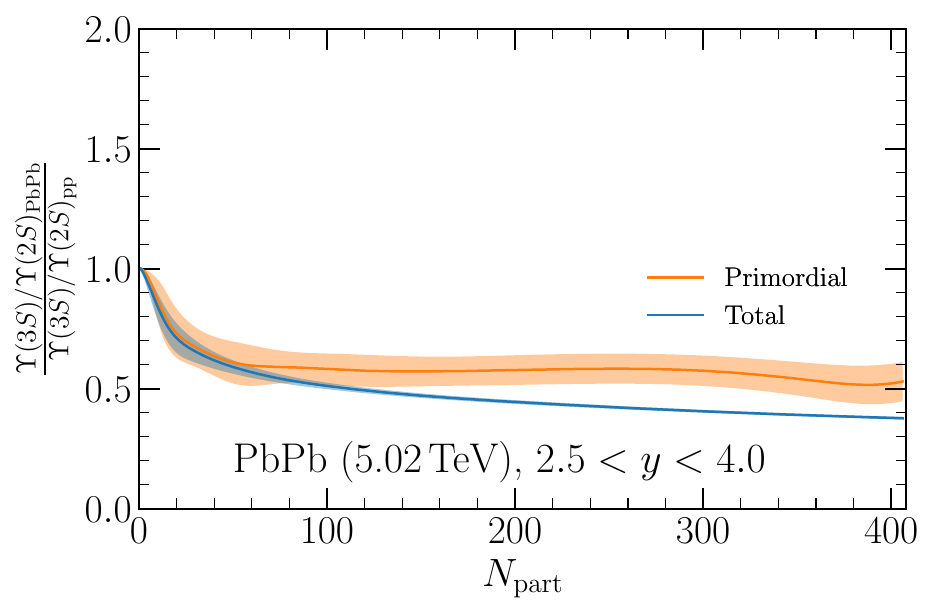 }
        \caption{
           Double ratios of $\Upsilon(2S)/\Upsilon(1S)$ (left panels), $\Upsilon(3S)/\Upsilon(1S)$ (middle panels), and $\Upsilon(3S)/\Upsilon(2S)$ (right panels) as a function of centrality in Pb-Pb (5.02\,TeV) collisions at mid-rapidity (top row) and forward rapidity (bottom row), compared to CMS~\cite{CMS:2018zza,CMS:2023lfu} and ATLAS~\cite{ATLAS:2022exb} data. Note that the upper (lower) panels include a $\pT<30(15)$\,GeV cut. 
        }
        \label{fig:double_ratio}
     \end{figure}

To further scrutinize the interplay of the relative modification of the different bottomonium states, we display double ratios in Fig.~\ref{fig:double_ratio}, defined as
\begin{equation}
\frac{\raa[\Upsilon(2S)]}{\raa[\Upsilon(1S)]}\,,
\quad
\frac{\raa[\Upsilon(3S)]}{\raa[\Upsilon(1S)]}\,,
\quad\text{and}\quad
\frac{\raa[\Upsilon(3S)]}{\raa[\Upsilon(2S)]}\,.
\end{equation}
From an experimental perspective, these observables reduce common normalization uncertainties and may provide a more sensitive probe of the interplay between suppression and regeneration. 
In particular, suppression alone tends to reduce the excited-to-ground-state ratios because of the weaker binding of the excited states, whereas regeneration tends to modify this hierarchy, especially toward more central collisions.
At mid-rapidity, the double ratios involving the $\Upsilon(1S)$ denominator decrease rapidly from peripheral to mid-central collisions.
For $\Upsilon(2S)/\Upsilon(1S)$, suppression alone tends to underestimate this ratio, while the full result with regeneration lies in the upper range of the data.
An even stronger reduction is obtained for $\Upsilon(3S)/\Upsilon(1S)$, with a  preference for the full over the suppression-only calculation. 
The difference is less pronounced for the $\Upsilon(3S)/\Upsilon(2S)$ double ratio, but the data do not yet allow for a definite discrimination.
In sufficiently central collisions, where the regeneration components of $2S$ and $3S$ approach their equilibrium limits, the total $\Upsilon(3S)/\Upsilon(2S)$ double ratio becomes a direct measure of regeneration.
For both mid- and forward rapidity, regeneration increases the double ratios with the $\Upsilon(1S)$ in the denominator,
while for $\Upsilon(3S)/\Upsilon(2S)$ a (modest) opposite trend occurs. 
This indicates that regeneration enhances the $\Upsilon(2S)$ yield more strongly, in relative terms, than the $\Upsilon(3S)$ yield.

\begin{figure}[b]
    \centering
        \includegraphics[width=0.32\textwidth]{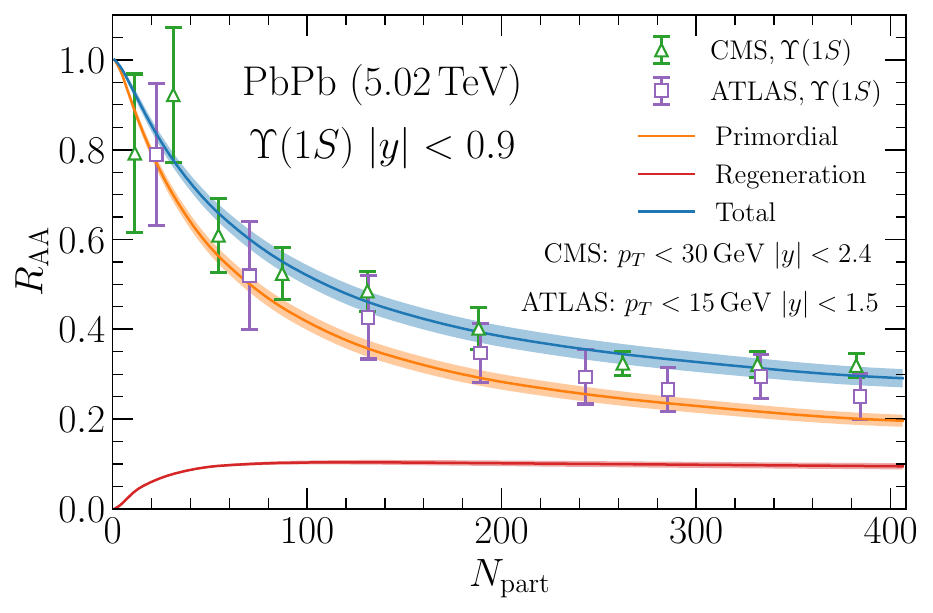 }
        \includegraphics[width=0.32\textwidth]{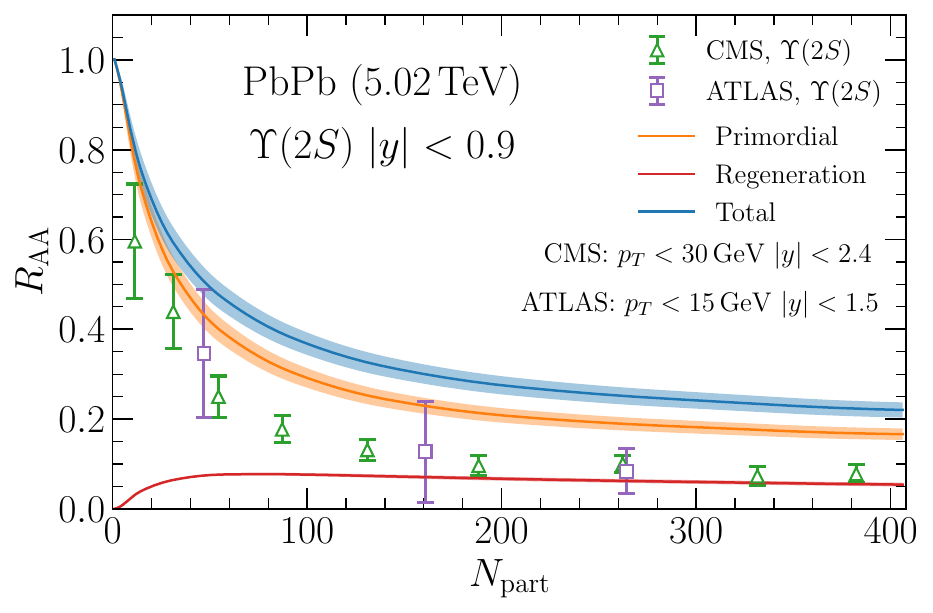 }
        \includegraphics[width=0.32\textwidth]{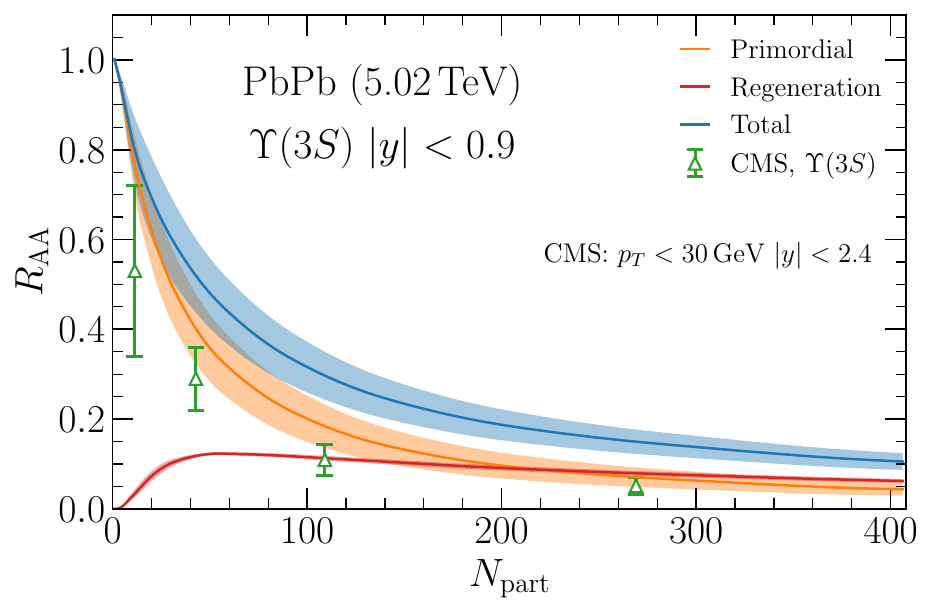 }
        \caption{Results for the $\raa$ of $\Upsilon(1S)$ (left panel), $\Upsilon(2S)$ (middle panel), and $\Upsilon(3S)$ (right panel) 
        in mid-rapidity Pb-Pb (5.02\,TeV) collisions using previous reaction rates with a perturbative coupling to the QGP medium~\cite{Du:2017qkv}.
        The bands represent a shadowing range of up to 15-35\% in central collisions, and the results are compared to ATLAS~\cite{ATLAS:2022exb} and CMS~\cite{CMS:2018zza,CMS:2023lfu} data.}
        \label{fig:npart_pert} 
\end{figure}

To illustrate how the changes from previously used reaction rates (based on a perturbative coupling to the QGP medium with binding energies obtained from an internal-energy potential)~\cite{Du:2017qkv} to the $T$-matrix rates~\cite{Wu:2025hlf} affect the interpretation of observables, we show the results from the former, implemented into the same hydrodynamic background, in Fig.~\ref{fig:npart_pert}. 
The main difference is that the much smaller magnitude of the previous reaction rates generally leads to less suppression and less regeneration than what we find from the $T$-matrix rates. While  this works out reasonably well in comparisons to $\Upsilon(1S)$ data, a notable lack of overall suppression for the  $\Upsilon(2S)$ and $\Upsilon(3S)$ is found (this was not the case in Ref.~\cite{Du:2017qkv}, which can be traced back to the longer QGP lifetimes in the fireball model compared to the hydro evolution). This is a clear indication that the perturbative rates for small binding energies, which approach twice the free heavy-quark scattering rate, are not large enough. This is reassuring to see since the large values of the $T$-matrix reaction rates are directly related to the heavy-quark transport coefficients in the QGP, and the predicted interaction strength is required to describe the $\raa$ and $v_2$ of charm hadrons at the LHC~\cite{Krishna:2025bll}.

\subsubsection{Transverse-momentum spectra}
\label{sssec:pT}
\begin{figure}[t]
   \includegraphics[width=0.32\textwidth]{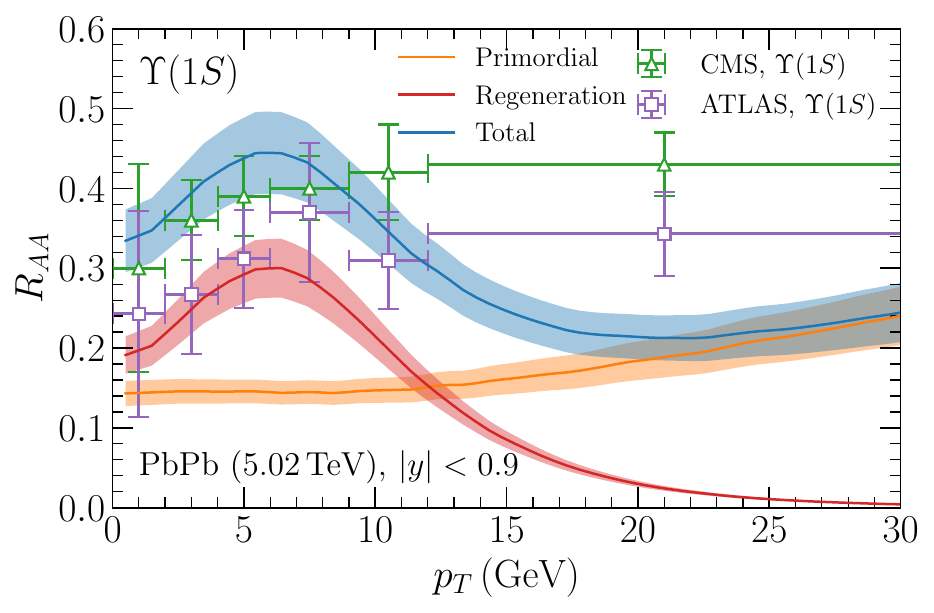 }
   \hfill
   \includegraphics[width=0.32\textwidth]{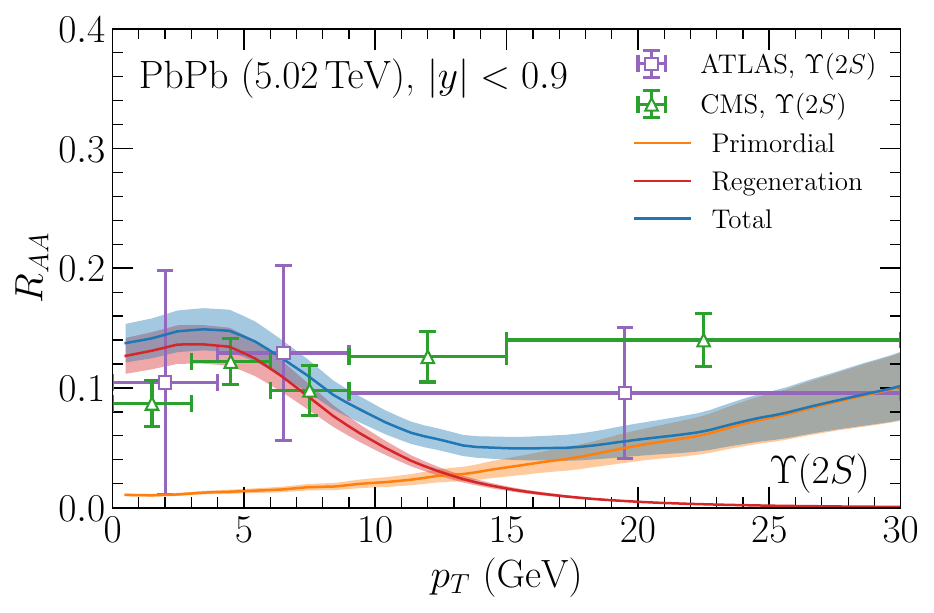 }
   \hfill
   \includegraphics[width=0.32\textwidth]{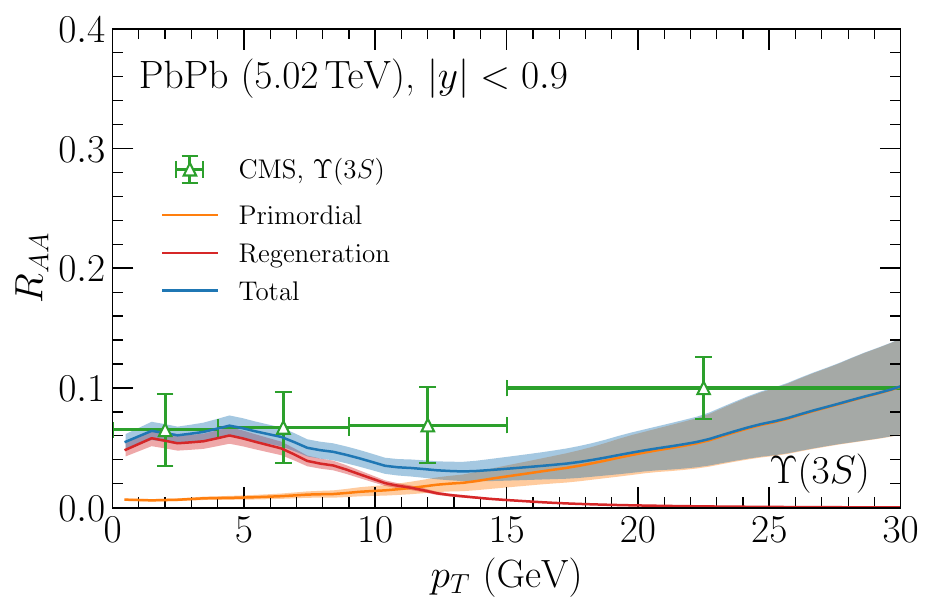 }
   \hfill
   \includegraphics[width=0.32\textwidth]{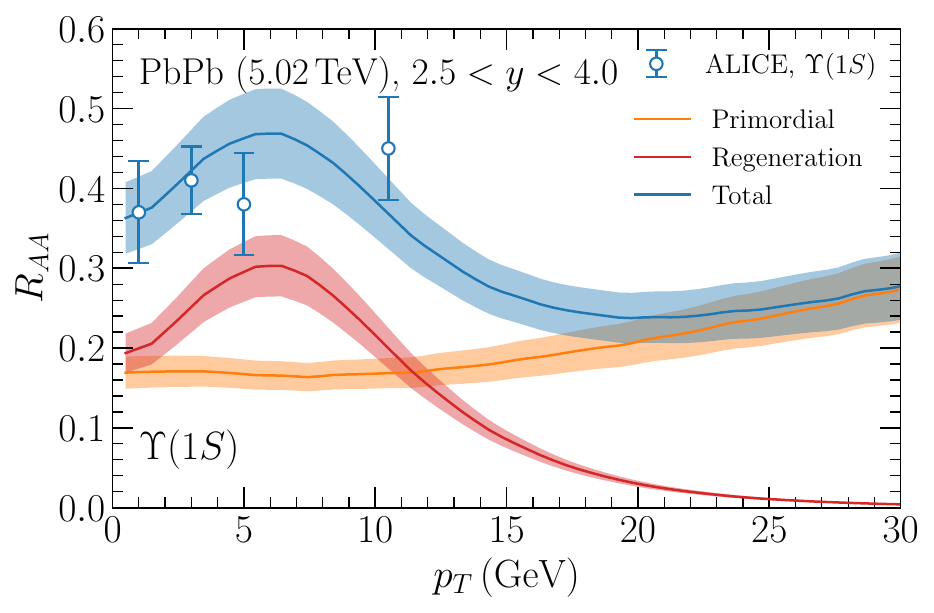 }
   \hfill
   \includegraphics[width=0.32\textwidth]{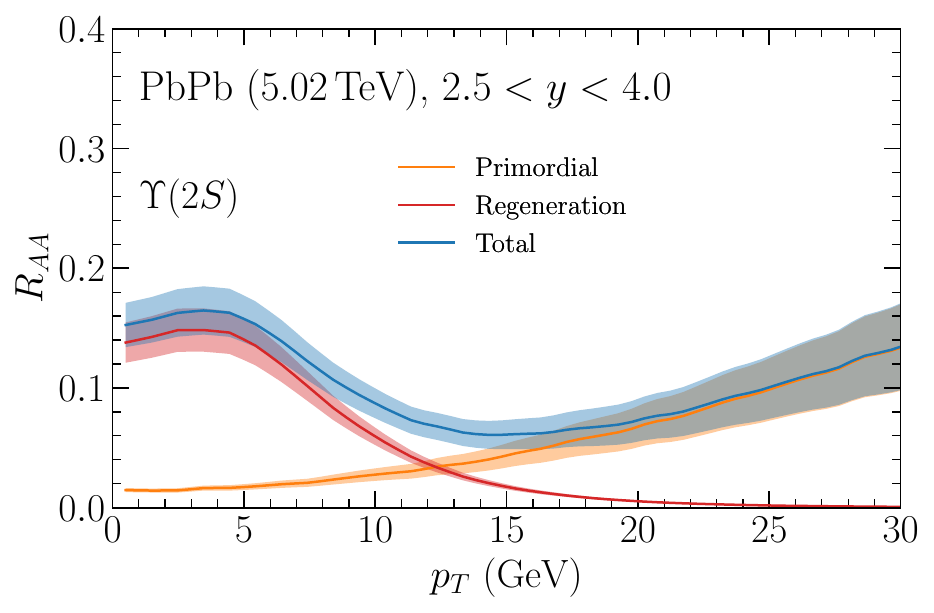 }
   \hfill
   \includegraphics[width=0.32\textwidth]{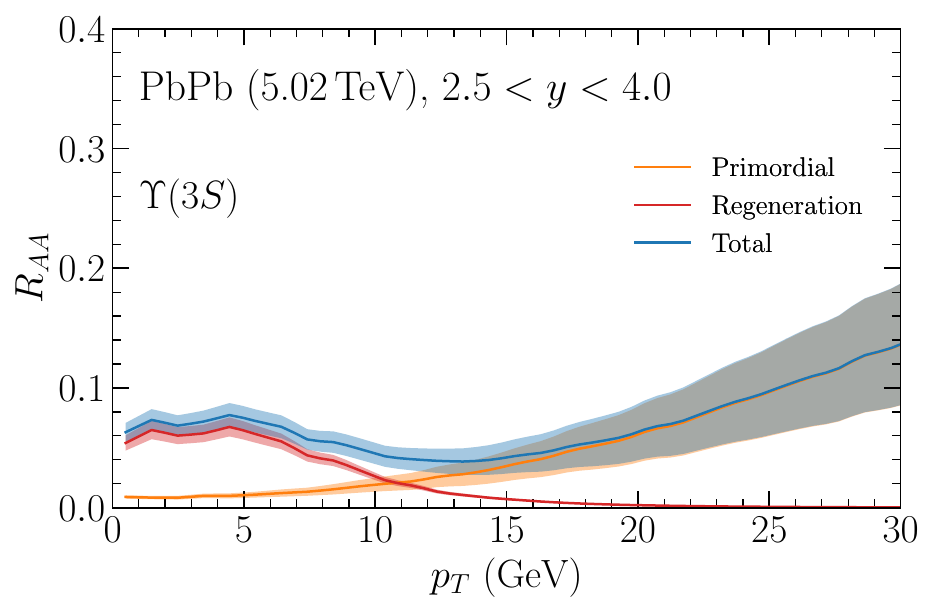 }
   \caption{
      Nuclear modification factors vs.~transverse momentum for the total (blue), primordial (orange) and regenerated (red) contributions to $\Upsilon(1S)$ (left panels), $\Upsilon(2S)$ (middle panels), and $\Upsilon(3S)$ (right panels) production at mid- (upper row) and forward rapidity (lower row) in minimum-bias Pb-Pb (5.02\,TeV) collisions.
   The bands represent up to 15-35\% shadowing at mid-rapidity and 20-40\% at forward rapidity.
   }
\label{fig:LHC_pt}
\end{figure}
The calculation of the $\pT$ dependence of the two production sources has been described in detail above. For the primordial source, the $\pT$-dependent reaction rates are deployed along the trajectories of each $Y$ state (augmented by formation time effects early on). For regeneration we use the ICM to compute the $\pT$ spectra from transported $b$ and $\bar b$ quarks at an average $Y$ production time.
Our combined results for the nuclear modification factor, $\raa(\pT)$,
for inclusive $1S$, $2S$, and $3S$ bottomonia at mid-rapidity and forward rapidity in minimum-bias Pb-Pb collisions are summarized in Fig.~\ref{fig:LHC_pt} and compared to available data.
At low $\pT\lesssim 10$~GeV, regeneration is the leading source for all three states, to varying degrees: it is comparable for the $1S$, and increasingly dominant for the $2S$ and $3S$, relative to suppressed primordial production. The shape of the regeneration contribution becomes softer when going from $1S$ via $2S$ to $3S$ (\ie, with larger accumulation at small $\pT$) indicating subsequently later production times at which the $b$ quark spectra have softened due to diffusion. 
\begin{figure}[t]
     \centering
   \includegraphics[width=0.48\textwidth]{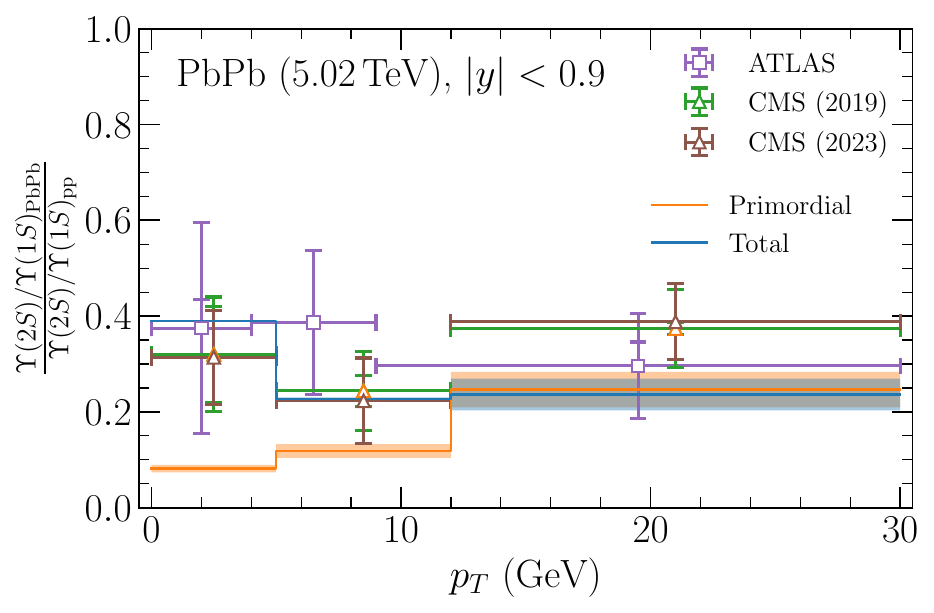 }
   \includegraphics[width=0.48\textwidth]{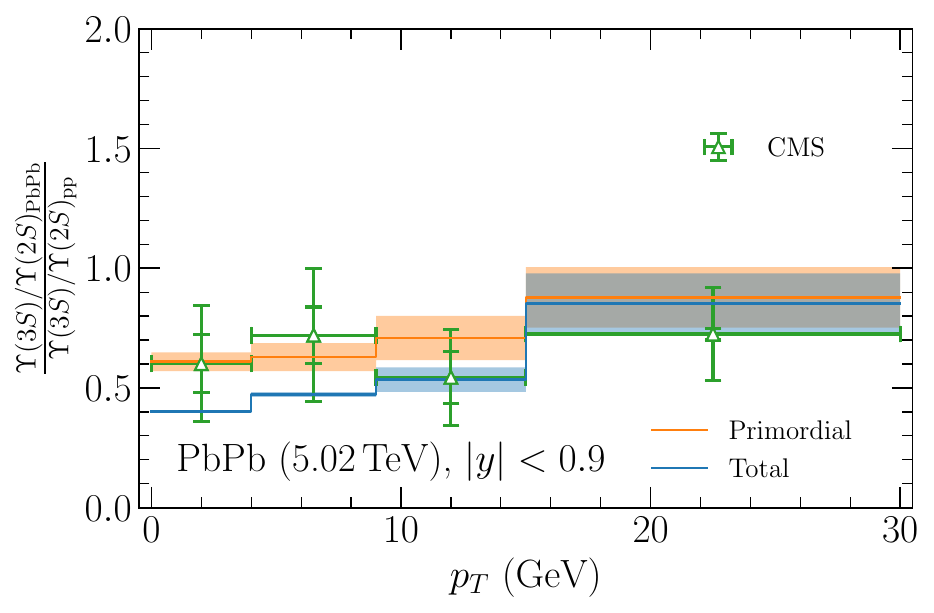 }
   \includegraphics[width=0.48\textwidth]{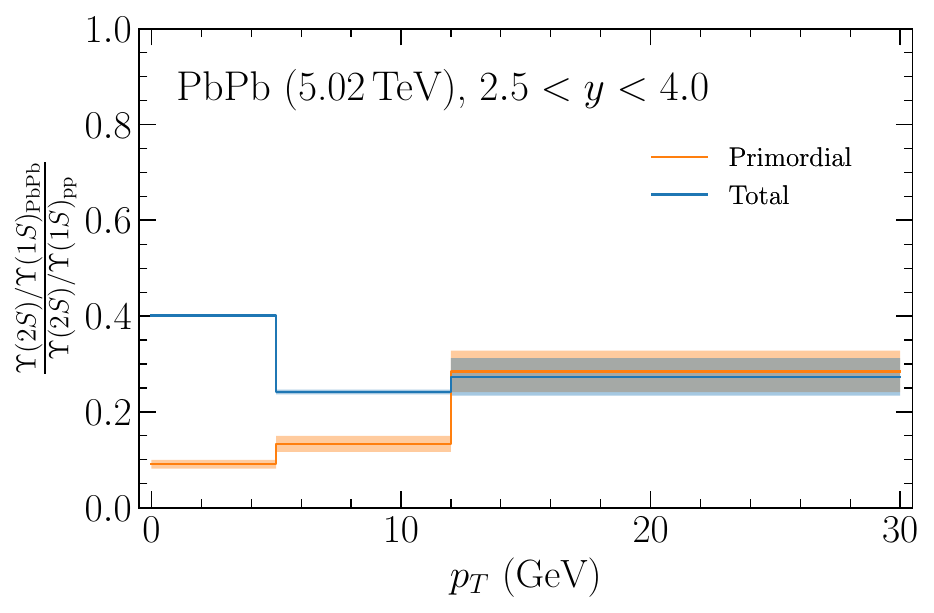 }
   \includegraphics[width=0.48\textwidth]{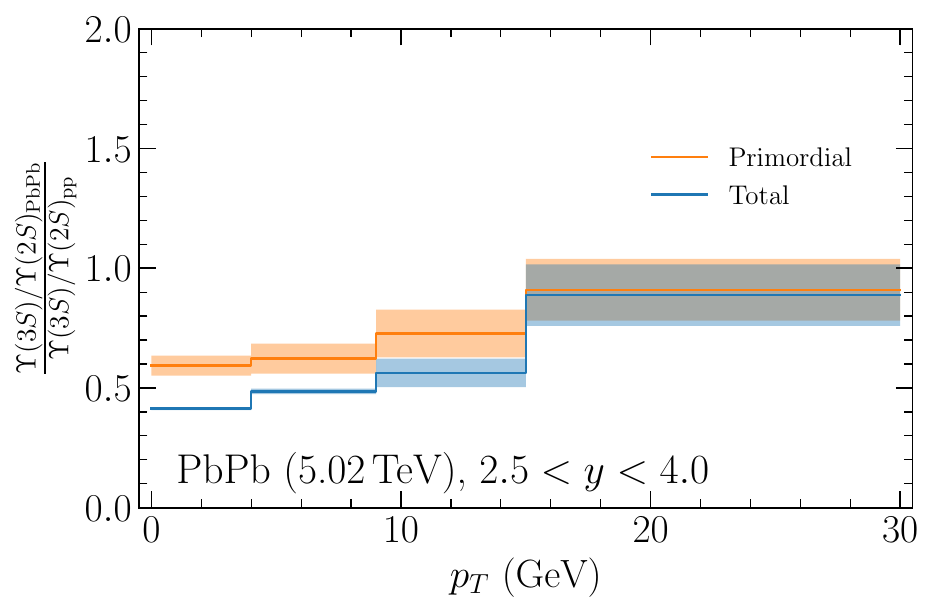 }
   \caption{
      Transverse-momentum-dependent double ratios for $\Upsilon(2S)/\Upsilon(1S)$ (left panels) and $\Upsilon(3S)/\Upsilon(2S)$ (right panels) in 
      minimum-bias Pb-Pb (5.02\,TeV) collisions at mid- (upper panels) and forward rapidity (lower panels) for the total (blue lines) and primordial (orange lines) components.
   The bands represent up to 15-35\% (20-40\%) shadowing at  mid- (forward) rapidity.}
   \label{fig:LHC_double_pt}
\end{figure}
This is presumably also the reason why the crossing of the regenerated and primordial components occurs at roughly the same $\pT$ for all states. 
The primordial curves exhibit a rise with increasing $\pT$, especially for the excited states, which is a combination of the $\pT$-dependent rates (which fall off with momentum, especially for small binding, see Fig.~16 in Ref.~\cite{Wu:2025hlf}) and formation-time effects which time-dilate the suppression mechanism. For both excited states, there is hardly any primordial yield left at low $\pT$. For all states, the primordial yield is the sole source for $\pT\gsim 2m_Y$. However, as emphasized in our previous work~\cite{Wu:2025hlf}, some tension exists in that the calculations underestimate the data for $\pT\gsim 10$\,GeV. We believe that the most likely reason is the production mechanism, which may change from the direct $\bbb$ production and subsequent wave-packet evolution to (gluon) fragmentation processes where the $\raa$ is rather determined from the suppression of the parent gluon in the medium. Also, space-momentum correlations between recombining $b$ and $\bar b$ pairs likely extend the $\pT$ reach of recombination, although probably not enough to resolve the discrepancies. 
The forward and mid-rapidity results share essentially the same features, even at a rather quantitative level.

Next, we analyze the $\pT$-dependent double ratios, plotted in Fig.~\ref{fig:LHC_double_pt}. We note that $2S$ regeneration plays a key role in the low-$\pT$ $2S$/$1S$ ratio (which was already apparent from the $\raa$), while for the $3S$/$2S$ case the situation is less conclusive. 
For both double ratios, the high-$\pT$ region is well described. For the $2S/1S$ case, the ratio of $\sim$0.3 suggests that bound-state structure effects (\eg, binding energy and/or size) are still relevant, while for the $3S/2S$ case, with a double-ratio near 1, this is not the case.

\begin{figure}[t]
   \centering
   \includegraphics[width=0.48\textwidth]{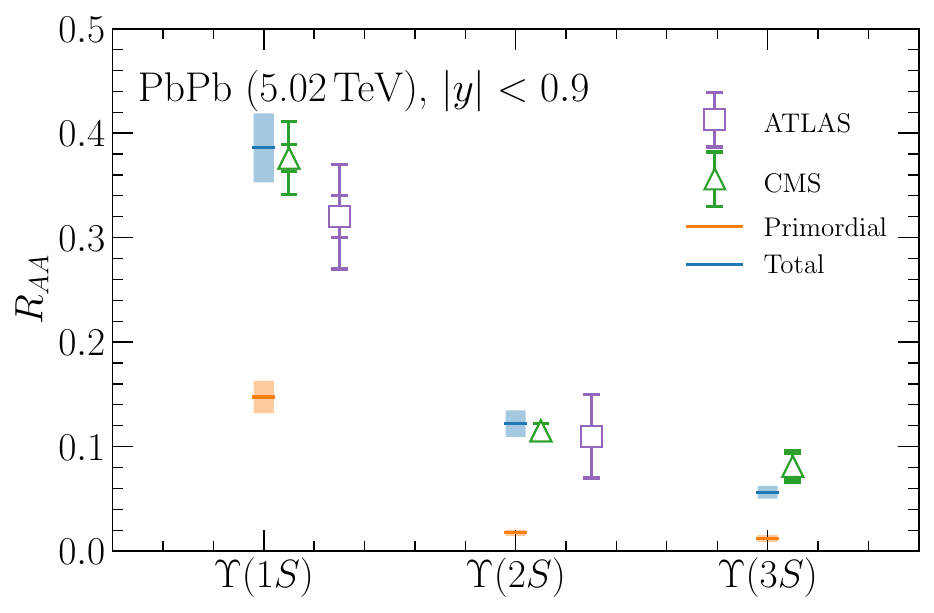 }
   \includegraphics[width=0.48\textwidth]{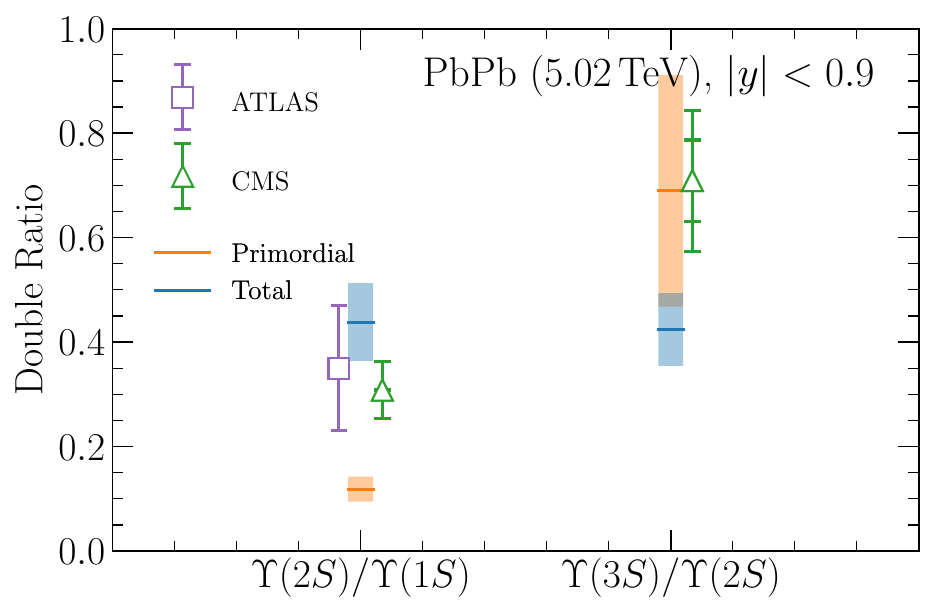 }
   \includegraphics[width=0.48\textwidth]{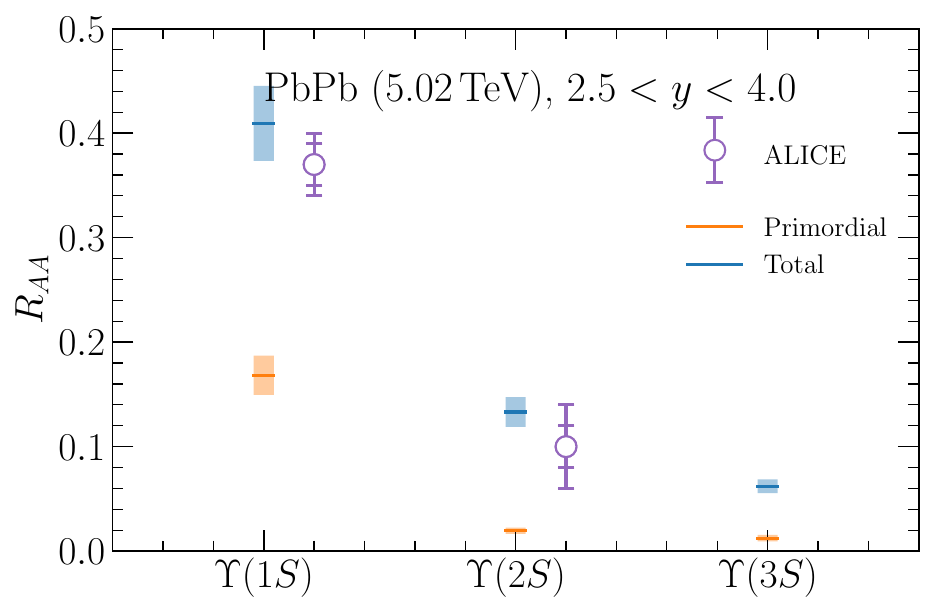 }
   \includegraphics[width=0.48\textwidth]{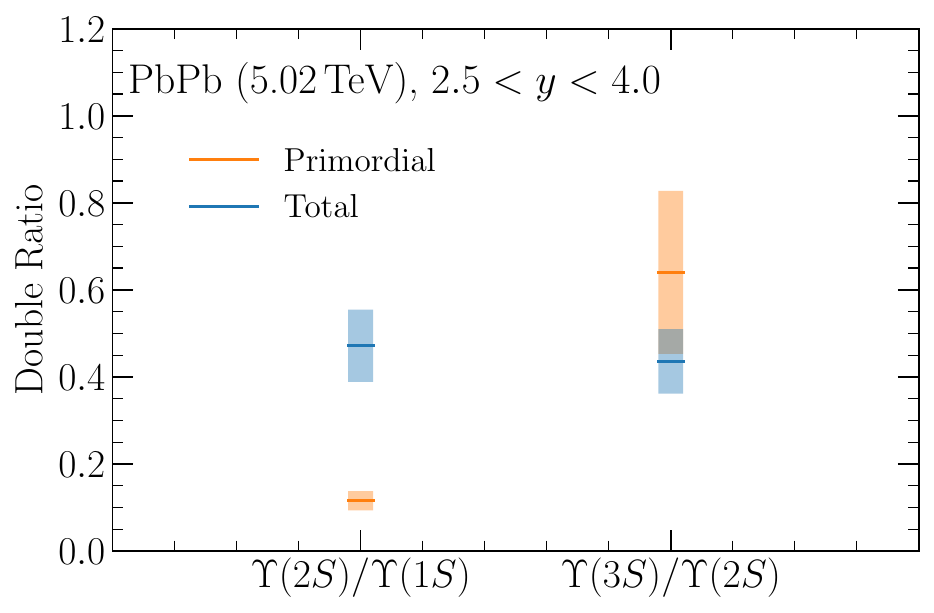 }
   \caption{Transverse-momentum-integrated nuclear modification factors (left panels) and double ratios (right panels) for the total (blue lines) and primordial (orange lines) components in minimum-bias Pb-Pb (5.02\,TeV) collisions at mid- (upper panels) and forward rapidity (lower panels). 
   The bands represent an up to 15-35\% shadowing at mid-rapidity (upper panels) and 20-40\% at forward rapidity (lower panels).}
   \label{fig:pt_integrated}
\end{figure}

In Fig.~\ref{fig:pt_integrated} we compile the nuclear modification factors (left panels) for the $\pT$-integrated yields of $\Upsilon(1S)$, $\Upsilon(2S)$, and $\Upsilon(3S)$ and their ratios in minimum-bias Pb-Pb (5.02\,TeV) collisions. 
At mid-rapidity, the $\raa$ data exhibit a sequential suppression pattern, which, however, in the calculations is strongly (yet roughly uniformly) influenced by regeneration contributions which are critical for the agreement with the data. 
In the $3S/2S$ double ratio one observes some tension in terms of an underestimation in the calculations; recalling  the $\pT$-dependent $\raa$'s from Fig.~\ref{fig:LHC_pt}, this appears to be mostly driven by the overestimate of the low-$\pT$ $2S$ spectra, and by a slight underestimate of the $3S$ data at intermediate $\pT$.

\subsection{Bottomonium observables in Au-Au collisions at RHIC}
\label{ssec:rhic}
\begin{figure}[t]
        \includegraphics[width=0.48\textwidth]{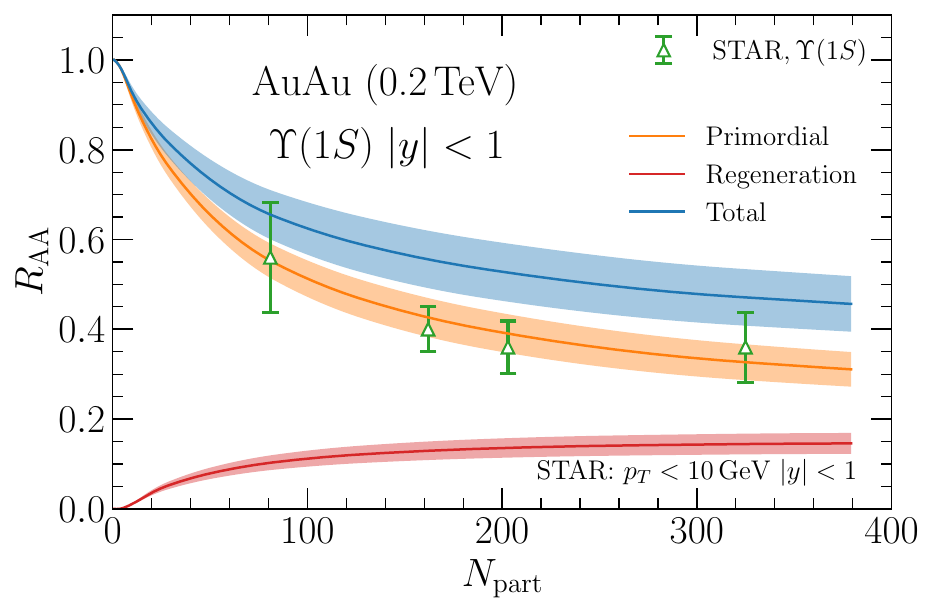 }
        \hfill
        \includegraphics[width=0.48\textwidth]{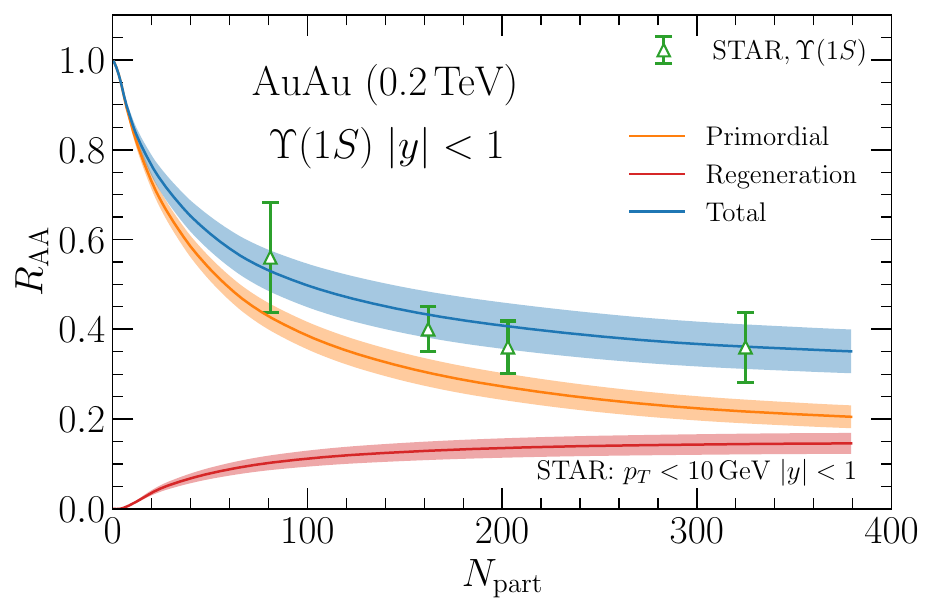 }
       \hfill
      \includegraphics[width=0.48\textwidth]{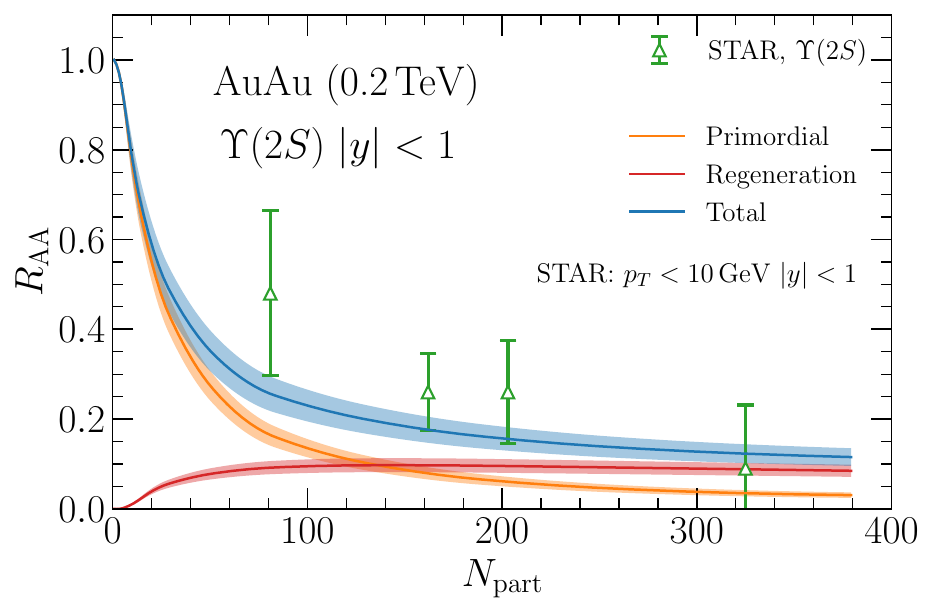 }
        \hfill
       \includegraphics[width=0.48\textwidth]{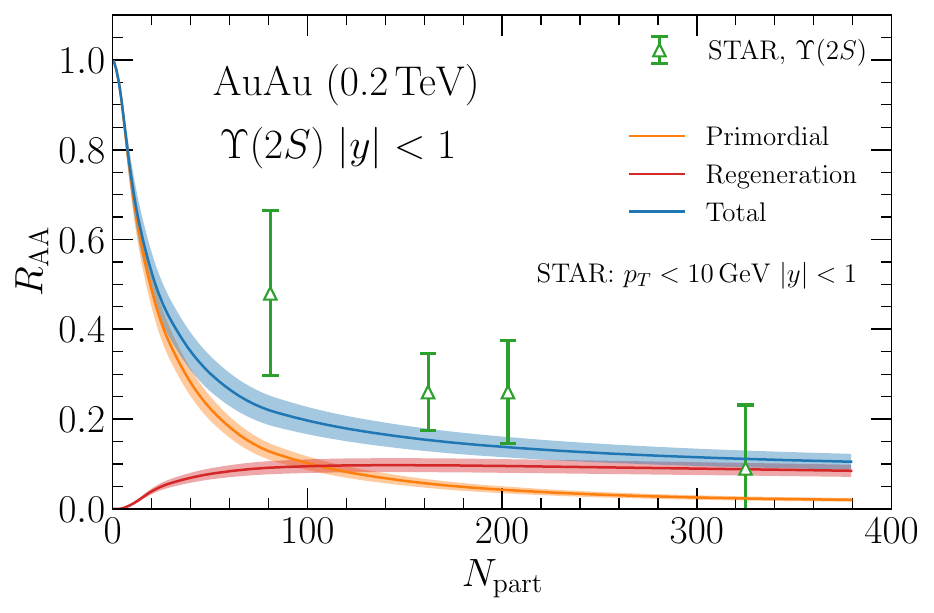 }
       \hfill
       \includegraphics[width=0.48\textwidth]{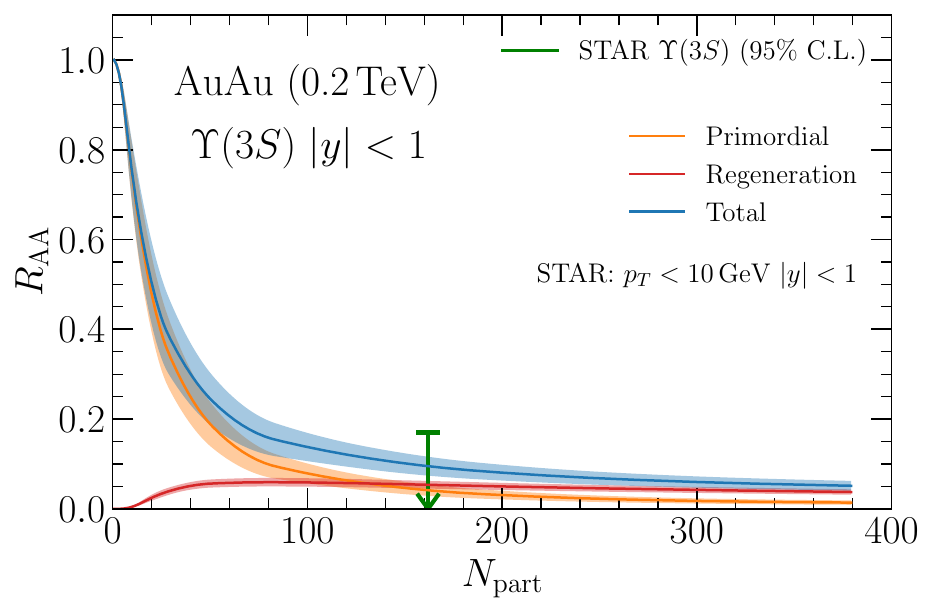 }
       \hfill
        \includegraphics[width=0.48\textwidth]{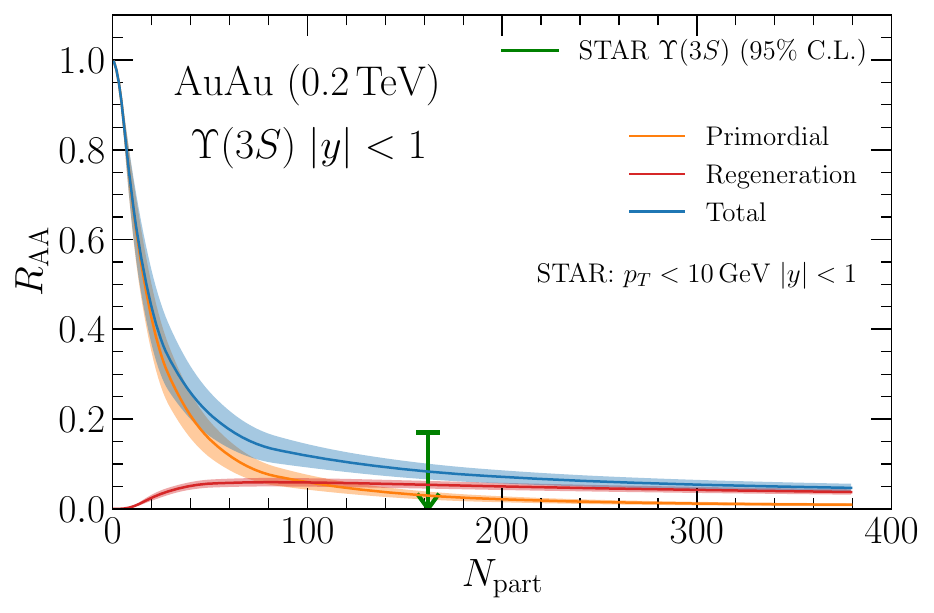 }
        \caption{Centrality-dependent $\raa$ of $\Upsilon(1S)$ (top row), $\Upsilon(2S)$ (middle row), and $\Upsilon(3S)$ (bottom row) in Au-Au (200\,GeV) collisions at mid-rapidity without nuclear absorption (left column) and with a nuclear absorption cross section of $\sigma_{\rm abs}=3\,$mb (right column).
        The bands represent an up to 10-30\% shadowing.
        }
        \label{fig:rhic_npart}
     \end{figure}
We finally turn to our results for Au-Au collisions at RHIC, starting with the centrality dependence of the three $\Upsilon$ states
shown in Fig.~\ref{fig:rhic_npart}. Overall, one notices that both suppression and regeneration are less pronounced than at the LHC, which is simply due to the cooler and shorter-lived fireball (containing roughly half of the entropy compared to LHC conditions). As a result, the primordial production remains the dominant source for $\Upsilon(1S)$ at all centralities, while for the excited states regeneration takes over in mid-central collisions and becomes the dominant source in central collisions. Focusing first on a scenario without nuclear absorption (left column in Fig.~\ref{fig:rhic_npart}), the total $1S$ $\raa$ remains above the STAR data. In central collisions, the data level off at $\sim$0.35, which is only slightly larger than for the LHC data at $\sim$0.3. The theoretical results give 0.35 plus 0.1 from primordial and regeneration sources, respectively, at RHIC, compared to 0.1 and 0.2 at the LHC, \ie, a good part of the much stronger suppression at the LHC is recovered from regeneration.
\begin{figure}[thb]
   \includegraphics[width=0.48\textwidth]{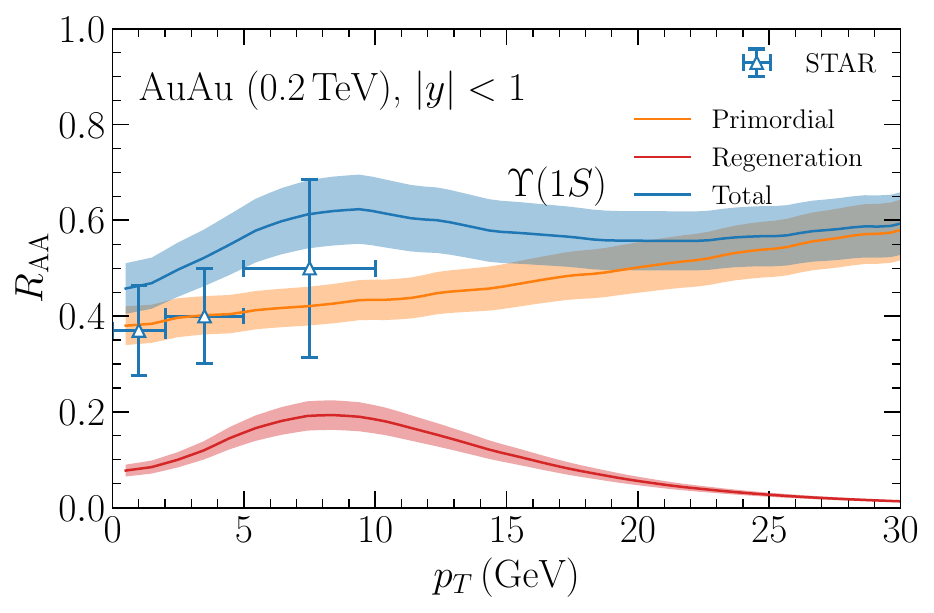 }
   \hfill
   \includegraphics[width=0.48\textwidth]{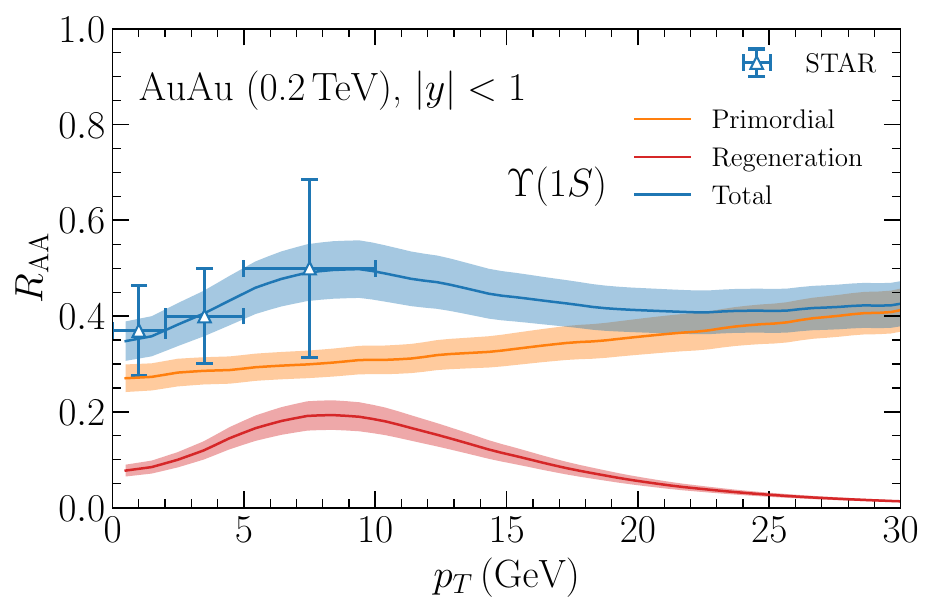 }
   \hfill
   \includegraphics[width=0.48\textwidth]{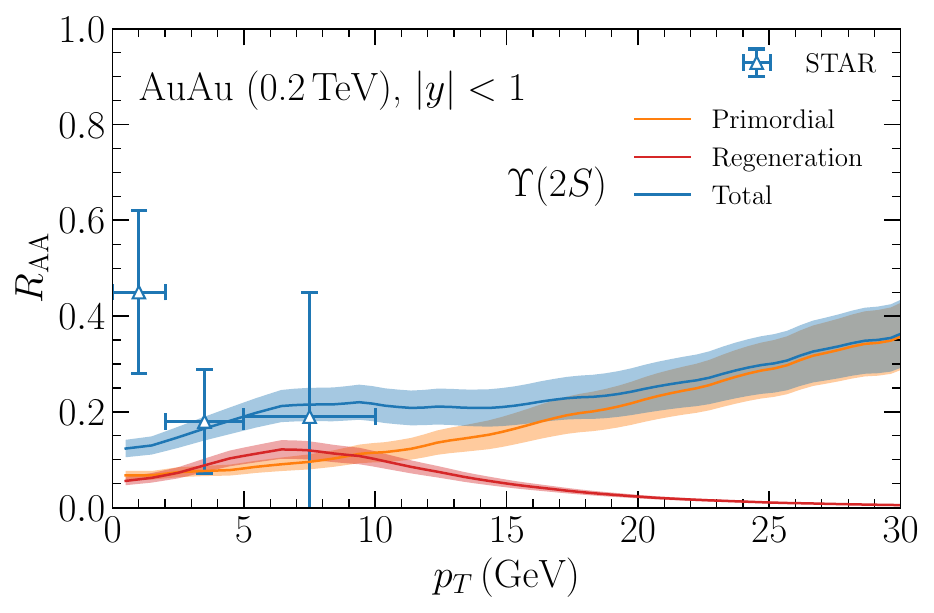 }
   \hfill
   \includegraphics[width=0.48\textwidth]{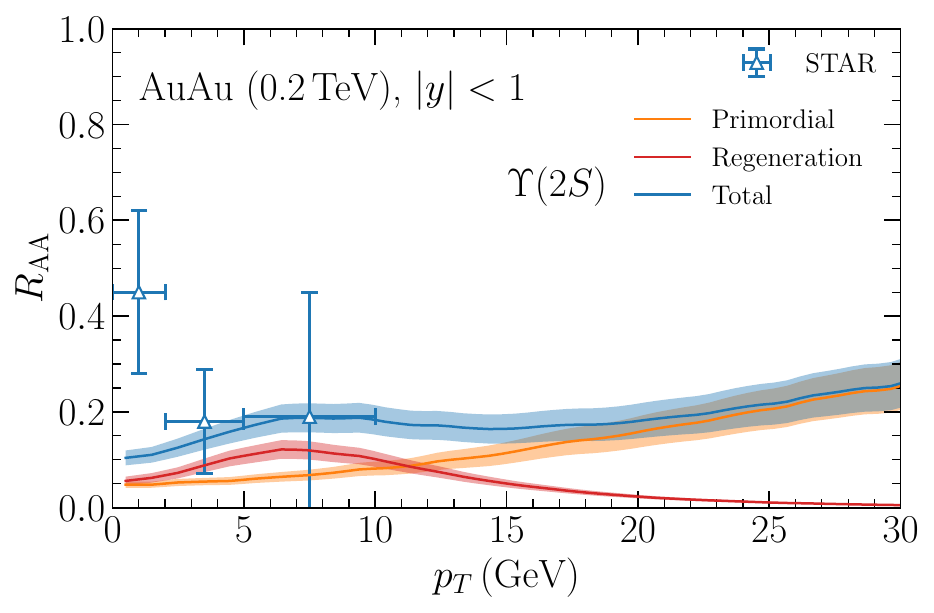 }
   \hfill
   \includegraphics[width=0.48\textwidth]{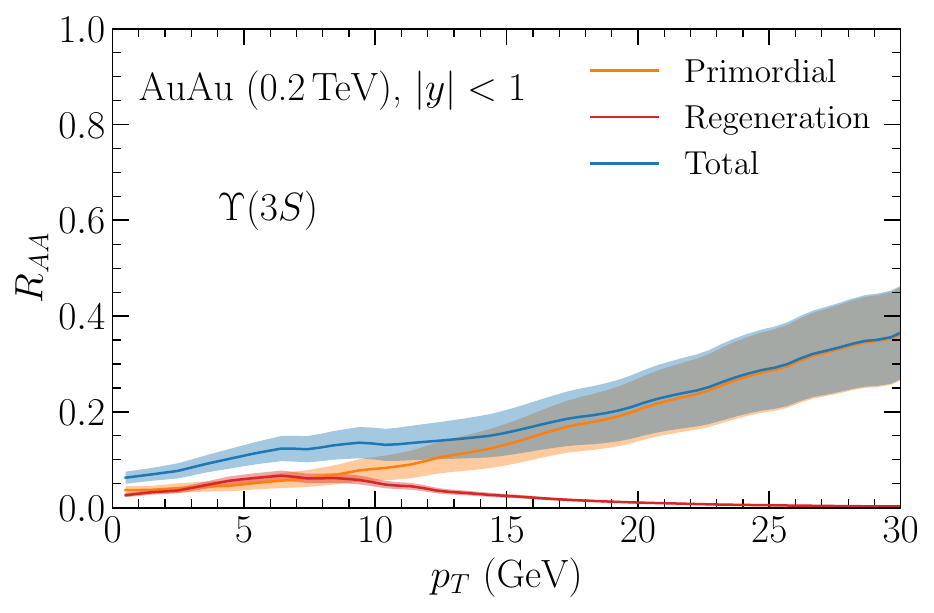 }
   \hfill
   \includegraphics[width=0.48\textwidth]{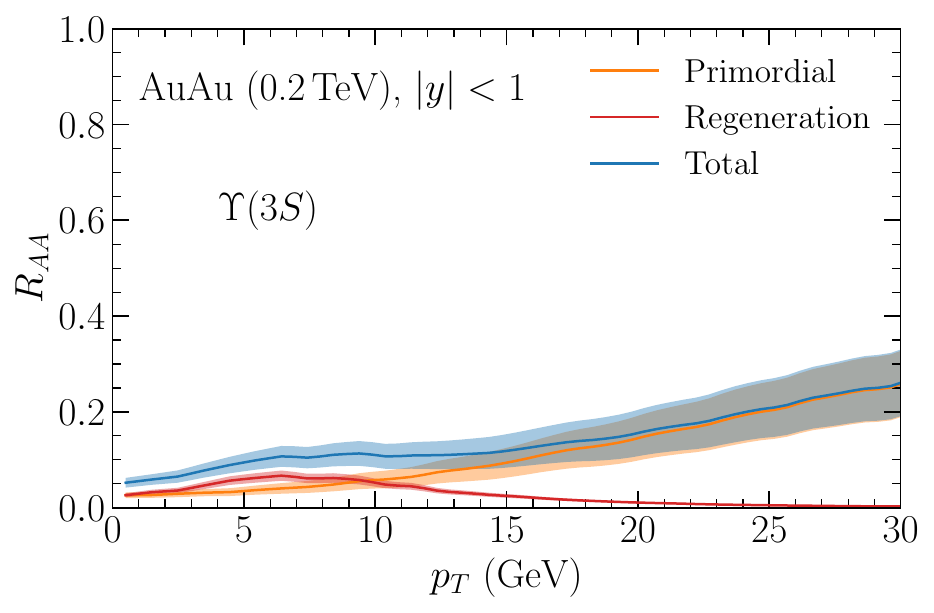 }
   \caption{
       $\raa$ of $\Upsilon(1S)$ (top row), $\Upsilon(2S)$ (middle row), and $\Upsilon(3S)$ (bottom row)
             without nuclear absorption (left) and with nuclear absorption ($\sigma_{\rm abs}=3\,$mb; right panels) as a function of $\pT$.
           The bands represent 10\%-30\% shadowing.
   }
\label{fig:rhic_pt}
\end{figure}
The situation further improves when allowing for a moderate nuclear absorption at RHIC (which is, as we argued above, compatible with p+Au and d+Au data), reducing the primordial $\raa$ by about 0.11 and resulting in a good overall agreement with the STAR data, and thus resolving the apparent puzzle of why the $\Upsilon(1S)$ $\raa$ is comparable at RHIC and the LHC, despite the much hotter medium at the latter. In this interpretation, it is caused by an approach toward the equilibrium limit, rather than a ``temperature-sequential'' suppression as the collision energy and temperature increase from RHIC to the LHC. The immediate prediction is that this trend continues to still higher collision energies and eventually leads to further enhancement when recombination from multiple $\bbb$ pairs (grand-canonical limit) in the fireball takes over.
On the other hand, the large rates for the excited states are already operative at the lower temperatures created at RHIC and thus lead to strong suppression and a relatively small but dominant regeneration component, similar to the LHC.

Moving on to the transverse-momentum dependence, a weak signature of regeneration appears in our results for the $\Upsilon(1S)$ $\raa$ as a rising trend at low $\pT$ (even a shallow maximum near $\pT\simeq 10$\,GeV), which is not inconsistent with data. The role of nuclear absorption appears to be less significant than in the centrality dependence.  The regeneration signature appears to be less pronounced for the excited states.

\section{Conclusion}
\label{sec:concl}

We have conducted a comprehensive study of bottomonium production in Pb-Pb ($5.02\,\mathrm{TeV}$) and Au-Au ($200\,\mathrm{GeV}$) collisions
by coupling a kinetic rate equation to a (3+1)D quasiparticle anisotropic hydrodynamic model. This framework improves on previous calculations, where a more schematic fireball evolution and/or perturbative $Y$ reaction rates have been used.

By decomposing the general solution to the rate equation into a sum of the homogeneous (representing suppression of primordial $Y$ production) and inhomogeneous solutions (which is driven by regeneration), we have evaluated the former component by sampling $Y$ eikonal trajectories through the hydrodynamically evolving QGP, utilizing recently developed nonperturbative reaction rates which were constrained by lattice-QCD data through quantum many-body theory (thermodynamic $T$-matrix approach). 
Initial conditions have been implemented via an optical Glauber model and cold-nuclear matter effects (such as nuclear shadowing and absorption) and followed by quantum-formation time effects (estimated through a reduced rate related to an expanding wave packet), while final-state feed-down from excited states has been based on experimental branching ratios.

Bottomonium regeneration has been evaluated with the same reaction rates as the suppression (as dictated by detailed balance), but the calculation of the equilibrium limit -- as the second main transport coefficient -- required an extension that accounts for the temperature gradients present in the hydrodynamic medium at any given proper time. To this end, we have computed the total entropy contained within the thermodynamically active medium to evaluate the thermal-equilibrium number for time-dependent trajectory-averaged temperatures. Further corrections included correlation-volume effects arising from the initial point-like pair production, thermal relaxation times associated with the diffusion of $b$ quarks,  as well as their escape from the fireball. In addition, the same melting temperatures (defining the onset of regeneration) and formation-time delays as in the suppression part have been incorporated.

While significant uncertainties in our framework persist, its major components, \ie, the nonperturbative reaction rates and the hydrodynamic bulk medium evolution, do not involve tunable parameters. Moreover, these components constitute a mutually consistent implementation of the strongly coupled QGP where the microscopic reaction rates are evaluated in an interacting medium with a universal potential that, in turn, can describe the lattice-QCD equation of state used in the hydrodynamic evolution.
Our results at the LHC suggest that regeneration contributes substantially to the total yields and transverse-momentum spectra
of $\Upsilon(1S)$, $\Upsilon(2S)$, and $\Upsilon(3S)$, and significantly affects their double ratios.
The computed nuclear modification factors as a function of transverse momentum and collision centrality are in fair agreement with measurements reported by the CMS, ATLAS, and ALICE collaborations, with a thus far unresolved tension at transverse momenta larger than the $Y$ mass, for all states.
On the other hand, the rather large regeneration caused by the large reaction rates offers a possible resolution of the $\Upsilon(1S)$ puzzle at RHIC, \ie, a suppression level very similar to that at the LHC: significantly less regeneration, together with a moderate nuclear absorption cross section 
results in a reasonable description of the $\raa$  of the observed $\Upsilon$ states at RHIC.

A number of further developments are in order. The schematic thermal relaxation approximation for $b$ quark thermalization should be improved by implementing a coupled transport approach between open and hidden heavy-flavor states, along the lines of the calculation carried out in Ref.~\cite{Fu:2026ous} in the charm/onium sector. Part of the challenge will be to achieve this in a hydrodynamic background. A more accurate calculation of the equilibrium limit is desirable, utilizing more direct constraints from lattice QCD. The discrepancies in our current calculations at high $\pT$ need to be addressed. Whether this can be achieved through a more explicit quantum evolution or rather through a change in the production mechanism, such as gluon fragmentation, is presently not clear. Work in these directions is in progress.

\section*{Acknowledgments}
We thank S.~Thapa and R.~Vogt for helpful comments on the manuscript.
This work is supported by the U.S. National Science Foundation under grant no. PHY-2514775 and the Topical Collaboration in Nuclear Theory on \textit{Heavy-Flavor Theory (HEFTY) for QCD Matter} under award no.~DE-SC0023547.

\bibliographystyle{JHEP}
\bibliography{refcnew}

\end{document}